# SAT.STFR.FRQ (UWA) DETAIL DESIGN REPORT (LOW)

S.W. Schediwy: sascha.schediwy@uwa.edu.au
D.R. Gozzard: david.gozzard@research.uwa.edu.au

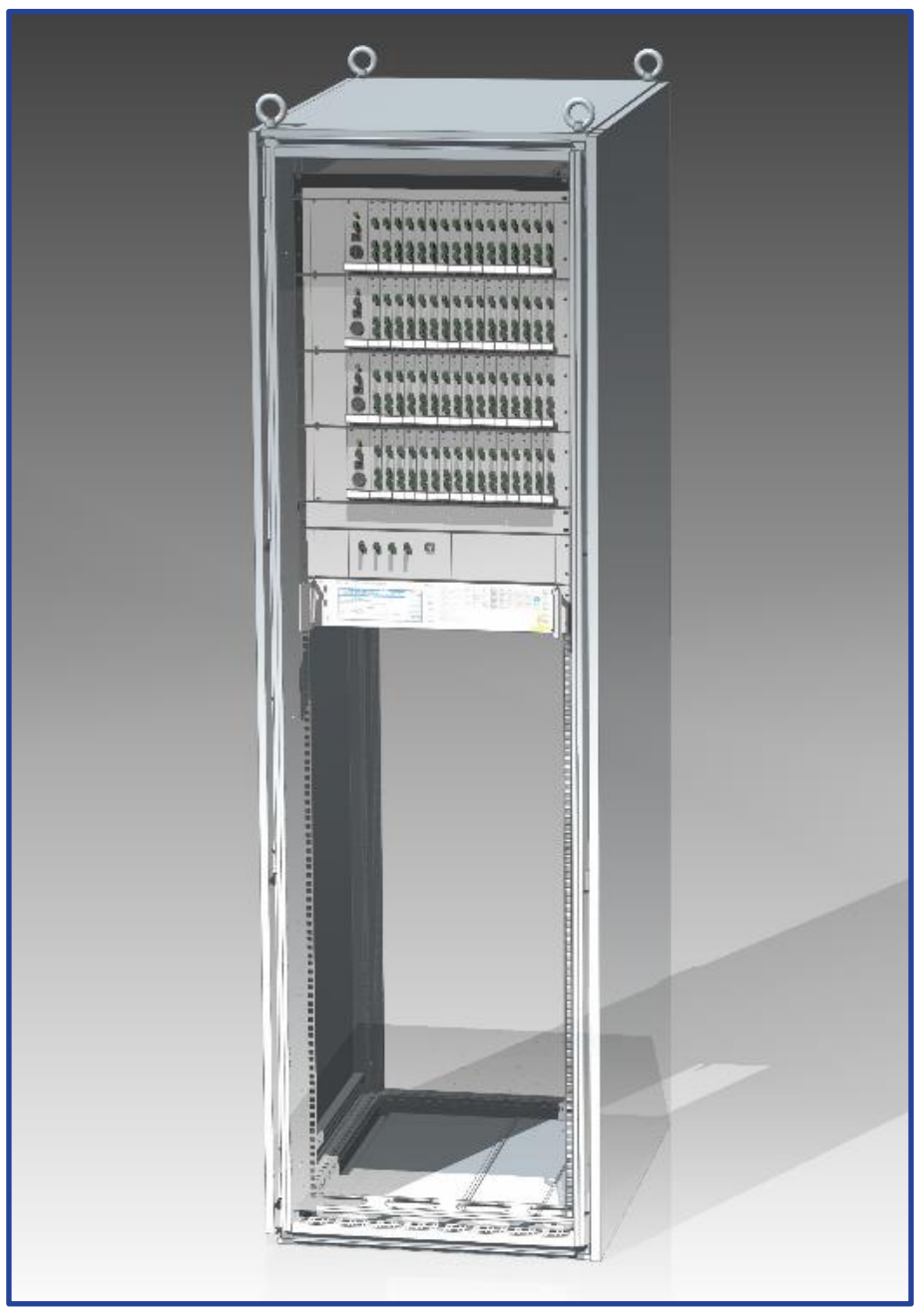
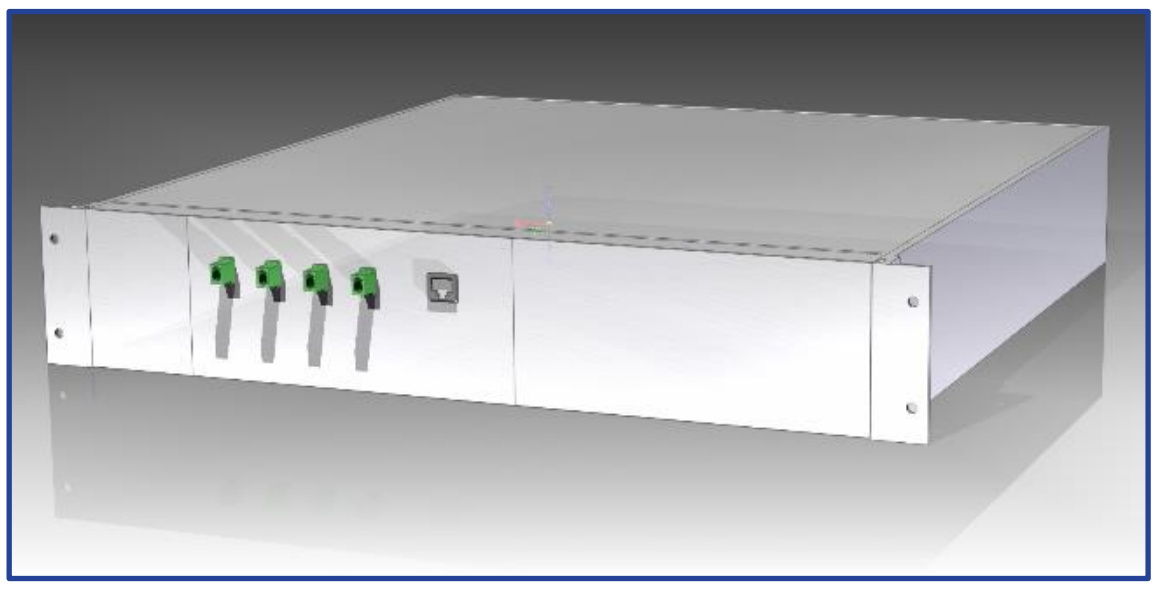
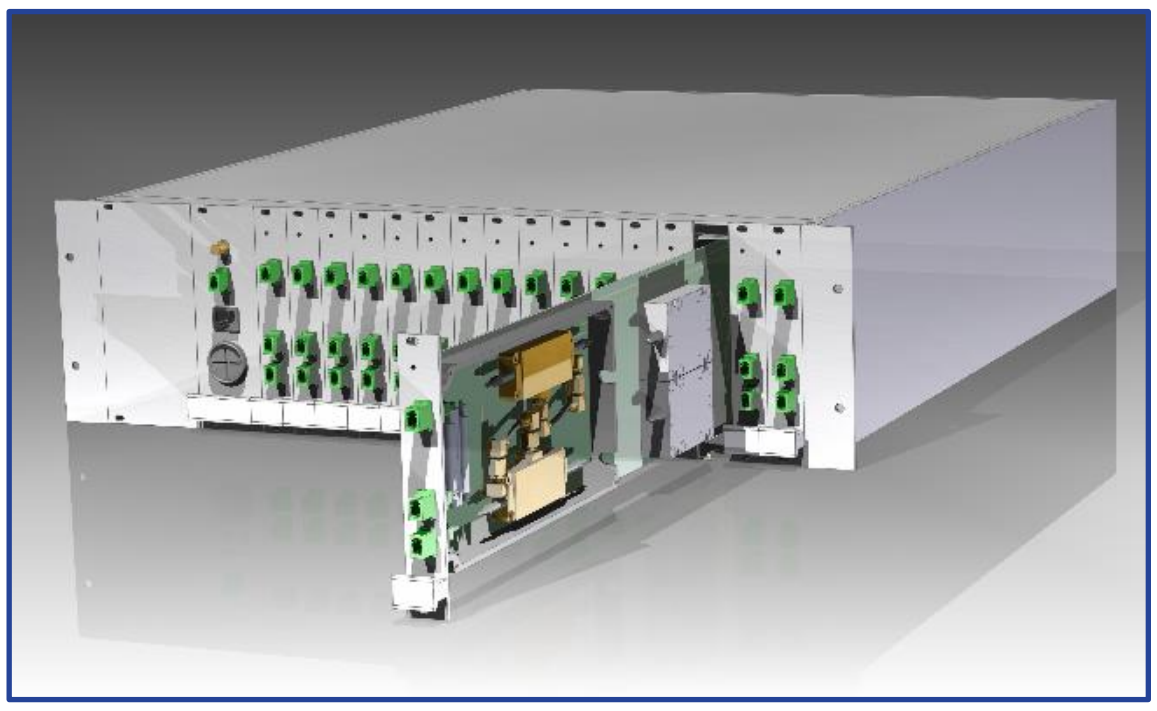
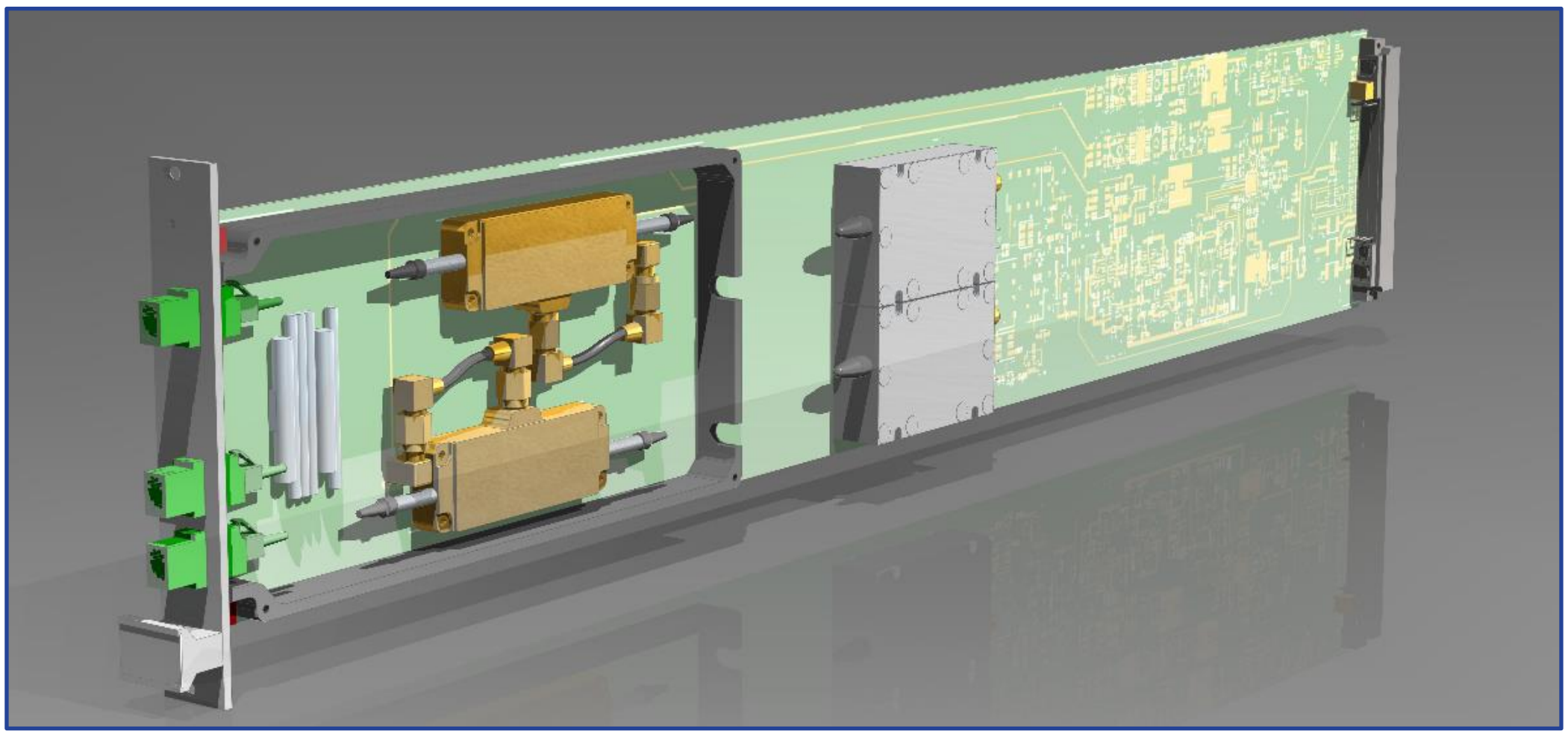

THE UNIVERSITY OF WESTERN AUSTRALIA

International Centre for Radio Astronomy Research

# SAT.STFR.FRQ (UWA) DETAIL DESIGN REPORT (LOW)

| | |
|---|---|
| Document Number | SKA-TEL-SADT-0000380-SADT.STFR.FRQ-DDD-01 |
| Revision | [Comments] |
| Author | Sascha Schediwy;David Gozzard |
| Date | 2018-08-10 |
| Status | Issued |

| Name | Designation | Affiliation | Date | Signature |
|---|---|---|---|---|
| Owned by: | | | | |
| Sascha Schediwy | SADT.SAT.STFR. FRQ (UWA) Lead Designer | University of Western Australia | 2018-08-10 | |
| Approved by: | | | | |
| Jill Hammond | SADT Project Manager | University of Manchester | 2018-08-10 | |
| Released by: | | | | |
| Rob Gabrielczyk | SADT System Engineer | University of Manchester | 2018-08-10 | |

## DOCUMENT HISTORY

| Revision | Date Of Issue | Engineering Change Number | Comments |
|---|---|---|---|
| 01 | 2018-08-10 | - | First release. |

## DOCUMENT SOFTWARE

| | Package | Version | Filename |
|---|---|---|---|
| Wordprocessor | Word | 2013 | SKA-TEL-SADT-0000380-SADT.STFR.FRQ-DDD-01-SAT.STFR.FRQ (UWA) Detail Design Report (LOW).docx |
| Block diagrams | PowerPoint | 2013 | |
| Other | SolidEdge<br>DesignSpark<br>Excel<br>Java | ST4<br>8.0<br>2013<br>SE 8 | |

## ORGANISATION DETAILS

| **Name** | University of Manchester |
|---|---|
| Registered Address | Jodrell Bank Centre for Astrophysics<br>Alan Turing Building<br>The University of Manchester Oxford Road<br>Manchester, UK<br>M13 9PL |
| Fax. | +44 (0)161 275 4247 |
| Website | www.manchester.ac.uk |

## INTELLECTUAL PROPERTY DISCLAIMER

The design work for this phase synchronisation system was performed by the UWA as a member of the SaDT consortium contributing to the SKA project and is subject to the SKA IP policy.

TABLE OF CONTENTS

LIST OF FIGURES

LIST OF TABLES

# 1 EXECUTIVE SUMMARY

The Square Kilometre Array (SKA) project [RD1] is an international effort to build the world's most sensitive radio telescope operating in the 50 MHz to 14 GHz frequency range. Construction of the SKA has been divided into phases, with the first phase (SKA1) accounting for the first 10% of the telescope's receiving capacity. During SKA1, a low-frequency aperture array comprising over a hundred thousand individual dipole antenna elements will be constructed in Western Australia (SKA1-low), while an array of 197 parabolic-dish antennas, incorporating the 64 dishes of MeerKAT, will be constructed in South Africa (SKA1-mid).

Radio telescope arrays such as the SKA require phase-coherent reference signals to be transmitted to each antenna site in the array. In the case of the SKA, these reference signals will be generated at a central site and transmitted to the antenna sites via fibre-optic cables up to 175 km in length [RD2]. Environmental perturbations affect the optical path length of the fibre and act to degrade the phase stability of the reference signals received at the antennas, which has the ultimate effect of reducing the fidelity and dynamic range of the data [RD3]. Given the combination of long fibre distances and relatively high frequencies of the transmitted reference signals, the SKA will need to employ actively-stabilized frequency transfer technologies to suppress the fibre-optic link noise [RD4] in order to maintain phase-coherence across the array.

Since 2011, researchers at the University of Western Australia (UWA) have led the development of an *SKA phase-synchronisation system* designed specifically to meet the scientific needs and technical challenges of the SKA telescope. This system [RD5] is based on the transmission of actively stabilised phase-coherent reference signals generated at the central processing facility (CPF), and then transmitted via separate optical fibre links to each antenna site. The frequency transfer technique at the core of UWA's *SKA phase-synchronisation system* is an evolution of Atacama Large Millimeter Array's distributed 'photonic local oscillator system' [RD6], incorporating key advances made by the international frequency metrology community over the last decade [RD7-9], as well as novel innovations developed by UWA researchers [RD10, 11].

Two variants of UWA's *SKA phase-synchronisation system* have been designed, each one optimised specifically for its respective telescope. For SKA1-mid, the required microwave (MW) shift is generated using a dual-parallel Mach-Zehnder modulator (DPM), biased to generate single-sideband suppressed-carrier (SSB-SC) modulation [RD10]; while for SKA1-low the radio-frequency (RF) shift is generated using a simpler acousto-optic modulator (AOM) [RD11]. This results in two systems that easily meet the SKA functional performance requirements, as demonstrated by laboratory testing [RD10-13], overhead fibre field trials [RD14-16], and astronomical verification [RD17-19], yet maximise robustness and maintainability while keep complexity and costs to a minimum.

The key innovation of UWA's *SKA phase-synchronisation system* was finding a way to use AOMs as servo-loop actuators for radio- and microwave-frequency transfer [RD10, 11]. The large servo bandwidth and infinite feedback range these servo-loop AOMs ensures that the stabilisation system servo-loops never require integrator resets. UWA's *SKA phase-synchronisation system* also utilises AOMs to generate static frequency shifts at the antenna sites to mitigate against unwanted reflections that are inevitably present on real-world links. Reflection mitigation is absolutely essential for the *SKA phase-synchronisation system*, as there is no way to guarantee that all links will remain completely free of reflections over the lifetime of the project.

UWA's *SKA phase-synchronisation system* has the servo-loop electronics and the vast majority of all other optical and electronic components located at the CPF, greatly simplifying maintenance. A single high-quality frequency synthesiser, tied to the SKA master clock, is used to generate phase coherent reference signals, and these are distributed to the Transmitter Modules which are then used to transmit the optical signals across each fibre link. The Transmitter Modules incorporate the servo-loop AOMs, and these are able to add an independent and unique RF-scale frequency offset – in the optical domain – to the common transmission frequency for each link. This avoids any possibility of common frequencies at each antenna site to ensure any stray RF emissions will not be coherent if picked up by the receivers.

The Receiver Modules for UWA's *SKA phase-synchronisation system* have a very small form-factor and contain only a minimum number of simple optical and analogue electronic components, making them extremely robust to external environmental perturbation. In addition, they are designed to be capable of being mounted directly on the SKA1-mid antenna indexer alongside the receiver. Currently, the SADT interface with DISH is in the antenna pedestal, and the DISH Consortium are required to build a second frequency transfer system to transmit the reference signals up the cable wraps to the indexer. After a successful down-select, the DISH consortium and SKA Office have agreed to an engineering change request (ECP) to correct this inefficiency.

A small form -factor, industry standard, oven-controlled crystal oscillator (OCXO) is incorporated into the Receiver Module to provide phase coherence at timescales shorter than the light round-trip time of the fibre link. The OCXO is tied to the incoming reference signals using a simple, encapsulated phase-locked loop based on the proven design implemented by the Australian SKA Pathfinder (ASKAP). This is particularly important, as it has been shown that using multiple microwave-frequency synthesisers can easily lead to a significant loss of coherence, even if the transmission frequency is being successfully stabilised [RD18, 20].

UWA's *SKA phase-synchronisation system* is designed in such a way as to also stabilise the non-common optical fibre paths in the Transmitter Modules. This effectively stabilises the Transmitter Modules at the same time as the fibre link, making the equipment in the CPF extremely robust to external environmental changes. The optical phase sensing allows for the use of Faraday mirrors to give maximum detected signal at the servo photodetector without requiring any initial polarisation alignment, or any ongoing polarisation control or polarisation scrambling.

UWA's *SKA phase-synchronisation system* has been extensively tested using standard metrology techniques in a laboratory setting [RD12, 13], with signals transmitted over metropolitan fibre links and fibre spools under all required conditions; on 186 km of overhead fibre at the South African SKA site [RD14-16]; as well as astronomical verification with the Australian Telescope Compact Array (ATCA) for SKA1-mid [RD18, 20], and the ASKAP for SKA1-low [RD19]. This has demonstrated that UWA's *SKA phase-synchronisation system* is fully compliant with all SKA requirements, as well as demonstrating functionality of critical practical factors that are not captured by these requirements.

Furthermore, UWA researchers in partnership with MeerKAT and University of Manchester engineers, have developed the detailed designs into a set of mass-manufacture archetypes, effectively getting a head-start at addressing manufacturing issues that may be encounter by contractors during SKA Construction. The first set of mass-manufacture archetypes for SKA1-low were completed in Q2, 2016 [RD21]; and for SKA1-mid in Q1, 2017 [RD22]. All aspects of the mass-manufacture design are openly available and are provided with sufficient detail so that any firm with expertise in optical and electronic assembly can to reproduce these systems with minimal domain expert input. An optical technology consultancy firm was employed to provide an independent review of the labour costs associated with assembly and testing (see Appendix 7.9.5).

All sub-elements of UWA's *SKA phase-synchronisation system* have been designed to be hot-swappable, enabling simple installation and easy maintainability (especially as the vast majority are located at the CPF). The system is designed so that during commissioning, only one free parameter needs to be optimised per link.

The detailed design presented in this document has been critically assesses by three independent domain experts from the ASTRON Netherlands Institute for Radio Astronomy, the Jet Propulsion Laboratory, and Square Kilometre Array South Africa. The review has built confidence in the detailed design and ensured that UWA's *SKA phase-synchronisation system* is the best possible phase synchronisation solution for the SKA telescope.

# 2 INTRODUCTORY SECTIONS

## 2.1 Purpose

This document provides the detailed design of *SKA phase synchronisation system* for the SKA1-low telescope developed by the University of Western Australia (UWA). The design is used to derive a cost model for deploying the frequency dissemination equipment on both telescopes.

The detailed design report concerns the following aspects:

- Solution description — including the technical background, solution overview, and solution design justification, along with details of the hardware and software, safety and security, integration, interoperability, and costing;
- Evaluation — including testing and verification of the design, construction of a mass manufacture archetype, and independent assessment of solution;
- Conclusion, and recommendations for further development and procurement; and
- Statement of Compliance — indicating the system compliance with each of the SKA requirements.

The detailed design of *SKA phase synchronisation system* reflects the current baseline (Rev 2) of the SKA Programme [AD1] and Level 1 requirements as per Revision 10 [AD2] at the time of writing.

## 2.2 Scope

This report describes the full extent of UWA's *SKA phase synchronisation system* within the SKA's Signal and Data Transport (SADT) consortium's Synchronisation and Timing (SAT) network. The system receives an electronic reference signal from SAT.CLOCKS at the SKA1-low Central Processing Facility (CPF), and transfers the full stability of the reference signal across the SAT network to each Remote Processing Facility (RPF). At the RPF, an electronic copy of the reference signal is provided to LFAA. UWA's *SKA phase synchronisation system* is controlled and monitored using SAT.LMC with the required local infrastructure provided by SADT.LINFRA.

## 2.3 Intended Audience

This design report is to be used within the SADT consortium, by the SKAO, and other design consortia within SKA. It will form part of the Body of Evidence for the SADT Consortium down-select process of the *SKA phase synchronisation system*. If this design is selected, this report will form part of the documentation for the SADT Consortium Critical Design Review (CDR).

## 2.4 Applicable Documents

[AD1] P. Dewdney, *SKA1 System Baseline (v2) Description*. SKA Organisation, 2015. **SKA-TEL-SKO-0000308** (rev. 01).

[AD2] W. Turner. *SKA Phase 1 System (Level 1) Requirements Specification*. SKA Organisation, 2016. **SKA-TEL-SKO-0000008** (rev. 10).

[AD3] A. Wilkinson and M. Pearson. *STFR Frequency Dissemination System Down-Select Methodology*. SKA Signal and Data Transport Consortium, 2017. **SKA-TEL-SADT-0000524** (rev. 2.0): p. 31.

## 2.5 Reference Documents

[RD1] P. Dewdney, *SKA1 System Baseline V2 Description*. SKA Organisation, 2015. **SKA-TEL-SKO-0000308** (rev. 1): p. 58.

[RD2] R. Oberland, *NWA Model SKA1-MID*. SKA Signal and Data Transport Consortium, 2017. **SKA-TEL-SADT-0000523** (rev. 3.0): p. 14.

[RD3] J.F. Cliche and B. Shillue, *Precision timing control for radioastronomy: maintaining femtosecond synchronization in the Atacama Large Millimeter Array.* Control Systems, IEEE, 2006. **26**(1): p. 19-26.

[RD4] K. Grainge, et al., *Square Kilometre Array: The radio telescope of the XXI century.* Astronomy Reports, 2017. **61**(4): p. 288-296.

[RD5] S.W. Schediwy, et al., *A Phase Synchronization System For The Square Kilometre Array.* The Astronomical Journal, 2017.

[RD6] B. Shillue, S. AlBanna, and L. D'Addario. *Transmission of low phase noise, low phase drift millimeter-wavelength references by a stabilized fiber distribution system*. in *Microwave Photonics, 2004. MWP'04. 2004 IEEE International Topical Meeting on*. 2004.

[RD7] S.M. Foreman, et al., *Remote transfer of ultrastable frequency references via fiber networks.* Review of Scientific Instruments, 2007. **78**(2): p. 021101-25.

[RD8] O. Lopez, et al., *High-resolution microwave frequency dissemination on an 86-km urban optical link.* Applied Physics B: Lasers and Optics, 2010. **98**(4): p. 723-727.

[RD9] K. Predehl, et al., *A 920-Kilometer Optical Fiber Link for Frequency Metrology at the 19th Decimal Place.* Science, 2012. **336**(6080): p. 441-444.

[RD10] S.W. Schediwy, et al., *Stabilized microwave-frequency transfer using optical phase sensing and actuation.* Optics Letters, 2017. **42**(9): p. 1648-1651.

[RD11] S.W. Schediwy, et al., *Simple Stabilized Radio-Frequency Transfer with Optical Phase Actuation.* Photonics Technology Letters, IEEE, 2017.

[RD12] S.W. Schediwy and D.G. Gozzard, *Pre-PDR Laboratory Verification of UWA's SKA Synchronisation System*. SKA Signal and Data Transport Consortium, 2014. **SKA-TEL-SADT-0000616**: p. 16.

[RD13] S. Schediwy and D. Gozzard, *Pre-CDR Laboratory Verification of UWA's SKA Synchronisation System*. SKA Signal and Data Transport Consortium, 2017. **SKA-TEL-SADT-0000620**: p. 13.

[RD14] S.W. Schediwy and D.G. Gozzard, *UWA South African SKA Site Long-Haul Overhead Fibre Field Trial Report*. SKA Signal and Data Transport Consortium, 2015. **SKA-TEL-SADT-0000109**: p. 20.

[RD15] D.R. Gozzard, et al., *Characterization of optical frequency transfer over 154 km of aerial fiber.* Optics Letters, 2017. **42**(11): p. 2197-2200.

[RD16] D.R. Gozzard, et al., *Stabilized Modulated Photonic Signal Transfer Over 186 km of Aerial Fiber.* Transactions on Ultrasonics, Ferroelectrics, and Frequency Control, 2017.

[RD17] D.R. Gozzard, et al., *Astronomical verification of a stabilized frequency reference transfer system for the Square Kilometre Array* The Astronomical Journal, 2017.

[RD18] D. Gozzard and S. Schediwy, *SKA-mid Astronomical Verification*. SKA Signal and Data Transport Consortium, 2016. **SKA-TEL-SADT-0000524** (rev. 3.1): p. 29.

[RD19] S. Schediwy and D. Gozzard, *SKA-low Astronomical Verification*. SKA Signal and Data Transport Consortium, 2015. **SKA-TEL-SADT-0000617**: p. 21.

[RD20] D.R. Gozzard, et al., *Astronomical Verification of a Stabilized Frequency Reference Transfer System for the Square Kilometer Array.* The Astronomical Journal, 2017. **154**(1): p. 9.

[RD21] S. Schediwy, et al., *SKA Phase Synchronisation: Design for Mass Manufacture*. 2016: p. 1.

[RD22] S. Stobie, S.W. Schediwy, and D.R. Gozzard, *Design-for-Manufacture of the SKA1-Mid Frequency Synchronisation System*. SKA Signal and Data Transport Consortium, 2017. **SKA-TEL-SADT-0000618** (rev. 1): p. 57.

[RD23] S.W. Schediwy, *A Clock for the Square Kilometre Array.* Proceedings of Science, 2013. **170**: p. 13.

[RD24] S.W. Schediwy and D.R. Gozzard, *Simultaneous transfer of stabilized optical and microwave frequencies over fiber.* IEEE Photonics Technology Letters, 2017.

[RD25] D. Gozzard, *Notes on Calculating the Relationship between Coherence Loss and Allan Deviation*. SKA Signal and Data Transport Consortium, 2017. **SKA-TEL-SADT-0000619** (rev. 1): p. 5.
[RD26] R. Oberland, *NWA Model SKA1-LOW*. SKA Signal and Data Transport Consortium, 2017. **SKA-TEL-SADT-0000522** (rev. 3.0): p. 10.
[RD27] S.T. Garrington, et al. *e-MERLIN*. 2004.
[RD28] R. McCool, et al., *Phase transfer for radio Astronomy interferometers, over Installed fiber networks, using a roundtrip correction system.* 40th Annual Precise Time and Time Interval (PTTI) Meeting, 2008.
[RD29] A. Wootten and A.R. Thompson, *The Atacama Large Millimeter/Submillimeter Array.* Proceedings of the IEEE, 2009. **97**(8): p. 1463-1471.
[RD30] S. Schediwy and D. Gozzard, *SAT.STFR.FRQ (UWA) Detail Design Report (MID)*. SKA Signal and Data Transport Consortium, 2017. **SKA-TEL-SADT-0000390**.
[RD31] S. Schediwy, *A Clock for the Square Kilometre Array.* Proceedings of Science, 2013. **170**: p. 1-13.
[RD32] B. Wang, et al., *Square Kilometre Array Telescope—Precision Reference Frequency Synchronisation via 1f-2f Dissemination.* Scientific Reports, 2015. **5**: p. 13851.
[RD33] L. Primas, G. Lutes, and R. Sydnor. *Fiber optic frequency transfer link*. in *Proceedings of the 42nd Annual Frequency Control Symposium, 1988.* 1988.
[RD34] Y. He, et al., *Stable radio-frequency transfer over optical fiber by phase-conjugate frequency mixing.* Optics Express, 2013. **21**(16): p. 18754-18764.
[RD35] L.-S. Ma, et al., *Delivering the same optical frequency at two places: accuratecancellation of phase noise introduced by an optical fiber or other time-varyingpath.* Optics Letters, 1994. **19**(21): p. 1777-1779.
[RD36] S.T. Dawkins, J.J. McFerran, and A.N. Luiten, *Considerations on the measurement of the stability of oscillators with frequency counters.* Ultrasonics, Ferroelectrics and Frequency Control, IEEE Transactions on, 2007. **54**(5): p. 918-925.
[RD37] P. Lesage, *Characterization of Frequency Stability: Bias Due to the Juxtaposition of Time-Interval Measurements.* IEEE Transactions on Instrumentation and Measurement, 1983. **32**(1): p. 204-207.
[RD38] O. Lopez, et al., *Frequency and time transfer for metrology and beyond using telecommunication network fibres.* Comptes Rendus Physique, 2015. **16**(5): p. 531-539.
[RD39] W. Shillue, et al. *The ALMA Photonic Local Oscillator system*. in *General Assembly and Scientific Symposium, 2011 XXXth URSI*. 2011.
[RD40] B. Alachkar, P. Boven, and A. Wilkinson, *SKA1 Level1 Synchronisation and Timing Requirements Analysis and Verification*. SKA Signal and Data Transport Consortium, 2017. **SKA-TEL-SADT-0000499** (rev. 0.1): p. 25.
[RD41] M.M. Yamada, et al., *Phase Stability Measurement of an Optical Two-Tone Signal Applied to a Signal Reference Source for Millimeter and Sub-Millimeter Wave Interferometer.* Publications of the Astronomical Society of Japan, 2006. **58**(4): p. 787-791.
[RD42] N. Kawaguchi, *Coherence Loss and Delay Observation Error In Very-Long-Baseline Interferometry.* J. Radio Research Laboratories, 1983. **30**(129): p. 59-87.
[RD43] H. Kiuchi, *Coherence Estimation for Measured Phase Noise In Allan Standard Deviation*. National Radio Astronomy Observatory, 2005. **ALMA memo 530**: p. 12.
[RD44] A.R. Thompson, J.M. Moran, and G.W. Swenson, *Interferometry and Synthesis in Radio Astronomy*. 2 ed. 2008, Hoboken, NJ: Wiley.
[RD45] P.A. Williams, W.C. Swann, and N.R. Newbury, *High-stability transfer of an optical frequency over long fiber-optic links.* J. Opt. Soc. Am. B, 2008. **25**(8): p. 1284-1293.
[RD46] R. Gabrielczyk, *SADT-DSH_ICD*. SKA Signal and Data Transport Consortium, 2017. **300-0000000-026_02**.
[RD47] S. Gregory, *SADT SAT Network Integration and Test Report*. SKA Signal and Data Transport Consortium, 2017. **SKA-TEL-SADT-0000579** (rev. 0.8): p. 73.
[RD48] S. Droste, et al., *Optical-Frequency Transfer over a Single-Span 1840 km Fiber Link.* Physical Review Letters, 2013. **111**(11): p. 110801.

## 2.6 Relevant Publications

The journal papers that relate to the content of this report are summarised in Table 1.

**Table 1 – Journal Papers**

| Journal Paper Information | Appendix # | Reference # |
|---|---|---|
| K. Grainge, B. Alachkar, S. Amy, et al. *Square Kilometre Array: the Radio Telescope of the XXI Century.* Astronomy Reports **61** (2017) 288–296. | Appendix 7.2.1 | Reference [RD4] |
| S.W. Schediwy, *A Clock for the Square Kilometre Array*. Proceedings of Science **170** (2012) 1-13. | Appendix 7.2.2 | Reference [RD23] |
| S.W. Schediwy, D.R. Gozzard, R. Whitaker, et al. *A Phase Synchronisation System for the Square Kilometre Array*. Submitted to Publications of the Astronomical Society of Australia (2018). | Appendix 7.2.3 | Reference [RD5] |
| D.R. Gozzard, S.W. Schediwy, and K. Grainge. *Simultaneous transfer of stabilized optical and microwave frequencies over fiber*. Photonics Technology Letters **30** (2018) 87. | Appendix 7.2.4 | Reference [RD24] |
| D.R. Gozzard, S.W. Schediwy, R. Whitaker, and K. Grainge. *Simple Stabilized Radio-Frequency Transfer with Optical Phase Actuation*. Photonics Technology Letters **30** (2018) 258. | Appendix 7.2.5 | Reference [RD11] |
| S.W. Schediwy, D.R. Gozzard, S. Stobie, J.A. Malan, and K. Grainge. *Stabilized microwave-frequency transfer using optical phase sensing and actuation*. Optics Letters **42** (2017) 1648. | Appendix 7.2.6 | Reference [RD10] |
| D.R. Gozzard, S.W. Schediwy, B. Wallace, et al. *Characterization of optical frequency transfer over 154 km of aerial fiber*. Optics Letters **42** (2017) 2197. | Appendix 7.2.7 | Reference [RD15] |
| D.R. Gozzard, S.W. Schediwy, B. Wallace, et al. *Stabilized Modulated Photonic Signal Transfer Over 186 km of Aerial Fiber*. Submitted to Transactions on Ultrasonics, Ferroelectrics, and Frequency Control (2017). | Appendix 7.2.8 | Reference [RD16] |
| D.R. Gozzard, S.W. Schediwy, R. Dodson, et al. *Astronomical verification of a stabilized frequency reference transfer system for the Square Kilometre Array*. The Astronomical Journal **154** (2017) 1. | Appendix 7.2.9 | Reference [RD17] |

The SADT Reports that relate to the content of this report are summarised in Table 2

**Table 2 – SADT Reports**

| SADT Report Information | Appendix # | Reference # |
|---|---|---|
| S.W. Schediwy and D.G. Gozzard, *Pre-PDR Laboratory Verification of UWA's SKA Synchronisation System*. SADT Report **616** (2014) 1-16. | Appendix 7.3.1 | Reference [RD12] |
| S.W. Schediwy and D.G. Gozzard, *Pre-CDR Laboratory Verification of UWA's SKA Synchronisation System*. SADT Report **620** (2017) 1-79. | Appendix 7.3.2 | Reference [RD13] |
| S.W. Schediwy and D.G. Gozzard, *UWA South African SKA Site Long-Haul Overhead Fibre Field Trial Report*. SADT Report **109** (2015) 1-20. | Appendix 7.3.3 | Reference [RD14] |
| S.W. Schediwy and D.G. Gozzard, *SKA-low Astronomical Verification*. SADT Report **617** (2015) 1-21. | Appendix 7.3.4 | Reference [RD19] |
| D.G. Gozzard and S.W. Schediwy, *SKA1-mid Astronomical Verification*. SADT Report **524** (2016) 1-29. | Appendix 7.3.5 | Reference [RD18] |
| D.G. Gozzard, *Notes on Calculating the Relationship between Coherence Loss and Allan Deviation*. SADT Report **619** (2017) 1-5. | Appendix 7.3.6 | Reference [RD25] |
| S. Stobie, S.W. Schediwy, and D.R. Gozzard, *Design-for-Manufacture of the SKA1-Mid Frequency Synchronisation System*. SADT Report **618** (2017) 1-57. | Appendix 7.3.7 | Reference [RD22] |

## 2.7 Glossary of Terms

**Table 3 – Glossary of Terms**

| Term | Acronym Long Form | Definition |
|---|---|---|
| AAVS1 | Aperture Array Verification System 1 | Prototype aperture array on the Australian SKA site, comprising many in-development SKA sub-systems, for the purpose of on-site verification. |
| ADC | Analogue-to-digital converter | Key device within the LFAA and DISH receiver used to sample an incoming analogue waveform, using the SADT.SAT.STFR reference signals as its clock. |
| ALMA | Atacama Large Millimeter Array | Radio telescope array operating between 31 and 1,000 GHz situated in the high mountains of Chile. |
| AOM | Acousto-optic modulator | Optoelectronic device that can be used to apply a fixed or variable radio-frequency shift onto a transmitted optical signal. |
| ARC | Australian Research Council | Australia's national science funding body. |
| ASKAP | Australian SKA Pathfinder | Australian SKA pathfinder radio telescope array. |
| ATCA | Australia Telescope Compact Array | Australia's current workhorse radio telescope array. |
| CCD | Command and Control Device | Unit integrated within the Sub Rack of UWA's *SKA phase-synchronisation system* for controlling and recording signals on the attached Transmitter Modules. |
| CDR | Critical Design Review | Second-stage review conducted by the SKAO on SKA Consortia design elements. |
| CIN | Configuration Identification Number | Unique identification number assigned to each SADT line replaceable unit. |
| COTS | Commercial off-the-shelf | Equipment that can be purchased in their entirety as a single entity from a commercial supplier. |
| CPF | Central Processing Facility | Environmentally controlled structure which houses (amongst other things) all the SADT.SAT.STFR.FRQ (UWA) elements other than the Receiver Modules and Optical Amplifier. |
| DDS | Direct digital synthesiser | 4-channel printed circuit board surface mount chip located on each Transmitter Module, which takes a radio-frequency reference signal from Signal Generator, to produce the servo-loop local oscillator signal, as well as other ancillary signals required by the Transmitter Module. |
| DISH | N/A | SKA Consortium responsible for producing the main physical elements of the SKA1-mid telescope. |
| DPM | Dual-parallel Mach-Zehnder modulator | Optoelectronic device that can be used to apply a series of complex modulation states (up to microwave frequencies) onto a transmitted optical signal. |
| ECP | Engineering Change Proposal | Formal process for varying elements of the SKA telescope design. |
| EICD | External Interface Control Document | Document defining an interface between an SADT sub-element and a sub-element belonging to another SKA consortia. |
| e-MERLIN | Enhanced Multi Element Radio Linked Interferometry Network | Radio telescope array, situated in the UK and linked by optical fibre, with similar maximum baselines as the SKA1-mid (but comprising only seven dishes). |
| EMI | Electromagnetic Interference | Self-generated unwanted radio- or microwave-frequency signals. |
| FM | Faraday mirror | Optical device that reflects incoming light with a 90° turn of polarisation. |
| FS | Frequency Synthesiser | SADT.SAT.STFR.FRQ (UWA) hardware element: A high-quality microwave-frequency signal source, referenced to the SKA clock enable, which is used to provide a static signal to the Microwave Shift line replaceable unit. |
| H-maser | Hydrogen maser | A high-precision frequency reference. |
| LFAA | Low-Frequency Aperture Array | SKA Consortium responsible for producing the main physical elements of the SKA1-low telescope. |
| LO | Local oscillator | The radio frequency signal within each Transmitter Module that is the reference against which the incoming optical link radio-frequency signal is compared to |
| | | produce the servo-loop error signal. |
| LRU | Line replaceable unit | Single reproducible hardware building block of the SADT work package elements; can be commercial off-the-shelf or bespoke. |
| MeerKAT | More Karoo Array Telescope | South African SKA pathfinder radio telescope array. |
| MI | Michaelson interferometer | Optical interferometer, where one arm comprises the entire fibre link, and the other short arm contained within the Transmitter Module, provides the optical reference signals that for the link stabilisation servo loop. |
| MS | Microwave Shift | SADT.SAT.STFR.FRQ (UWA) hardware element: Applies an electronic signal from Microwave Synthesiser to optical signal from Optical Source to produce two optical signals separated by the microwave-frequency reference signal on two separate optical fibres. |
| MW | Microwave | For the purposes of this document; frequencies between 1 MHz and 1 GHz. |
| MRO | Murchison Radioastronomy Observatory | This is a designated radio quiet zone located near Boolardy station in Western Australia that currently host two main instruments; the Murchison Widefield Array and the Australian Square Kilometre Array Pathfinder. It is also the Australian site for the Square Kilometre Array. |
| MZI | Mach-Zehnder interferometer | Fiberised two-arm optical interferometer localised at the Central Processing Facility which is used to produce the microwave (SKA1-mid) or radio (SKA-low) frequency separation between the two transmitted optical signals. |
| NPL | National Physical Laboratory | The United Kingdom’s national measurement institute. Design authority of SADT.SAT.CLOCK |
| NTFN | National Time and Frequency Network | Australian ARC-funded project with the aim to develop continental-scale time and frequency transfer technology. |
| OA | Optical Amplifier | SADT.SAT.STFR.FRQ (UWA) hardware element: Bi-directional erbium-doped fibre amplifier, used to amplify optical signals on the longest SKA1-mid links. |
| OCXO | Oven-controlled crystal oscillator | Ultra-low noise, thermally stabilised oscillator located with the Receiver Module to provide coherence for UWA’s *SKA Phase Synchronisation System* at timescales shorter than the light round trip time of the fibre link. |
| OS | Optical Source | SADT.SAT.STFR.FRQ (UWA) hardware element: A single high-coherence 1552.52 nm laser that is the maser source for all optical signals. |
| OTDR | Optical time-domain reflectometry | Technique for identifying the locations of large unwanted reflections in optical fibre links, as well as their total optical loss. |
| PBS | Product Breakdown Structure | Hierarchical structure that identifies all line replaceable units within the SADT Consortium using unique Configuration Identification Numbers. |
| PCB | Printed circuit board | Mechanical support for electronic components, conductive tracks, pads and other features etched from copper sheets on a non-conductive substrate. |
| PDR | Preliminary Design Review | First-stage review conducted by the SKAO on SKA Consortia design elements. |
| PIC | Peripheral Interface Controller | A family of microcontrollers made by the company Microchip Technology. |
| PLL | Phase-locked loop | A simple electronic circuit that includes a voltage-driven oscillator which constantly adjusts to match the frequency of an input signal. |
| ICRAR | International Centre for Radio Astronomy Research | World-class astronomy institute joint venture between the University of Australia and Curtin University. |
| IICD | Internal Interface Control Document | Document defining an interface between two SADT sub-elements. |
| RD | Rack Distribution | SADT.SAT.STFR.FRQ (UWA) hardware element: Used to distribute optical-, microwave-, and radio-frequency signals from their sources to the Sub Racks. |
| RF | Radio frequency | For the purposes of this document; frequencies between 1 and 100 GHz (as per the common definition within radio engineering). |
| RFI | Radio frequency interference | Unwanted radio- or microwave-frequency signals caused by external sources. |
| RM | Receiver Module | SADT.SAT.STFR.FRQ (UWA) hardware element: Hot-swappable, small form-factor enclosed module which houses fiberised optics and electronics (including a clean-up OCXO in a PLL) to convert the stabilised signal received across the fibre link, into an electronic signal which is passed to the LFAA or DISH. |
| RPF | Remote Processing Facility | Environmentally controlled structure which houses (amongst other things) the SADT.SAT.STFR.FRQ (UWA) Receiver Modules and LFAA analogue-to-digital receivers. |
| SADT | Signal and Data Transport | The signal and data transport element of phase 1 of the SKA telescope. |
| SADT Consortium | N/A | Collection of organisations, including UWA, led by the UoM to design the SADT element of the SKA Phase 1 telescope. |
| SADT.SAT.STFR. CLOCK | N/A | SADT work-package that includes the SKA hydrogen maser ensemble and timescale. |
| SADT.SAT.STFR. FRQ | N/A | SADT work-package which incorporates the frequency dissemination technology at the core of the UWA's *SKA phase-synchronisation system*. |
| SADT.SAT.STFR. FRQ (UWA) | N/A | UWA's *SKA phase-synchronisation system*. |
| SADT.SAT.LMC | N/A | SADT local management and control for the SAT network. |
| SADT.SAT.STFR. UTC | N/A | SADT work-package responsible for disseminating absolute time. |
| SAT | Synchronisation and Timing | SADT network used to disseminating references signals and absolute time. |
| s.d. | standard deviation | In statistics the standard deviation is a measure that is used to quantify the amount of variation or dispersion of a set of data values |
| SG | Signal Generator | SADT.SAT.STFR.FRQ (UWA) hardware element: Source of radio-frequency signals used to reference the DDS chips in each Transmitter Module. |
| SKA1-low | N/A | The low frequency SKA Phase 1 telescope located in Australia. |
| SKA1-mid | N/A | The mid frequency SKA Phase 1 telescope located in South Africa. |
| SKAO | Square Kilometre Array Office | The organisation that is the design authority for the SKA telescope and is the client of the SADT Consortium. |
| SR | Sub Rack | SADT.SAT.STFR.FRQ (UWA) hardware element: 3U, 19" rack mount enclosure that houses 16 hot-swappable Transmitter Modules; as well as a common power supply, the command and control module, and internal optical-, RF-, and MW-distribution units. |
| SSB-CS | Single-sideband suppressed-carrier | A modulation type where only one sideband is generated and all other optical signals, including the carrier, are suppressed. |
| STFR | Station Time and Frequency Reference | SADT sub-element that comprises the *SKA Phase Synchronisation System* and the SKA system for disseminating absolute time. |
| THU | Tsinghua University | Design authority of the alterative *SKA Phase Synchronisation System* in the SADT technology down select. |
| TM | Transmitter Module | SADT.SAT.STFR.FRQ (UWA) hardware element: Hot-swappable printed circuit board, mounted within the Sub Rack, that includes fiberised optics and the servo-loop electronics for stabilising the transmitted reference signals to be transmitted across the fibre link. |
| TOR | Top of Rack | The primary Ethernet switch for each rack cabinet. |
| UoM | University of Manchester | Lead institute of the SADT Consortium. |
| UWA | University of Western Australia | Design authority for UWA's *SKA Phase Synchronisation System*, the detailed design of which is described in this document. |
| VCO | Voltage-controlled oscillator | Frequency source, the output frequency of which can be controlled given a range of input voltages. |

# 3 SOLUTION DESCRIPTION

## 3.1 Technical Background

### 3.1.1 SKA Telescope Phase Synchronisation

The Square Kilometre Array (SKA) project [RD1] is an international effort to build the world's most sensitive radio telescope operating in the 50 MHz to 14 GHz frequency range. Construction of the SKA has been divided into phases, with the first phase (SKA1) accounting for the first 10% of the telescope's receiving capacity. During SKA1, a low-frequency aperture array comprising over a hundred thousand individual dipole antenna elements will be constructed in Western Australia (SKA1-low), while an array of 197 parabolic-dish antennas, incorporating the 64 dishes of MeerKAT, will be constructed in South Africa (SKA1-mid).

Radio telescope arrays such as the SKA require phase-coherent reference signals to be transmitted to each antenna site in the array. In the case of the SKA-low, these reference signals will be generated at a central site and transmitted to the antenna sites via fibre-optic cables up to 58 km in length [RD26]. Environmental perturbations affect the optical path length of the fibre and act to degrade the phase stability of the reference signals received at the antennas, which has the ultimate effect of reducing the fidelity and dynamic range of the data [RD3]. Given the combination of long fibre distances and relatively high frequencies of the transmitted reference signals, the SKA will need to employ actively-stabilized frequency transfer technologies to suppress the fibre-optic link noise [RD4] in order to maintain phase-coherence across the array. Phase-synchronisation systems have been successfully used on other radio telescope arrays including e-MERLIN in the UK [RD27, 28] and the Atacama Large Millimetre Array (ALMA) [RD6, 29].

In this report, we describe the detailed design of the SKA-low variant of UWA's *SKA phase synchronisation system* based on actively stabilised radio-frequency transfer via optical fibre [RD10]. A separate report describes the detailed design of the SKA-mid variant [RD30].

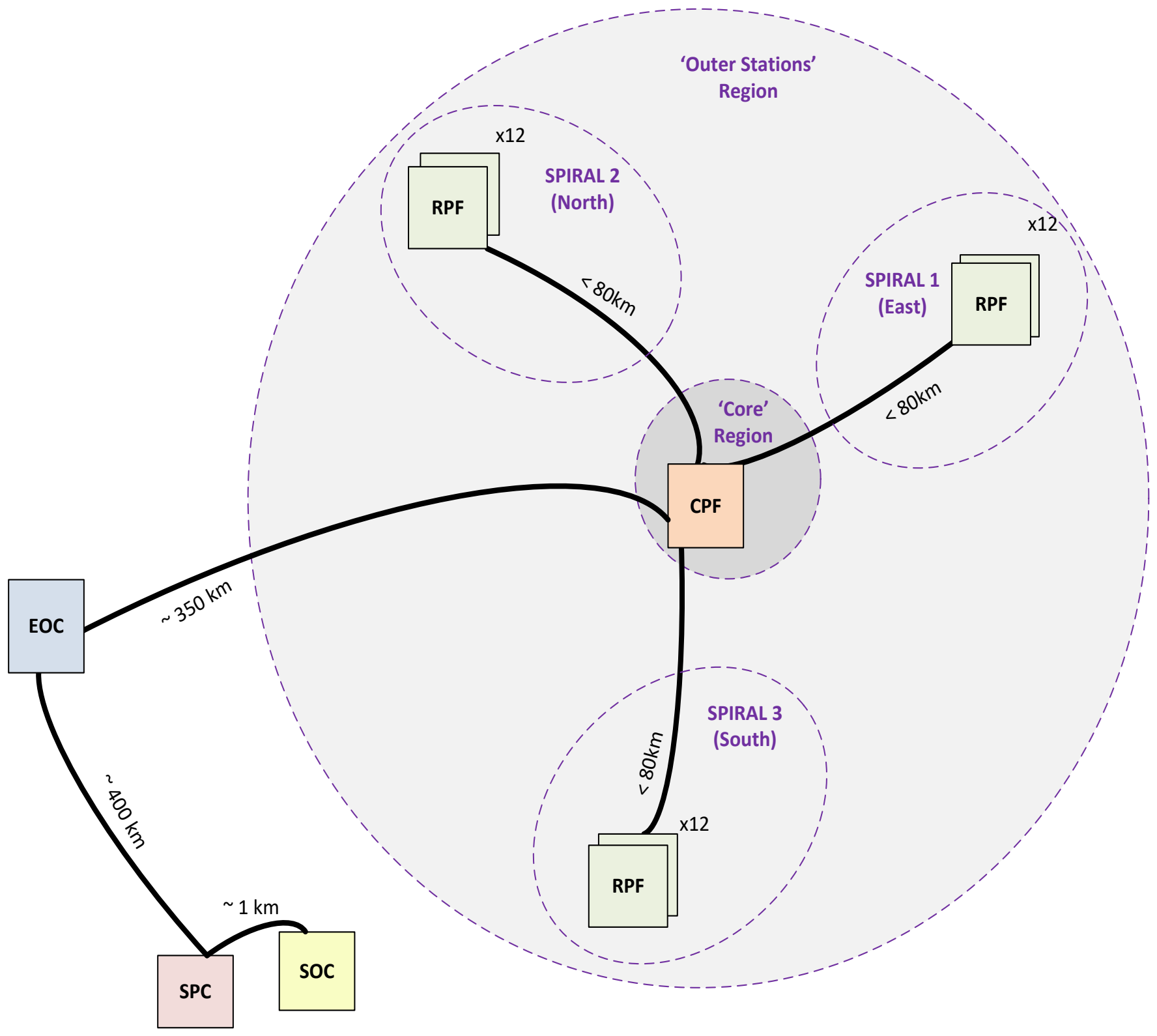

**Figure 1 – SKA1-low network layout. From [RD26].**

### 3.1.2 Actively Stabilised Frequency Transfer via Optical Fibre

In both SKA telescopes, the master reference signals originate from an ensemble of three hydrogen masers, situated in each telescopes' central processing facility (CPF). For SKA1-low, the reference signals are transmitted from the Australian CPF to 36 remote processing facilities (RPFs), as shown in Figure 1.

The reference signals are converted into the optical domain and then transmitted over the optical fibre link where they acquire phase noise due to environmental disturbances on the link. This degrades the phase-stability and thus the coherence of the reference signals, which has the ultimate effect of reducing the fidelity and dynamic range of the data.

The SKA telescope will employ actively stabilise frequency transfer technology to suppress these environmental disturbances and ensure phase coherence across the array. A simplified diagram of a generalised stabilised frequency transfer system is shown in Figure 2.

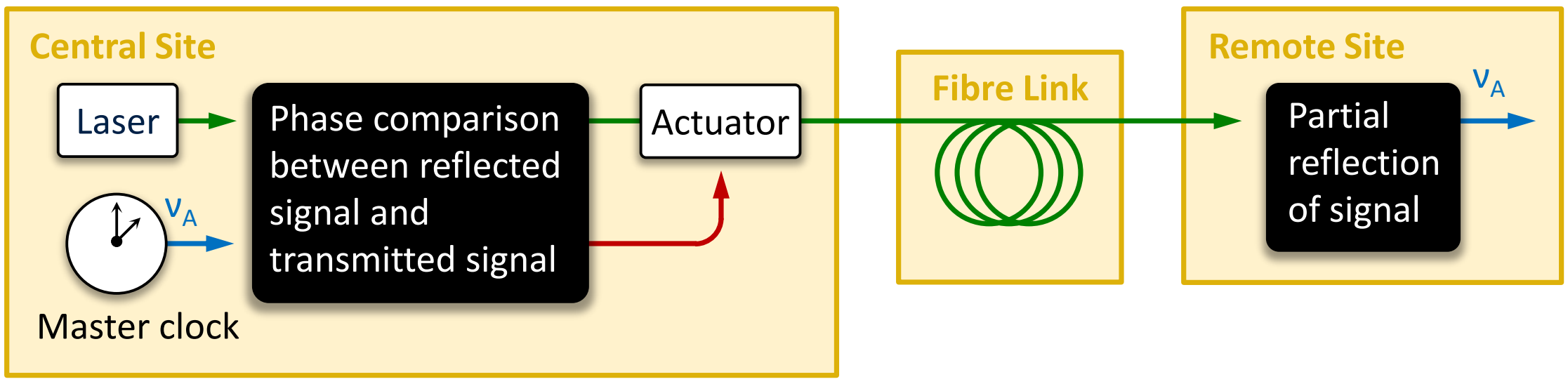

**Figure 2 – Simplified schematic of a generalised stabilised frequency transfer system.**

The reference signals are transmitted from the central site across the fibre link to the remote site. At the remote site a partial reflection of the transmitted reference signal (having made a round-trip through the link) is compared against a copy of the original transmitted reference signal. The environmental disturbances on the link are encoded as the phase difference between these two signals. This comparison is used to generate a correction signal, which is then used to control an actuator that modifies the outgoing signal in such a way as to supress the effect of the environmental disturbances. At the remote site, the now stabilised reference signals are converted into the electronic domain, and supplied the telescope's analogue-to-digital converters.

### 3.1.3 Designing to SKA Requirements

UWA's *SKA phase-synchronisation system* must meet a series of performance and operational requirements to allow the SKA to operate at maximum capacity. The functional performance requirements, listed in Section , concern the coherence loss, phase drift, and jitter requirements that must be met for optimal operation of the telescope. The normal operating conditions (Section 6.2) and key additional requirements (6.3) concern the climatic condition the system will be subjected to, conformance with the design and layout of the telescope, and other practical considerations such as radio-frequency interference and system monitoring. The full list of relevant requirements is given in the Statement of Compliance in Section 6.

## 3.2 Solution Overview

### 3.2.1 Solution Development

Since 2011, researchers at the University of Western Australia (UWA), have led the development of an *SKA phase-synchronisation system* designed specifically to meet the scientific needs and technical challenges of the SKA telescope. A concept document (Appendix 7.1.1) and paper [RD31] were developed as part of the Australian National Time and Frequency Network (NTFN) project; and this led to a collaboration with the University of Manchester (UoM) and the UK's National Physical Laboratory (NPL) from mid-2012. The collaboration was formalised through the funding of a UWA collaboration grant in 2013.

The UWA-UoM-NPL proto-consortium produced a document in mid-2013 (Appendix 7.1.2) that compared the pros and cons of four classes of phase synchronisation solutions for the SKA. These three techniques where then down-selected by the UWA-UoM-NPL proto-consortium to the one that would evolve to become UWA's *SKA phase-synchronisation system* described in this report.

With the formation of the UoM-led Signal and Data Transport (SADT) consortium in December 2013, several other members, including Tsinghua University (THU), joined the consortium. Researchers from THU then proposed an alternative phase-synchronisation system for the SKA [RD32] based on a 'phase-conjugation' technique first demonstrated in 1988 [RD33], and then further developed for radio astronomy by UWA's NTFN colleagues [RD34].

An SADT consortium-coordinated technology down-select process will be used to select the most suitable technology for the SKA1-mid and SKA1-low telescopes.

### 3.2.2 Solution Description Summary

UWA's *SKA phase-synchronisation system* [RD5] is based on the transmission of actively stabilised phase-coherent reference signals generated at the central processing facility (CPF), and then transmitted via separate optical fibre links to each antenna site. The star-shaped network topology of such a phase-synchronisation system, conveniently matches the fibre topology of the SKA's data network, which transmits the astronomical data from the antenna sites to the CPF. The frequency transfer technique at the core of UWA's *SKA phase-synchronisation system* is an evolution of ALMA's distributed 'photonic local oscillator system' [RD6], incorporating key advances made by the international frequency metrology community over the last decade [RD7-9], as well as novel innovations developed by UWA researchers [RD10, 11].

Just as with the ALMA system, UWA's *SKA phase-synchronisation system* transmits the reference signals as a sinusoidal optical modulation encoded as the difference between two optical-frequency signals. Given the lower operating frequency of the SKA compared to ALMA, the two optical-frequency signals can be generated using a single laser and applying a frequency shift in one arm of a Mach-Zehnder interferometer (MZI); thereby avoiding the differential phase noise from offset-locking two independent lasers as per the ALMA system.

Two variants of UWA's *SKA phase-synchronisation system* have been designed, each one optimised specifically for each SKA telescope. For SKA1-mid, the required microwave (MW) shift is generated using a dual-parallel Mach-Zehnder modulator (DPM) biased to generate single-sideband suppressed-carrier (SSB-SC) modulation [RD10]; while for SKA1-low the radio-frequency (RF) shift is generated using a simpler acousto-optic modulator (AOM) [RD11]. This results in two systems that easily meet the SKA coherence requirements, as demonstrated by laboratory testing [RD10-13], overhead fibre field trials [RD14-16], and astronomical verification [RD17-19], yet keep complexity and costs to a minimum.

The basis of the frequency transfer technique at the core of UWA's *SKA phase-synchronisation system* is that the optical signal from one laser is split into two arms of a Mach-Zehnder interferometer (MZI). In one arm either a static microwave-frequency shift (SKA1-mid) or static radio-frequency shift (SKA1-low) is applied to the optical signal. When the two arms are recombined at the output of the MZI, this results in two optical signals on a single fibre with either microwave- or radio-frequency separation. This signal is transmitted over the optical fibre link to the remote telescope sites where it is translated into the electronic domain by a photodetector. Part of the signal is reflected back to the transmitter site, where it is mixed with a copy of the transmitted optical signals, and the frequency modulation is extracted with a local photodetector. In the case of the SKA1-mid system, this electronic signal is then mixed with a copy of the microwave signal, to produce an error signal that has encoded the fluctuations of the link. Applying this to the drive signal of an acousto-optic modulator closes the servo loop and effectively cancels the link noise for the remote site.

The colours for signal types used in all schematic diagrams throughout the document are shown in Table 4.

**Table 4 – Figure signal colour codes**

| Colour | Signal |
|---|---|
| Green | Optical frequencies |
| Blue | Microwave frequencies |
| Red | Radio frequencies |
| Black | Power/DC/audio frequencies |
| Orange | Absolute time signal |
| Pink | Ethernet/serial |

### 3.2.3 Analytical Derivation of Solution

This section provides the detailed analytical derivation of the radio-frequency transfer technique at the core of UWA's *SKA phase-synchronisation system* for SKA1-low. Further information on this technique is provided in [RD11].

As shown in Figure 3, an optical signal with frequency $\nu_L$, is generated by a laser located at the **SKA-low Central Processing Facility**. Just as is the case in standard stabilized optical transfer techniques [RD35], the optical signal enters an imbalanced MI via an optical isolator (to prevent reflections returning to the laser). The short arm of the MI provides the physical reference for the optical phase sensing. The optical reference signal $\nu_{ref}$ at the photodetector is

$$\nu_{\text{ref}} = \nu_L + \frac{1}{2\pi}(2\Delta\dot{\phi}_{MI}),$$ Equation 1

where $\Delta\phi_{MI}$ is the undesirable phase noise picked up by the optical signals passing through the MI reference arm.

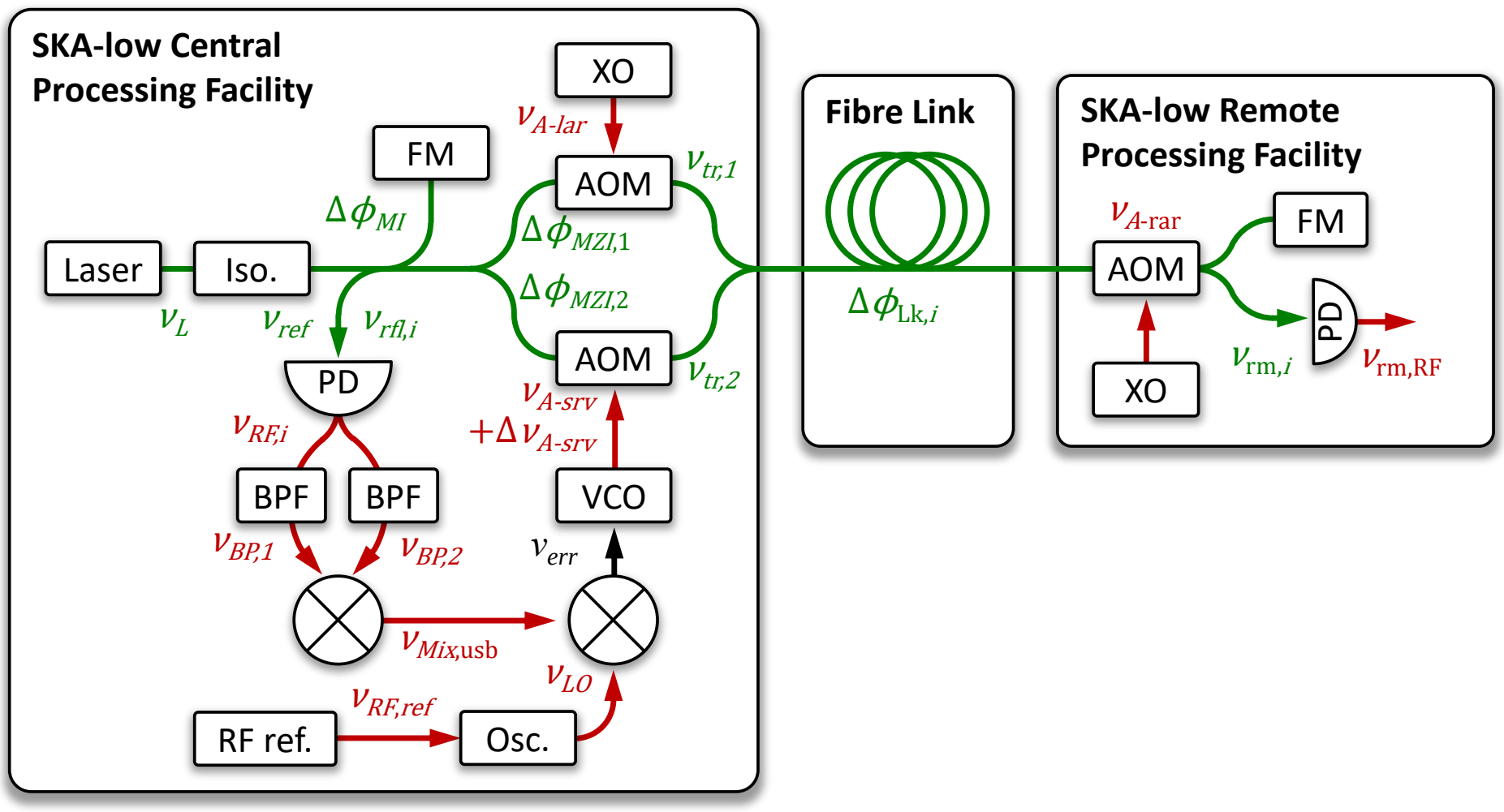

**Figure 3 – Simplified schematic of UWA's SKA1-low stabilized radio-frequency transfer technique.**

In the long arm of the MI the optical signal is then split into two arms of a Mach-Zehnder interferometer (MZI), each arm of which contains an AOM. In the case shown in Figure 3, the bottom arm of the MZI contains the servo AOM, which has a nominal frequency $\nu_{A-srv}$, and also applies the frequency correction $\Delta\nu_{A-srv}$. The frequency fluctuations due to optical path length changes in the bottom arm of the MZI are represented by $\frac{1}{2\pi}\Delta\dot{\phi}_{MZI,2}$. The top arm of the MZI contains the local 'anti-reflection' AOM, with frequency $\nu_{A-lar}$. The frequency fluctuations in this arm of the MZI are given by $\frac{1}{2\pi}\Delta\dot{\phi}_{MZI,1}$. A combination of up- and down-shifting AOMs gives the greatest RF separation, but any unique combination can be used. (Note — the anti-reflection AOM is not essential for this technique. It serves to provide reflection mitigation as well as increase the frequency separation.)

At the output of the MZI, the two optical signals to be transmitted are now:

$$\nu_{tr,1} = \nu_L + \nu_{A-lar} + \frac{1}{2\pi}\Delta\dot{\phi}_{MZI,1}, \qquad \text{Equation 2}$$

and

$$\nu_{tr,2} = \nu_L + (1+\Delta)\nu_{A-srv} + \frac{1}{2\pi}\Delta\dot{\phi}_{MZI,2}. \qquad \text{Equation 3}$$

As the two optical signals pass through the **Fibre Link**, they pick-up frequency fluctuations, $\Delta\dot{\phi}_{Lk,i}$, due to optical path length changes in the link that are unique to their specific transmitted frequency.

At the **Remote Processing Facility**, the two optical signals pass through a remote anti-reflection AOM (again, this AOM is not essential for the technique, but allows rejection of unwanted optical reflections) to give:

$$\nu_{rm,1} = \nu_L + \nu_{A-lar} + \nu_{A-rar} + \frac{1}{2\pi}\left(\Delta\dot{\phi}_{MZI,1} + \Delta\dot{\phi}_{Lk,1}\right), \qquad \text{Equation 4}$$

and

$$\nu_{rm,2} = \nu_L + (1+\Delta)\nu_{A-srv} + \nu_{A-rar} + \frac{1}{2\pi}\left(\Delta\dot{\phi}_{MZI,2} + \Delta\dot{\phi}_{Lk,2}\right). \qquad \text{Equation 5}$$

At the **Remote Processing Facility**, the signal is split with one part going to a photodetector. The electronic signal $\nu_{rm,e}$ from the beat of $\nu_{rm,1}$ and $\nu_{rm,2}$ is

$$\nu_{rm,e} = \frac{1}{2\pi}\left(\Delta\dot{\phi}_{MZI,2} - \Delta\dot{\phi}_{MZI,1} + \Delta\dot{\phi}_{Lk,2} - \Delta\dot{\phi}_{Lk,1}\right) + (1+\Delta)\nu_{A-srv} - \nu_{A-lar}. \qquad \text{Equation 6}$$

A Faraday mirror (FM) reflects the two optical signals back through the link to the **Central Processing Facility** where they then pass back though the MZI. The returning reflected optical signals reaching the photodetector for the MI are then:

$$\nu_{rfl,1} = \nu_L + 2\left(\nu_{A-lar} + \nu_{A-rar} + \frac{1}{2\pi}\left(\Delta\dot{\phi}_{MZI,1} + \Delta\dot{\phi}_{Lk,1}\right)\right), \qquad \text{Equation 7}$$

$$\nu_{rfl,2} = \nu_L + 2\left((1+\Delta)\nu_{A-srv} + \nu_{A-rar} + \frac{1}{2\pi}\left(\Delta\dot{\phi}_{MZI,2} + \Delta\dot{\phi}_{Lk,2}\right)\right), \qquad \text{Equation 8}$$

and

$$\nu_{rfl,3j} = \nu_L + (1+\Delta)\nu_{A-srv} + \nu_{A-lar} + 2\nu_{A-rar} + 2\Delta\dot{\phi}_{Lk,\mathrm{j}} + \frac{1}{2\pi}\left(\Delta\dot{\phi}_{MZI,1} + \Delta\dot{\phi}_{MZI,2}\right). \qquad \text{Equation 9}$$

where $j$ is 1 or 2 corresponding to the signals on the link. The three signals are at unique frequencies as long as $\nu_{A-srv}$ does not equal $\nu_{A-lar}$. At the photodetector these optical frequencies mix with $\nu_{ref}$ to give the following RF signals.

$$\nu_{RF,1} = 2\left(\nu_{A-lar} + \nu_{A-rar} + \frac{1}{2\pi}\left(\Delta\dot{\phi}_{MZI,1} + \Delta\dot{\phi}_{Lk,1} - \Delta\dot{\phi}_{MI}\right)\right), \qquad \text{Equation 10}$$

$$\nu_{RF,2} = 2\left((1+\Delta)\nu_{A-srv} + \nu_{A-rar} + \frac{1}{2\pi}\left(\Delta\dot{\phi}_{MZI,2} + \Delta\dot{\phi}_{Lk,2} - \Delta\dot{\phi}_{MI}\right)\right), \quad \text{Equation 11}$$

and

$$\nu_{RF,3j} = (1+\Delta)\nu_{A-srv} + \frac{1}{2\pi}\left(\Delta\dot{\phi}_{MZI,1} + \Delta\dot{\phi}_{MZI,2} + 2\Delta\dot{\phi}_{Lk,j}\right) + \nu_{A-lar} + 2\nu_{A-rar}, \quad \text{Equation 12}$$

as well as intermodulation signals. In the electronic domain, the signals are split and bandpass filtered. The RF bandpass filter values are set to $\nu_{BP,1} = 2(\nu_{A-lar} + \nu_{A-rar})$ and $\nu_{BP,2} = 2(\nu_{A-srv} + \nu_{A-rar})$.

The bandpass filters eliminate $\nu_{\mathrm{RF,3j}}$, the intermodulation signals, and the opposing RF signal. The filtered signals are then mixed together, producing the following upper- and lower-sideband frequency products:

$$\nu_{Mix,1} = 2\left((1+\Delta)\nu_{A-srv} + \nu_{A-lar} + \frac{1}{2\pi}\left(\Delta\dot{\phi}_{MZI,1} + \Delta\dot{\phi}_{MZI,2} + \Delta\dot{\phi}_{Lk,1} + \Delta\dot{\phi}_{Lk,2} - 2\Delta\dot{\phi}_{MI}\right)\right),$$

Equation 13

$$\nu_{Mix,2} = 2\left((1+\Delta)\nu_{A-srv} - \nu_{A-lar} + \frac{1}{2\pi}\left(\Delta\dot{\phi}_{MZI,1} - \Delta\dot{\phi}_{MZI,2} + \Delta\dot{\phi}_{Lk,1} - \Delta\dot{\phi}_{Lk,2}\right)\right). \quad \text{Equation 14}$$

Note that in $\nu_{Mix,2}$ the frequency perturbation $\Delta\dot{\phi}_{MI}$ cancels out. A bandpass filter with the center frequency at $2 \times (\nu_{A-srv} - \nu_{A-lar})$ is used to reject $\nu_{Mix,1}$, before mixing $\nu_{Mix,2}$ with the servo local oscillator $\nu_{LO}$ also set at $\nu_{LO} = 2(\nu_{A-srv} - \nu_{A-lar})$ to produce an error signal of

$$\nu_{err} = 2\left(\Delta\nu_{A-srv} + \frac{1}{2\pi}\left(\Delta\dot{\phi}_{MZI,2} - \Delta\dot{\phi}_{MZI,1} + \Delta\dot{\phi}_{Lk,2} - \Delta\dot{\phi}_{Lk,1}\right)\right). \quad \text{Equation 15}$$

When the servo is engaged, the error signal is driven to zero, $\nu_{err}$ = 0, so:

$$\nu_{A-srv} = -\frac{1}{2\pi}\left(\Delta\dot{\phi}_{MZI,2} - \Delta\dot{\phi}_{MZI,1} + \Delta\dot{\phi}_{Lk,2} - \Delta\dot{\phi}_{Lk,1}\right). \quad \text{Equation 16}$$

Substituting this into (6) gives:

$$\nu_{rm,e*} = \nu_{A-srv} - \nu_{A-lar}, \quad \text{Equation 17}$$

where $\nu_{rm,e*}$ is the electronic remote signal with the servo engaged.

This derivation demonstrates analytically that UWA's radio-frequency transfer technique [RD11] is effective at transferring the stability of a reference signal from the **Central Processing Facility** to the **Remote Processing Facility** (within the light round-trip bandwidth and other practical gain limitations). However, the design schematic in Figure 3 shows only a point-to-point link. The following section describes how this radio-frequency transfer technique is engineered to create a practical realisation of UWA's *SKA phase-synchronisation system* that is able to provide reference signals to all 36 SKA1-low **Remote Processing Facilities**.

### 3.2.4 Practical Realisation of Solution

This section describes how the frequency transfer technique described above, is able to be engineered into a practical realisation of UWA's *SKA phase-synchronisation system* for SKA1-low.

UWA's *SKA phase-synchronisation system* is an element within the SADT Consortium's Synchronisation and Timing (SAT) network. The full SADT element name of UWA's *SKA phase-synchronisation system* is SADT.SAT.STFR.FRQ (UWA), where STFR is an acronym of 'Station Time and Frequency Reference' and FRQ is an abbreviation of 'Frequency'. The other elements within the SAT network are SAT.CLOCK, the SKA's hydrogen maser clock ensemble and timescale; SAT.UTC, the system for disseminating absolute time; and SAT.LMC, the local monitor and control system for the SAT network. The output of SADT.SAT.STFR.FRQ (UWA) is passed to the DISH Consortium. A simplified schematic of the SADT.SAT network architecture, showing the interrelationships between the various SADT.SAT network elements and DISH, is given in Figure 4.

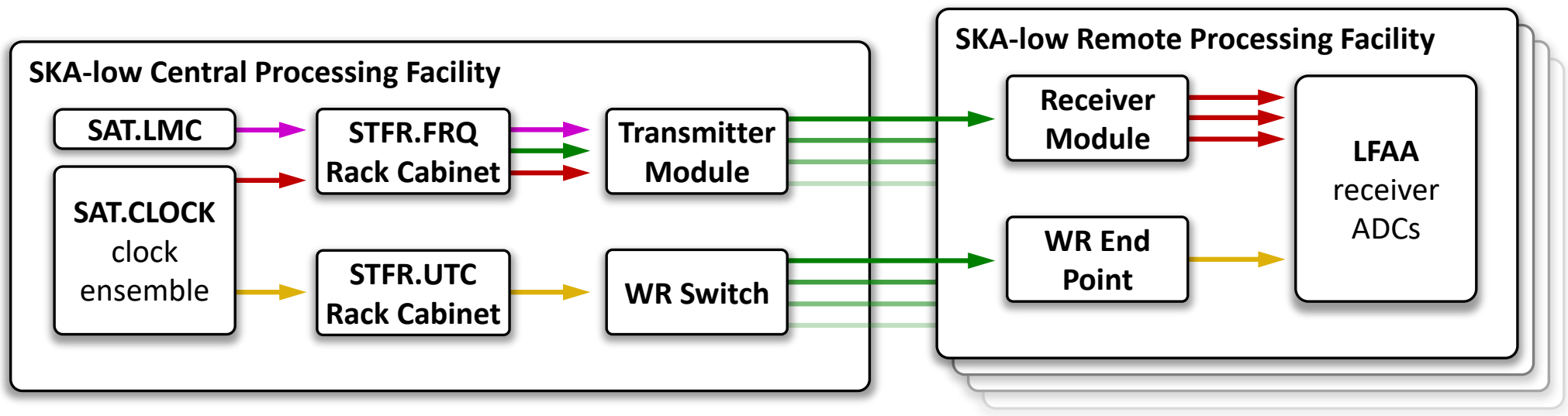

**Figure 4 – Simplified schematic of the SKA1-low Synchronisation and Timing (SAT) network.**

UWA's *SKA phase-synchronisation system* for SKA1-low, is made up of six hot-swappable, line replaceable units (LRU) as defined in the SADT product breakdown structure (PBS). These, along with their unique Configuration Identification number (CIN) are given below:

- Rack Cabinet (141-022900)
- Optical Source (141-022400)
- Signal Generator (141-023100)
- Sub Rack (141-022700)
- Transmitter Module (141-022100)
- Receiver Module (141-022300)

The combination of these LRUs comprise the entire STFR.FRQ system. The Transmitter Module (141-022100) and Receiver Module (141-022300) were already shown in Figure 4. Figure 5 shows the simplified schematic of the hardware housed in the STFR.FRQ Rack Cabinet.

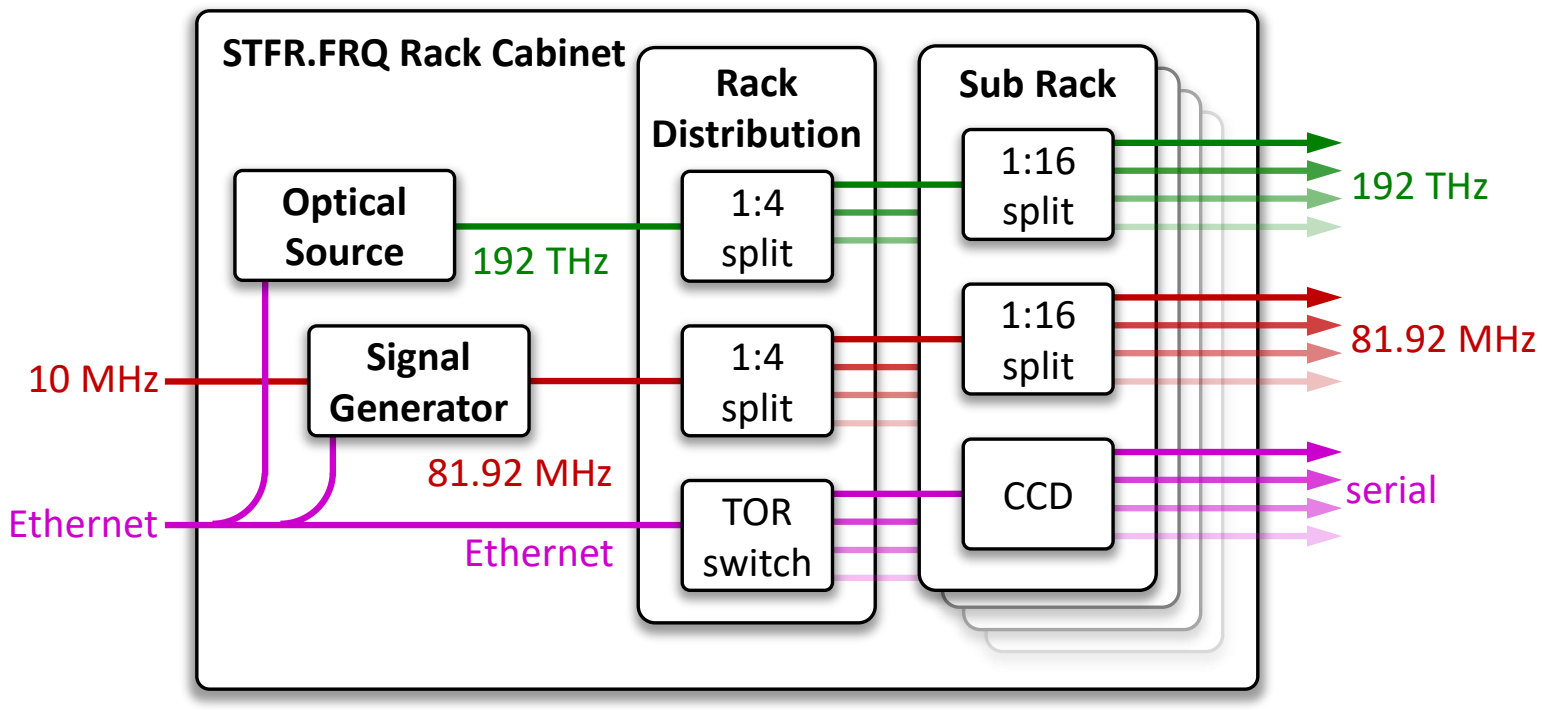

**Figure 5 – Simplified schematic of the SKA1-low STFR.FRQ Rack Cabinet.**

The SKA1-low STFR.FRQ Rack Cabinet incorporates Optical Source (141-022400), Signal Generator (141-023100), Rack Distribution (141-022800), and Sub Rack (141-022700). Note, Rack Cabinet (141-022900) is a collection of rack cabinet accessories including equipment heat management and cable management, SADT.LINFRA is responsible for supplying the actual rack cabinet hardware.

The STFR.FRQ Rack Cabinet takes input from SAT.CLOCKS and SAT.LMC, and produces a series of outputs that are passed to the Transmitter Module. The Transmitter Module and Receiver Module simplified schematics are shown in Figure 6.

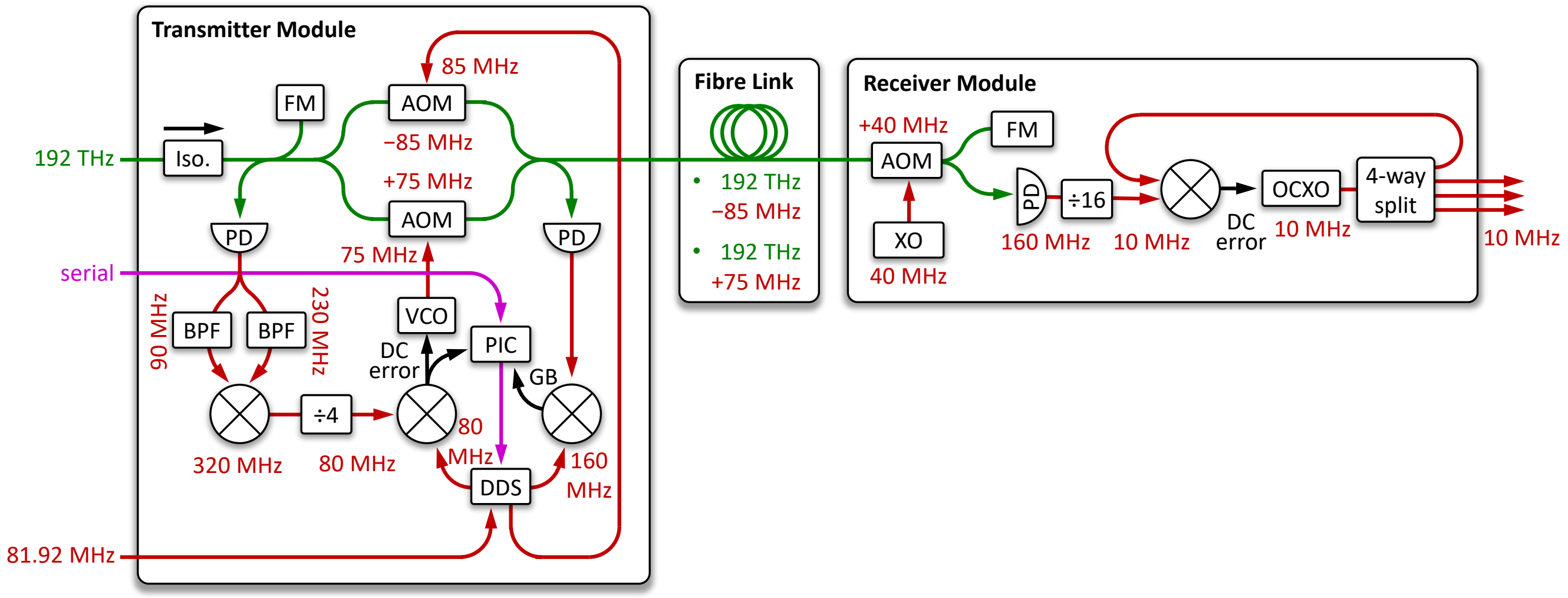

**Figure 6 – Simplified schematic of the SKA1-low Transmitter Module, Fibre Link, and Receiver Module.**

The Transmitter Module transmits the reference signal across the Fibre Link to the Receiver Module. The Transmitter Module contains the servo-loop electronics and all monitor and control hardware. The Receiver Module includes a clean-up oscillator in a simple phase-locked loop, with the output then passed onto LFAA. Note, the optical amplifiers are not required for SKA1-low as the longest Fibre Link is 58 km [RD26]. The detailed design of UWA's *SKA phase-synchronisation system* hardware for SKA1-low is provided in greater detail in Section 3.4.

## 3.3 Solution Design Justification

This section provides justifications for why certain design elements of UWA's *SKA Phase Synchronisation System* are the way they are. Where possible, these justifications are supported by verification and measurement. However, this section only includes a summary of the key results from a limited sub-set of measurements, the full set of campaigns conducted to demonstrate that UWA's *SKA Phase Synchronisation System* is fully compliant with the SKA requirements is described in Section 4.

UWA's *SKA phase-synchronisation system* has been extensively tested using standard metrology techniques in a laboratory setting [RD12, 13], with signals transmitted over metropolitan fibre links under all required conductions; on 186 km of overhead fibre at the South African SKA site [RD14-16]; as well as astronomical verification with the Australian Telescope Compact Array (ATCA) for SKA1-mid [RD18, 20], and the Australian SKA Pathfinder (ASKAP) for SKA1-low [RD19]. This has demonstrated that UWA's *SKA phase-synchronisation system* is fully compliant with all SKA requirements, as well as demonstrated functionality of critical practical factors that are not captured by the requirements. The full list of all relevant requirements is displayed in a series of tables in Section 6 Statement of Compliance.

The principal requirements of the *SKA Phase Synchronisation System* dictate that the distributed reference signals to be stable enough, over all relevant timescales, to ensure a sufficiently low coherence loss of the telescope array. Therefore, measurements of the frequency transfer stability of UWA's *SKA Phase Synchronisation System* form the core data set against which these Functional Performance Requirements (listed in Section 3.3.2) are evaluated.

Before describing these design justifications, two critical aspects relating to the measurement methods and equipment used to estimate frequency stability, are highlighted in Section 3.3.1 below.

### 3.3.1 **Measurement Methods and Equipment**

#### 3.3.1.1 Impact of the Measurement Device on the Estimate of Phase Coherence

The first comment relates to the primary equipment traditionally used in laboratory testing by the global metrology community to estimate frequency stability, the frequency counter. A frequency counter records the value of frequency as a function of time. From these data, a range of statistical algorithms (for example, Allan deviation) can be used to calculate estimates of frequency stability [RD36]. However, different models of frequency counters (and related similar equipment) can produce data sets with various biases; and the aforementioned statistical algorithms must be selected according to the bias of the dataset in order to ensure accurate estimates.

Figure 7 shows two estimates of fractional frequency stability of the UWA's *SKA Phase Synchronisation System* derived from two measurements made concurrently using an Agilent 53132A frequency counter (Λ-weighted measurements with dead-time, dark blue circles) and a Microsemi 5125A phase noise test set (Π-weighted, dead-time free, light blue triangles).

The stability measurement using the Agilent counter follows a power-law close to $\tau^{-1/2}$ while the Microsemi measurement is closer to $\tau^{-1}$. The reason for this discrepancy is that measurements made with dead-time (in the manner we have done at UWA) are known to bias the fractional frequency stability from $\tau^{-1}$ to $\tau^{-1/2}$ [RD37]. It is common practice in frequency metrology to use dead-time counters such as the Agilent (although they do not produce Allan deviation measurements) to produce fractional frequency stability estimates as long as the measurement setup is clearly stated.

For fractional frequency stability measurements made with dead-time, both white phase noise and white frequency noise follow power-laws of $\tau^{-1}$ [RD37]. In the same way that Allan deviation measurements cannot distinguish between white phase noise and flicker phase noise (both having a slope of $\tau^{-1/2}$), measurements made using a counter with dead-time cannot distinguish between white phase and white frequency noise.

The dead-time free Microsemi measurements (which directly output Allan deviation values) show the noise processes to be predominantly white phase noise. For the UWA system, Microsemi (dead-time free) measurements must be used to accurately calculate the coherence loss instead of Agilent frequency counter measurements; however, measurements made with other frequency counters are still totally valid for comparative purposes. More detail on this point is provided in [RD25].

It should also be noted that the functional performance of UWA's *SKA Phase Synchronisation System* was also independently verified using direct measurements with existing astronomical radio interferometers, including the Australia Telescope Compact Array (ATCA) [RD18, 20] and the Australian SKA Pathfinder (ASKAP) [RD19].

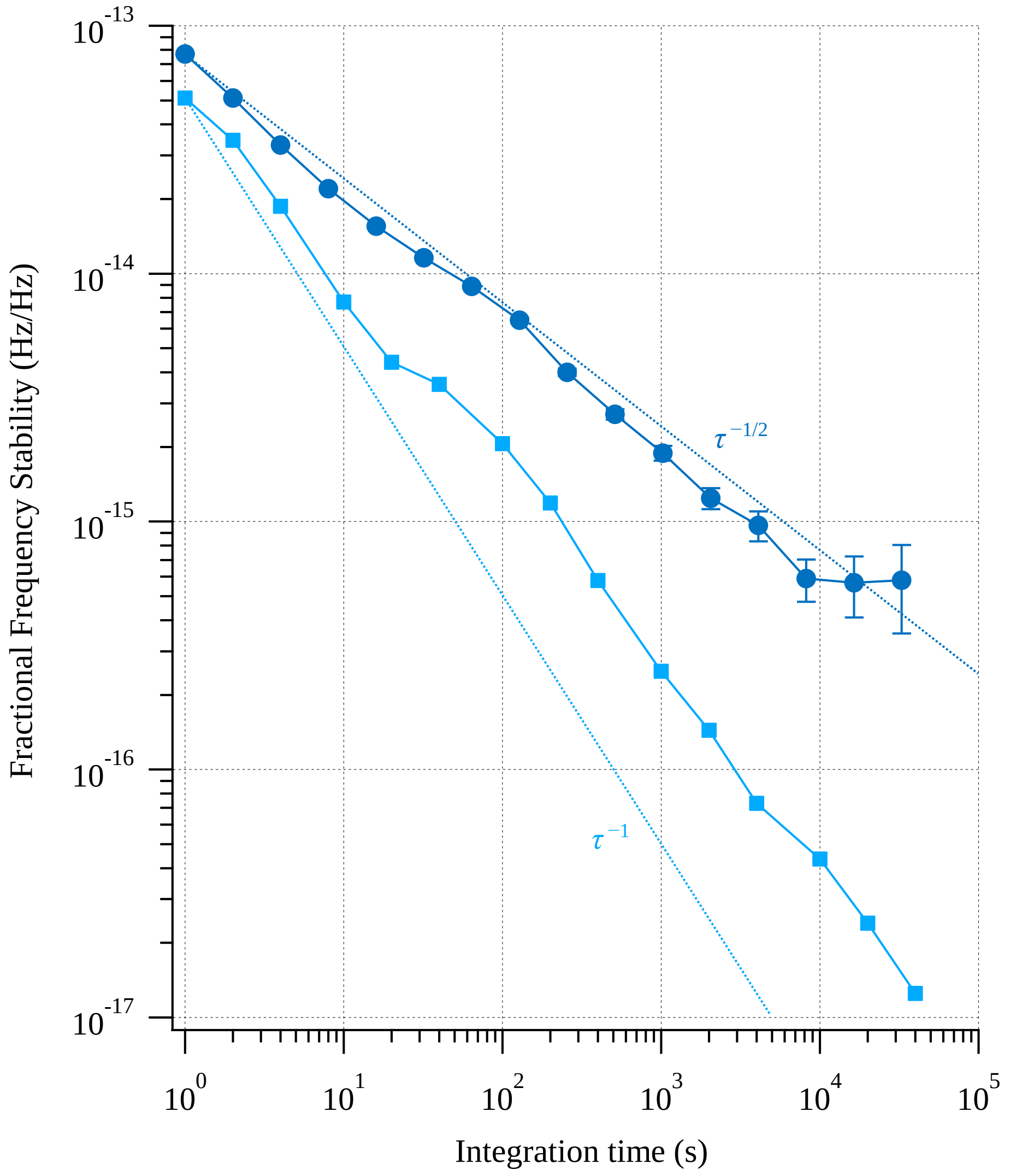

**Figure 7 – Fractional frequency stability of UWA's *SKA Phase Synchronisation System*. Figure 1 of [RD25].**

### 3.3.1.2 Impact of Transmission Frequency on Absolute Frequency Stability

The second aspect of measurement methods and equipment relates to the selection of frequencies for the transmitted reference signals. Due to practical constraints, including the availability of electronic and optical components, as well as the interface requirements of the radio telescopes used in field trials, it was not always possible to transmit the exact reference signals frequency called for in UWA's *SKA Phase Synchronisation System* design. However, as outlined below, the absolute frequency transfer stability of UWA's *SKA Phase Synchronisation System* is largely insensitive to the frequency of the transmitted reference signal [RD16].

Figure 8 shows the absolute frequency stability of the 20 MHz transfer over 144 km of ground fibre (taken from reference [RD16]) extrapolated to the predicted performance of a 166 km link (orange). This is compared to 160 MHz transfer over 166 km of metropolitan fibre link (red) using the modulated photonic signal transfer system from [RD11] on which the 20 MHz transfer system is based. The stability is also compared to 8,000 MHz transfer over the 166 km link (blue) using a related design (from [RD10]).

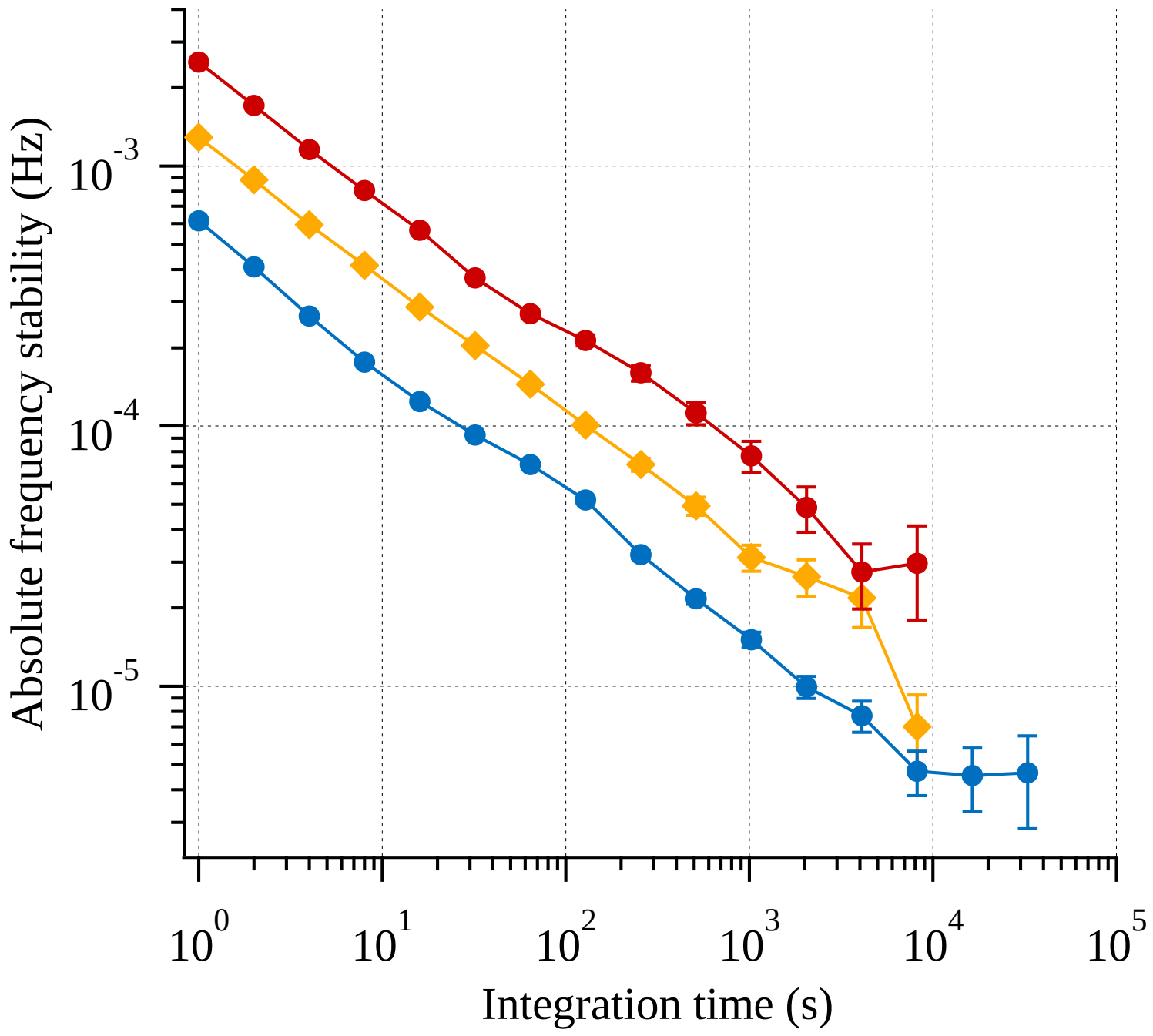

**Figure 8 – Absolute frequency stability of transmissions over 166 km of buried fibre. Figure 5 of [RD16].**

Even though the highest and lowest frequencies differ by a factor of 400, the absolute frequency transfer stabilities are within less than a factor of four of each other. The remaining differences in the absolute stability levels shown by the three traces can be attributed to the fact that the experimental setups were not identical. Although based on the same design, the 20 MHz and 160 MHz transmission systems used different models of photodetectors, filters, mixers, and other electronics, and transmitted over a differently configured link. The design of the 8,000 MHz stabilized transfer system, although related to the design transmitting the lower frequencies, utilized substantially different electronics to accommodate the higher frequencies involved. For the case of UWA's *SKA phase synchronisation system* at the transmitted stability levels reported here, the independence of the absolute frequency stability from the transmitted frequency value means that the fractional frequency stability increases with increasing transmission frequency. More detail on this point is provided in [RD16].

### 3.3.2 **Functional Performance Requirements**

The key innovation of UWA's *SKA phase-synchronisation system* was finding a way to use AOMs as servo-loop actuators for modulated frequency transfer [RD10, 11] – previously these devices had only been used as actuators for optical frequency transfer [RD38]. AOMs have large servo bandwidths as well an infinite feedback range (meaning the servo loop will never require an integrator reset, a common issue with practical implementation of most other techniques).

All 'standard' modulated frequency transfer techniques [RD8], including ALMA's [RD39], require group-delay actuation to compensate the physical length changes of the fibre link. For practical deployments, this usually involves a combination of fibre stretcher (medium actuation speed and very limited range) in series with at least one thermal spool (slow actuation speed and physically bulky). For the number of fibres, link distances, and servo loop actuation ranges encountered by the SKA, especially for SKA1-mid overhead fibre which required actuation ranges hundreds of times greater than comparable-length underground fibre [RD14, 15], standard modulated frequency transfer is totally impractical. Modulated 'phase conjugation' techniques [RD33] have been demonstrated over longer distances [RD34] than standard modulated frequency transfer, and this method is proposed by THU researchers for the SKA.

UWA's *SKA phase-synchronisation system* also utilises AOMs to generate static frequency shifts at the antenna sites to mitigate against unwanted reflections that are inevitably present on real-world links. This mitigation strategy cannot be implemented with modulated frequency transfer techniques, as the carrier and sidebands would be shifted by same frequency. Therefore, these modulated transfer techniques require the returned signal to be rebroadcast at either a different modulation frequency, optical wavelength, or on a different fibre core, to avoid frequency overlap from unwanted reflections on the link. These reflection mitigating methods can bring about additional complications, including those resulting from optical polarization and chromatic dispersion, which in turn requires further complexity. However, reflection mitigation is essential for the *SKA phase-synchronisation system*, as there is no way to guarantee that all links will remain completely free of reflections over the lifetime of the project.

These advantages ensure that UWA's *SKA phase-synchronisation system* can easily meet the functional performance requirements in practical realisation of the system while achieving maximum robustness, minimum cost, simple installation, and easy maintainability. The following sections describe the solution design justifications with respect to the coherence requirements (Section 3.3.2.1), the phase drift requirement (Section 3.3.2.2), and the jitter requirement (Section 3.3.2.3).

#### 3.3.2.1 Coherence Requirements

The coherence requirements for SKA1-low, as defined in [AD1], are:

- SAT.STFR.FRQ shall distribute a frequency reference with no more than 1.9% maximum coherence loss, within a maximum integration period of 1 second, and up to an operating frequency of 350 MHz. (SADT.SAT.STFR.FRQ_REQ-3242); and
- SAT.STFR.FRQ shall distribute a reference frequency with no more than 1.9% coherence loss for intervals of 1 minute, over an operating frequency range of between 350MHz and 13.8GHz (SADT.SAT.STFR.FRQ_REQ-3243).

These requirements are driven by the need for coherence over the correlator integration time (approximately 1 second) and the time for in-beam calibration (1 minute), and arise from random deviations in phase. They are expressed in terms of the coherence loss caused by the phase difference between the two frequency signals delivered at two receptors, or in other words, per baseline. We interpret the coherence requirement as a requirement to be met on the baseline with the worst-case stability, and not as an overall (averaged) coherence loss over the array [RD40]. Measurements are taken on a single frequency transfer link, but then adjusted to provide the coherence between a base-line formed by two antennas referenced by signals transmitted over two independent optical links.

The dominant factor affecting the value of coherence loss realisable with UWA's *SKA phase-synchronisation system* is the choice of frequency of the transmitted reference signals. Therefore, the system was designed to transmit the highest radio frequency practicably realisable given the constraints of its constituent key hardware elements. The reason for this, as described in Section 3.3.1.2, is that the absolute frequency transfer stability of UWA's *SKA Phase Synchronisation System* is largely insensitive to the frequency of the transmitted reference signal [RD16]. This means (all other things being equal) transmitting higher frequency reference signals results in greater fractional frequency stabilities, and therefore lower coherence loss values. Coherence loss scales with the square of fractional frequency stability [RD25, 40], so doubling the fractional frequency stability, increases the coherence loss by a factor of four. The moderating factor is that the LFAA interface specifies a 10 MHz interface (this a common, and therefore convenient, value from which to synthesise the 800 MHz LFAA receiver clock signal)[1].

Investigations of readily available radio-frequency electronic and optoelectronic components led to the conclusion that there seems to be a major relevant technology breaks at around 250-500 MHz (most notably, our preferred radio-frequency photodetector has a cut-off frequency of 250 MHz). As we wanted to keep the electronics complexity of the Receiver Module to a minimum, we only considered only divide-by-2*N* frequency conversions. In addition, our preferred supplier of AOMs produced suitable (low drive power) units at only ±40 MHz (±10%) and ±80 MHz (±10%). This led us to investigate transfer frequencies of 40 MHz, 80 MHz, and 160 MHz. As we wanted to minimise the risk of not meeting the SKA1-low Coherence Requirements, we settled on 160 MHz transfer[2].

Table 5 shows a summary of evaluated coherence loss using a variety of analyses and input data sets ranging from Laboratory Demonstration to Astronomical Verification.

**Table 5 – Compliance against SKA1-low Coherence Requirements**

| Analysis # | Analysis Method | Measuring Instrument | Key Relevant Measurement Parameters | Evaluated Coherence Loss at 1 sec | Excess over Requirement | Evaluated Coherence loss at 1 min | Excess over Requirement |
|---|---|---|---|---|---|---|---|
| Analysis 1 | Laboratory Demonstration; ALMA method [RD41-43] | Microsemi 5125A Test Set | 166 km urban conduit | $1.7\times10^{-5}$ (0.0017%) | 1118 | $9.7\times10^{-5}$ (0.0097%) | 195 |
| Analysis 2 | Laboratory Demonstration; Thompson, Moran, and Swenson method [RD44] | Microsemi 5125A Test Set | 166 km urban conduit | $7.5\times10^{-5}$ (0.0075%) | 253 | $2.2\times10^{-4}$ (0.022%) | 86 |
| Analysis 3 | Laboratory Demonstration; Thompson, Moran, and Swenson method [RD44] | Agilent 53132A frequency counter | 166 km urban conduit | $2.0\times10^{-4}$ (0.020%) | 95 | $1.0\times10^{-3}$ (0.10%) | 19 |

[1] The possibility of providing a frequency closer to 800 MHz (such as 200 or 400 MHz) was discussed with members of the LFAA consortium; however, this was turned-down as a 10 MHz reference signal provides much greater flexibility to vary the ADC clock frequency.
[2] 40 MHz transfer would have slightly simplified some detailed design elements, but the expected factor-of-sixteen degradation in coherence loss was judged to be too close not meeting the SKA1-low Coherence Requirements.

The coherence loss values demonstrates that regardless of which analysis is used to estimate the coherence loss, UWA's *SKA Phase Synchronisation System* is more than two orders-of-magnitude below the SKA1-low coherence loss for 1 second, and two orders-of-magnitude below the SKA1-low coherence loss for 1 minute, for the most challenging conditions to be faced for SKA1-low (highest operating frequency and over fibre links lengths longer than those that will be encountered)[3].

These outcome justifies the functional performance of UWA's *SKA Phase Synchronisation System.* While it does not achieve the same raw fractional frequency transfer performance of some other published radio-frequency transfer techniques, the technique is able to optimise other key design parameters (maximum robustness, minimum cost, simple installation, easy maintainability), while exceeding the functional performance needs of the SKA with ample margin

The detailed explanation of the experimental methods and frequency transfer performance results for Analysis 1 and Analysis 2 are given in:

- D.R. Gozzard, S.W. Schediwy, R. Whitaker, and K. Grainge. *Simple Stabilized Radio-Frequency Transfer with Optical Phase Actuation*. Submitted to Photonics Technology Letters (2017).
  Appendix 7.2.5, Reference [RD11]; and,
- S.W. Schediwy and D.G. Gozzard, *Pre-CDR Laboratory Verification of UWA's SKA Synchronisation System*. SADT Report **620** (2017) 1-79.
  Appendix 7.3.2, Reference [18].

As the Photonics Technology Letters journal paper was targeted at a frequency metrology audience, it does not include a discussion of the coherence loss. This is only covered in the SADT Report, and are also summarised in [RD5].

In addition, Astronomical Verification of the frequency transfer performance of SKA1-mid variant of UWA's *SKA Phase Synchronisation System* was conducted with ASKAP. The detailed explanation of the experimental methods and frequency transfer performance results are given in:

- S.W. Schediwy and D.G. Gozzard, *SKA-low Astronomical Verification*. SADT Report **617** (2015) 1-21.
  Appendix 7.3.4, Reference [RD19].

As the interface with ASKAP did not allow for the separation of instability between the frequency transfer system and the instability due to fluctuations of the atmosphere, this field trip confirmed the performance so validates the performance of the Laboratory Demonstrations, but could not be used to determine the coherence loss.

The experimental parameters outlined above form the baseline for all other related measurements described in this report. Any variation from these parameters that was required to be made in order to test against the full range of Normal Operating Conditions discussed in Section 3.3.3 and Key Additional Requirements discussed in Section3.3.4, are explicitly stated where relevant. Further details of experimental parameters for each of the 14 test campaigns are summarised in Section 4.

[3] The longest SKA1-low fibre link distance is only 58 km, so a single Perth fibre loop of 83 km would have been a more suitable measurement; however, we were also motivated in comparing the performance with the SKA1-mid microwave-frequency transfer which was measured over 166 km.

#### 3.3.2.2 Phase Drift Requirement

The phase drift requirement is [AD1]:

- SAT.STFR.FRQ shall distribute a reference frequency to a performance allowing a maximum of 1 radian phase drift for intervals up to 10 minutes, and up to an operating frequency of 350 MHz. (SADT.SAT.STFR.FRQ_REQ-3244).

This requirement is driven by the need for the phase on a calibrator, measured each side of an observation of up to 10 minutes, not to change by more than a radian during the observation. This is to ensure no wrap-round ambiguity in the phase solution. This is a requirement on phase drift, which includes both the systematic and random phase fluctuations [RD40].

The key parameter affecting phase drift is the accuracy of the frequency transfer technology. That is, if the frequency delivered is offset from the nominal value, then this will result in a linear phase drift; if that offset changes with time then so will the phase drift. UWA's *SKA phase-synchronisation system* is based on frequency transfer technology that is inherently phase-accurate. By essentially being a phase-locked loop, it is the phase delivered at the Antenna site that is stabilised (through actuation of the transmitted frequency). Other design choices, such as only having one master signal generator, ensures that the phase accuracy is maintained. It has been shown that using multiple synthesisers can easily lead to a significant loss of coherence, even if the transmission frequency being successfully stabilised [RD18, 20]. This is because for a synthesiser to maintain its output frequency relative to its reference, it must change the phase of its output as its internal temperature changes in response to variations in ambient temperature.

Table 6 shows a summary of evaluated phase drift using Laboratory Demonstration.

**Table 6 – Compliance against SKA1-low Phase Drift Requirements**

| Analysis # | Analysis Method | Measuring Instrument | Key Relevant Measurement Parameters | Evaluated Phase Drift at 10 mins | Excess over Requirement |
|---|---|---|---|---|---|
| Analysis 1 | Laboratory Demonstration, direct phase drift measurement | Agilent 34401A Multimeter | 83 km urban conduit | 8 µrad (1 s.d.) | 125,000 |

The phase drift results shows that UWA's *SKA Phase Synchronisation System* is more than five orders-of-magnitude below the SKA1-low phase drift requirement (REQ-2693). The 10 minute phase drift is dependent on both systematic and random phase fluctuations; however, as it is a long-timescale process, it should be dominated by the systematic effects of the transmitter and receiver units [RD45]. This means it should not vary with link length.

The detailed explanation of the experimental methods and results for Analysis 1 is given in:

- S.W. Schediwy and D.G. Gozzard, *Pre-CDR Laboratory Verification of UWA's SKA Synchronisation System*. SADT Report **620** (2017) 1-79.
  Appendix 7.3.2, Reference [18].

The results are also summarised in [RD5].

#### 3.3.2.3 Jitter Requirement

The jitter requirement is [AD1]:

- Jitter shall be equal to or less than 74 femtoseconds for Low as defined by EICD *100-0000000-026_03_SADTtoLFAA_ICD*.

Any frequency transfer technique is ultimately limited in the bandwidth of frequencies of noise which can suppressed by the light round-trip time of the fibre link. This is because random noise processes that occur at time-scales faster than the light round-trip time will have de-cohered between the time of the light going one way and then returning. This means the servo loop cannot sense this noise, and therefore cannot correct for it.

To overcome this limitation UWA's *SKA phase-synchronisation system* incorporates an industry-standard small form-factor oven-controlled clean-up oscillator (OCXO) into Receiver Module to provide phase coherence at timescales shorter than the light round-trip time of the fibre link. The OCXO is tied to the incoming reference signals using a simple encapsulated phase-locked loop. The circuit design and mechanical enclosure is based off the proven design used by ASKAP.

The jitter for the SKA1-low is calculated from the phase noise as given in the datasheet of our preferred supplier's OCXO, between 10 MHz and the maximum offset frequency (as per SADT guidelines given in Section 6.3 of [RD40]). Table 7 shows a summary of the evaluated jitter determined from the phase noise data of the SKA1-low oven-controlled clean-up oscillator.

**Table 7 – Compliance against SKA1-low Jitter Requirement**

| Analysis # | Analysis Method | Evaluated Jitter | Excess over Requirement |
|---|---|---|---|
| Analysis 1 | SADTtoLFAA_ICD; 74 fs [RD46] | 53 fs | 1.40 |

Analysis 1 shows that the jitter of our preferred supplier's OCXO falls is able to exceed the value of 74 fs stated in the LFAA ICD.

The phase noise data and jitter calculation are described in:

- S.W. Schediwy and D.G. Gozzard, *Pre-CDR Laboratory Verification of UWA's SKA Synchronisation System*. SADT Report **620** (2017) 1-79.
  Appendix 7.3.2, Reference [18].

These results are also summarised in [RD5].

We are confident that the supplier's datasheet phase noise is not only accurate, but can be realised in practice. This is because we also evaluated the timing jitter of proxy OCXO used in an early prototype system (described in Section 3 of [RD13]). For this OCXO, the phase noise was measured using two different instruments and methods, and then the timing jitter was evaluated from these two data sets (as well as the specification phase noise) using these two different calculators. All analyses were with within 4 fs of each other.

Furthermore, SKAO's own analysis of the jitter budget produces a value of 110 fs for the most stringent case of 10 cases considered; 800 MHz sampling frequency, max observing frequency (350 MHz) max effective number of bits (12).

The OCXO in UWA's current SKA1-low detailed design uses an industry standard Europack form-factor. If the requirement is relaxed in the future the OCXO could simply be replaced with a lower quality device, thereby significantly lowering the cost of the UWA's proposed design. In addition, the cross-over frequency of the phase-locked loop (that is, the boundary frequency between where the frequency transfer system is dominant compared with the clean-up oscillator) can be set to be optimised given the phase noise of the clean-up oscillator and the average (or worst) residual phase noise of the frequency transfer system (see [RD13]). While the phase noise of the oscillator is known, the residual phase noise of the frequency transfer system can depend on the particular noise characteristics of the fibre link infrastructure, which will not be completely known until after installation.

### 3.3.3 Normal Operating Conditions

As show in the previous section (Section 3.3.2), the functional performance of UWA's *SKA phase synchronisation system* has been proven using a number different measurement techniques. These techniques include: Allan deviation derived from a Microsemi 5125A phase noise test set; fractional frequency computed from data logged with an Agilent 53132A frequency counter; calculation of the coherence loss directly from phase noise data under Astronomical Verification conditions; direct measurements of the phase drift using an Agilent 34410A digital multimeter; phase drift analysis from Astronomical Verification data; and measurements of the short time-scale phase noise using a Microsemi 5125A phase noise test and a Rohde & Schwarz FSWP. These comparisons have provided confidence that any of these measurement techniques may be used to accurately evaluate the performance of UWA's *SKA phase synchronisation system*.

The full range of Normal Operating Conditions could not be evaluated using all of the measurement techniques outlined above. For a range of laboratory demonstration and field trials, the instruments that produce true Allan deviation or phase noise measurements were not always available, and so the much smaller and cheaper Agilent frequency counter was often selected as the most suitable instrument to make the measurements of the system performance.

#### 3.3.3.1 Ambient Temperature and Humidity Requirements

##### 3.3.3.1.1 Within the Remote Processing Facility

The ambient temperature and humidity requirements for the components of SKA1-low that are located within the Remote Processing Facility, as defined in [AD1], are:

- SAT.STFR.FRQ components sited within the Remote Processing Facility (RPF) shall withstand, and under normal operating conditions operate within specification, a fluctuating thermal environment between +18°C and +26°C. (SADT.SAT.STFR.FRQ_REQ-105-075).
- SAT.STFR.FRQ components sited within the Remote Processing Facility (RPF) shall withstand, and operate within, a fluctuating non-condensing, relative humidity environment between 40% and 60%. (SADT.SAT.STFR.FRQ_REQ-105-076).

As outlined in Section 3.2.4, the elements of UWA's *SKA phase-synchronisation system* that are present at this location are:

- Receiver Module (141-022300)

The Receiver Module will be located inside the climate controlled Remote Processing Facility. The Receiver Module is designed to be robust against the required temperature and humidity range. The Receiver Module is designed to use only a minimum number of electronic components, and these are housed inside a small form-factor metallic screening enclosure. All electronic and optical components in the Receiver Module have been specified to function properly in excess of this temperature and humidity range. The design of the Receiver Module also makes its functional performance insensitive to the environmental changes

The performance of the system has been verified by subjecting the Receiver Module and Optical Amplifier to temperature and humidity changes in excess of the SKA requirements. Table 8 shows a summary of evaluated coherence loss tested using a Laboratory Demonstration against the ambient temperature and humidity requirements for equipment located at the DISH pedestal.

**Table 8 – Compliance against SKA1-low Ambient Temperature and Humidity Requirements RPF**

| Analysis # | Analysis Method | Measuring Instrument | Key Relevant Measurement Parameters | Evaluated Coherence Loss at 1 sec | Excess over Requirement | Evaluated Coherence loss at 1 min | Excess over Requirement |
|---|---|---|---|---|---|---|---|
| Analysis 1 | Laboratory Demonstration; Thompson, Moran, and Swenson method [RD44] | Agilent 53132A frequency counter | temp. range +16°C to +32°C; temp. gradient ≤1.7°C/10 mins; 1 m fibre patch | $1.2\times10^{-5}$ (0.0012%) | 1,620 | $6.0\times10^{-5}$ (0.006%) | 317 |

The system continued to perform to specification, exceeding the 1 s and 60 s coherence loss requirements by three orders-and-magnitude and two orders-of-magnitude respectively. The detailed explanation of the experimental methods and results for Analysis 1 is given in:

- Section 5 of: S.W. Schediwy and D.G. Gozzard, *Pre-CDR Laboratory Verification of UWA's SKA Synchronisation System*. SADT Report **620** (2017) 1-79.
  Appendix 7.3.2, Reference [18].

Regarding the impact on phase drift; the raw frequency offset time traces in [18] show absolute no indication of phase drift with changes in temperature or humidity.

In addition, we have tested a proxy SKA1-low OCXO (same company, same product range, different output frequency). The specification sheet indicates operation between −40 to +85°C. This is the same specification as the SKA1-mid OCXO selected for the detailed Design for Mass Manufacture of UWA's *SKA phase-synchronisation system*.

#### 3.3.3.1.2 Within the Central Processing Facility

The ambient temperature and humidity requirements for the components of SKA1-low that are located within the Central Processing Facility (CPF), as defined in [AD1], are:

- SAT.STFR.FRQ components sited within the Central Processing Facility (CPF) shall withstand, and under normal operating conditions operate within specification, a fluctuating thermal environment between +18°C and +26°C (SADT.SAT.STFR.FRQ_REQ-105-079).
- SAT.STFR.FRQ components sited within the Central Processing Facility (CPF) shall withstand, and operate within, a fluctuating non-condensing, relative humidity environment between 40% and 60% (SADT.SAT.STFR.FRQ_REQ-105-080).

As outlined in Section 3.2.4, the elements of UWA's *SKA phase-synchronisation system* that are present at this location are:

- Rack Cabinet (141-022900)
- Optical Source (141-022400)
- Signal Generator (141-023100)
- Sub Rack (141-022700)
- Transmitter Module (141-022100)

The above listed equipment will be located inside the SKA1-low Central Processing Facility [RD26]. The moderate temperature range reflect that this equipment will be house inside a large environmentally controlled facility.

Nonetheless, from previous experience of other frequency transfer systems (see Section 3 of [RD19]) the conditions (particularly acoustic-frequency vibrations) even in these type of facilities can potentially impact the equipment. For this reason, UWA's *SKA phase-synchronisation system* is designed in such a way as to also stabilise the optical wavelength-scale perturbations of the non-common optical fibre paths across the CPF equipment (see Section3.2.3), by as much as 120 dB (see Section 4 of [RD19]). This makes the equipment in the CPF extremely robust to external environmental changes.

The performance of the system has been verified by subjecting complete transmitter system to temperature and humidity changes in excess of the SKA requirements. The tests and the performance of the system are described in detail in the [RD13] The test showed the system to be fully compliant with the requirements as shown below. Table 9 shows a summary of evaluated coherence loss tested using a Laboratory Demonstration against the ambient temperature and humidity requirements for equipment located at the Central Processing Facility.

**Table 9 – Compliance against SKA1-low Ambient Temperature and Humidity Requirements CPF**

| Analysis # | Analysis Method | Measuring Instrument | Key Relevant Measurement Parameters | Evaluated Coherence Loss at 1 sec | Excess over Requirement | Evaluated Coherence loss at 1 min | Excess over Requirement |
|---|---|---|---|---|---|---|---|
| Analysis 1 | Laboratory Demonstration; Thompson, Moran, and Swenson method [RD44] | Agilent 53132A frequency counter | temp. range +16°C to +32°C; temp. gradient ≤1.7°C/10 mins; 1 m fibre patch | $1.2{\times}10^{-5}$ (0.0012%) | 1,620 | $6.0{\times}10^{-5}$ (0.006%) | 317 |

The complete transmitter system was subjected to a temperature range of +16°C to +32°C with temperature gradients of 1.7°C/10 mins. The system continued to perform to specification, exceeding the 1 s and 60 s coherence loss requirements by three and two orders-of-magnitude respectively. The detailed explanation of the experimental methods and results for Analysis 1 is given in:

- Section 5 of: S.W. Schediwy and D.G. Gozzard, *Pre-CDR Laboratory Verification of UWA's SKA Synchronisation System*. SADT Report **620** (2017) 1-79.
  Appendix 7.3.2, Reference [18].

#### 3.3.3.1.3 Operating over the Fibre Link

The ambient temperature and humidity requirements for the components of SKA1-low that are located within the Central Processing Facility (CPF), as defined in [AD1], are:

- SAT.STFR.FRQ equipment and fibre located in non-weather protected locations shall be sufficiently environmentally protected to survive, and perform to specification for all ambient temperatures of between -5$^{o}$C and +50$^{o}$C (SADT.SAT.STFR.FRQ_REQ-3070).
- SAT.STFR.FRQ equipment and fibre located in non-weather protected locations shall be sufficiently environmentally protected to survive, and perform to specification for rates of change of ambient temperature of up to ±3$^{o}$C every 10 minutes (SADT.SAT.STFR.FRQ_REQ-3070).

As outlined in Section 3.2.4, there are no elements of UWA's *SKA phase-synchronisation system* that are located in non-weather protected locations – all fibre is buried.

### 3.3.3.2 Wind Speed Requirement

The wind speed requirement for SKA1-low, as defined in [AD1], is:

- SAT.STFR.FRQ equipment and fibre located in non-weather protected locations shall be sufficiently environmentally protected to survive and perform to specification under normal SKA telescope operating wind conditions up to wind speeds of 40km/hr. (SADT.SAT.STFR.FRQ_REQ-3070).

As outlined in Section 3.2.4, there are no elements of UWA's *SKA phase-synchronisation system* that are located in non-weather protected locations – all fibre is buried.

#### 3.3.3.3 Seismic Resilience Requirement

The seismic resilience requirement for SKA1-low, as defined in [AD1], is:

- SAT.STFR.FRQ components shall be fully operational subsequent to seismic events resulting in a maximum instantaneous peak ground acceleration of 1 m/s$^2$. Note: Seismic events include underground collapses in addition to earthquakes (SADT.SAT.STFR.FRQ_REQ-2798).

UWA's *SKA phase-synchronisation system* is designed in such a way as to also stabilise the optical wavelength-scale perturbations of the non-common optical fibre paths in the Transmitter Module. This effectively stabilises the Transmitter Module as well as the fibre link making the equipment in the CPF extremely robust to external environmental changes. The Receiver Module for UWA's *SKA phase-synchronisation system* has very small form-factor and contains only a minimum number of simple optical and analogue electronic components, making it extremely robust to extremal environmental perturbation.

Table 10 shows a summary of compliance against the SKA1-mid seismic resilience requirement.

**Table 10 – Compliance against SKA1-low Seismic Resilience Requirement**

| Analysis # | Analysis Method | Measuring Instrument | Key Relevant Measurement Parameters | Evaluated Acceleration | Equipment Functional | Excess over Requirement |
|---|---|---|---|---|---|---|
| Analysis 1 | Field Trial; car trip | 3-Axis Accelerometer | 10 ms integration | 4 m/s$^2$<br>3 m/s$^2$<br>10 m/s$^2$ | Yes | 4×<br>3×<br>10× |
| Analysis 2 | Rotating equipment | NA | We are located on Earth's surface | 9.81 m/s$^2$ | Yes | 9.81× |

UWA's SKA phase synchronisation system was shown to exceed the SKA1-low seismic resilience requirement by a factor of 10. The detailed explanation of the experimental methods and results for Analysis 1 and Analysis 2 are given in:

- Section 7 of: S.W. Schediwy and D.G. Gozzard, *Pre-CDR Laboratory Verification of UWA's SKA Synchronisation System*. SADT Report **620** (2017) 1-79.
  Appendix 7.3.2, Reference [18].

In addition, prototype equipment was driven 1,600 km by road (including over 550 km of unsealed roads) from Perth to the Murchison Radioastronomy Observatory return for Astronomical Verification with ASKAP [RD19]; and at least 7,500 kilometres from Perth to the Paul Wilde Observatory return for Astronomical Verification with ATCA [RD18, 20].

#### 3.3.3.4 Telescope Configuration Requirement

The telescope configuration requirement for SKA1-low, as defined in [AD1], is:

- SAT.STFR.FRQ shall disseminate the LOW Reference Frequency (the "Disseminated Reference Frequency") to 36 Remote Processing Facilities (RPFs) located on the LOW Spiral Arms as defined by *SKA-TEL-SKO-0000422 SKA1_Low Configuration Coordinates*. (SADT.SAT.STRF.FRQ_REQ-2142).

The SKA1-low telescope comprises a total of 36 remote antenna sites, separated from the Central Processing Facility (CPF) by up to 58 km [RD26] of optical fibre link. As described in greater detail in Section 3.2.4, at the CPF UWA's *SKA phase-synchronisation system* consists of 3 Sub Racks (plus one hot-swappable spare Sub Rack), each containing 16 Transmitter Modules. This system is therefore able to service all 36 optical fibre links, with twelve hot-swappable spare Transmitter Modules (plus an additional 16 Transmitter Modules in the spare Sub Rack).

The 36 actively-used Transmitter Modules then send phase-coherent reference signals via separate optical fibre links to each of the 36 RPF sites. The star-shaped network topology of such a phase-synchronisation system, conveniently matches the fibre topology of the SKA’s data network, which transmits the astronomical data from the RPF sites to the CPF.

As already demonstrated in Section3.3.2, UWA’s *SKA phase-synchronisation system* can deliver the reference signals to even the furthest antenna sites while meeting all the Functional Performance Requirements. This includes exceeding the SKA1-low coherence loss for 1 second requirement and 1 minute requirement by more than three and two orders-of-magnitude respectively; and exceeding the phase drift requirement by over five orders-of-magnitude. The jitter requirement is not dependent on the telescope configuration as the short-term phase noise is entirely determined by the clean-up oscillator in the Receiver Module.

### 3.3.4 **Key Additional Requirements**

#### 3.3.4.1 Monitoring Requirement

The monitoring requirement for SKA1-low, as defined in [AD1], is:

- At least the following STFR component parameters shall be monitored: the Lock signal (indicating that the STFR system is functioning correctly); the Control Voltage (giving an indication of how much control is still available to keep the STFR locked); and the Phase measurement (showing the corrections which have been applied to the frequency to compensate for the effects of changes in the fibre connecting the transmit and receive units of the STFR.FRQ system). (SADT.SAT.STFR.FRQ_REQ-2280).

The SKA telescope is monitored and controlled by the Telescope Monitor system, and this system is interfaced via all other subsystems of the SKA telescope via a monitor and control system 'local' to each SKA Consortium work package. For the UWA's SKA *phase synchronisation system* this SAT.LMC. As outlined in each of the simplified schematics layout figures in Section 3.4.1, SAT.LMC interfaces directly with each of the following SADT.SAT.STFR.FRQ (UWA) elements:

- Optical Source (141-022400)
- Signal Generator (141-023100)
- Sub Rack (141-022700)

For all elements, with the exception of Sub Rack, the SAT.LMC interface for Normal Operations calls for basic status and health monitoring (see Section 3.5.1 for more information).

The SAT.LMC interface with Sub Rack is with the Command and Control Device (CCD), which is a subsystem of the Sub Rack element. SAT.LMC interfaces with the CCD via a single Ethernet connector on the Sub Rack as shown in Figure 5. The primary CCD microcontroller then communicates with individual PIC microcontroller chips located on to each of the sixteen Transmitter Modules within the Sub Rack as shown in Figure 6. This communication is conducted using the RS485 serial transmission standard via the Sub Rack's backplane. The CCD is responsible for monitoring all the required STFR component parameters mentioned above.

As shown in Figure 6, a copy of the servo-loop error signal is fed into the PIC microcontroller chip. This is the basis of the required 'Lock signal'. As shown in Section 3.2.3, the servo-loop error signal is always driven to zero volts when the loop is locked. After low-pass filtering any high frequency leakage through the mixer, the locked error signal coming directly from the mixer typically has a residual plus-and-minus spread around zero volts of a few millivolts. With the servo-loop disengaged, the error signal will swing between the full range of the mixer output (typically a few hundred millivolts). The PIC microcontroller is programmed to report a positive Lock signal if the RMS voltage is below a specified threshold voltage (say 50 millivolts), and a negative Lock signal if the RMS is above this value.

As UWA's SKA *phase synchronisation system* use AOMs as the servo-loop actuators, a Control voltage does not exist. The AOMs suppresses the phase fluctuations of the reference signal transmitted on the fibre link, by actuating in the frequency of the reference signal. As phase is the integral of frequency, the combination of VCO and AOM effectively provide the servo-loop's integrator. This also explains the servo loop has an infinite feedback range and will never requires an integrator reset[4]. This part of the requirement is therefore not applicable.

The Phase measurement (sometime referred to as 'glass box') is conducted using an independent out-of-loop system incorporated into the Transmitter Module [5]. As shown in Figure 6, a photodiode is placed immediately after the output of Mach-Zehnder interferometer. This signal is then mixed with a reference signal provided by the DDS, using an IQ-mixer to produce two DC voltage signals (generated with a 90° phase offset) which are then recorded by PIC microcontroller chip. Using those two inputs, the microcontroller can then continuously determine the accumulated phase change applied by UWA's SKA *phase synchronisation system* without any phase ambiguity.

[4] The servo-loop is limited by the speed by which the fibre link is changing; that is the maximum magnitude of the Doppler shift that the AOM can apply to the reference signal. As the bandwidth of the AOM is several MHz, this is many of orders-of-magnitude greater than the most extreme situations encountered.

A mock-up implementation of both the Lock signal and Phase measurement monitoring were demonstrated to the SADT Consortium during the face-to-face meeting in Perth in 2016. Figure 9 is a photo of the SADT laboratory at UWA showing the mock-up monitoring set-up; a prototype Sub Rack of UWA's SKA *phase synchronisation system* can be seen on the optical table in the foreground, with the oscilloscope located on the shelf above the optical table shown recording one of the monitoring signals.

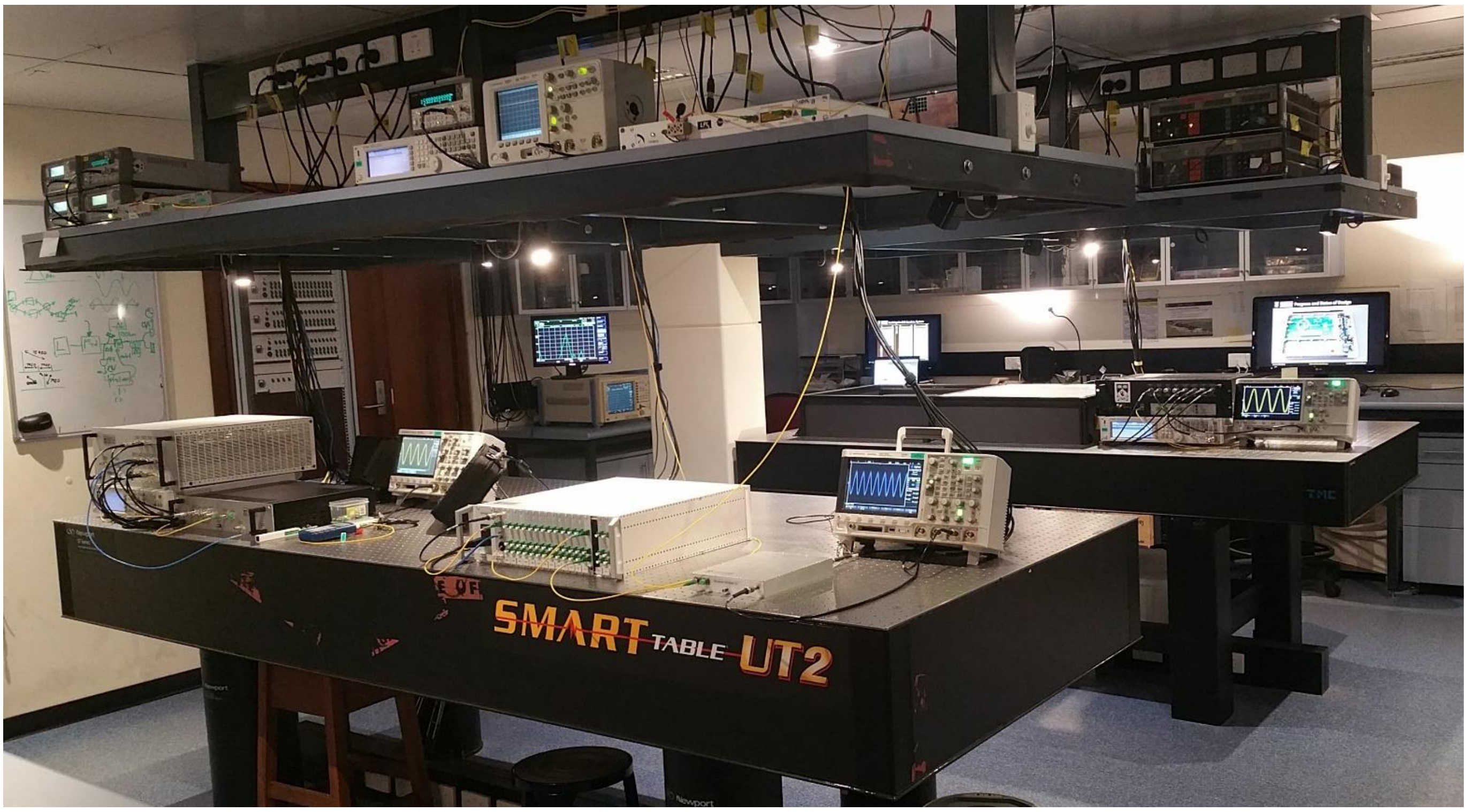

**Figure 9 – Photo of the SADT laboratory at UWA showing the mock-up monitoring set-up.**

[5] Using an independent out-of-loop system, provides a more accurate measurement of applied phase changes, compared to simply monitoring the servo-loop control voltage as is done with other stabilisation system. This is because the transfer function of an actuator (relationship between control voltage and applied correction) is subject to change with time and environmental parameters.

### 3.3.4.2 Radio Frequency Interference Requirement

The radio frequency interference requirement for SKA1-low, as defined in [AD1], are:

- SAT.STFR.FRQ components emitting electromagnetic radiation within frequency intervals for broad and narrow band cases shall be within the SKA RFI/EMI Threshold Levels as defined in *SKA-TEL-SKO-0000202-AG-RFI-ST-01* (SADT.SAT.STFR.FRQ_REQ-2462).

The purpose of these requirements is to ensure that no equipment which is part of the SKA produces radio frequency interference (RFI) at a level high enough to interact detrimentally with the operation and function of the telescope. UWA's *SKA phase synchronisation system* has several design elements which add to ensure low self-generated RFI emission.

Arguably the most critical hardware element of our system in terms of RFI is the Receiver Module. It is to be located within the SKA1-low Remote Processing Facility within close proximity of the surrounding six antenna stations. Our system is designed to use only a minimum number of electronic components at the Receiver Module, and these are housed inside a small form-factor metallic screening enclosure which is based on the proven design used on MeerKAT. The enclose design has a receded lid with many fasting points, shielded DC power feed-throughs, and FC optical connectors (the connector smallest non-metallic cross-section).

The Receiver Module also keeps the number of different electronics signals to a minimal; although it does have to include a clean-up oscillator in a phase-locked loop operating at a different frequency. This minimise any potential impact, this device is in its own encapsulated enclosure within the Receiver Module.

The Transmitter Module PCB uses coplanar waveguides for transmission of its radio-frequency signals on a four layer PCB (so two layers of ground shielding to the adjoining Transmitter Modules to limit cross talk). In addition, the Sub Rack mechanical enclosure can be upgraded with EMC baffles if required (these were not in place in the prototype EMC testing).

All these design choices, plus many more not described here, combine to provide a system that easily surpasses the SKA emitted electromagnetic radiation requirement SADT.SAT.STFR.FRQ_REQ-2462. Table 11 shows a summary of compliance against SKA1-low radio frequency interference requirements for a number of key frequencies that have been identified as being relevant for each location (see Section 3.2.4).

**Table 11 – Compliance against SKA1-low Radio Frequency Interference Requirements**

| Analysis # | Analysis Method | Measuring Instrument | Key Relevant Measurement Parameters | Relevant Frequency | Broad(Narrow)-Band Fieldstrength Threshold Mask | Maximum Measured Fieldstrength | Excess over Broad(Narrow)-Requirement |
|---|---|---|---|---|---|---|---|
| Analysis 1 | Field trail; Independent EMC Test; | Rohde % Schwarz ESU26; ETS-Lindgern semi-anechoic chamber | *CPF Hardware*: Optical Source Signal Generator Sub Rack Transmitter Module | 75 MHz | 97(114) dBµV/m | 48 dBµV/m | 49(66) dB |
| | | | | 85 MHz | 98(114) dBµV/m | 56 dBµV/m | 42(58) dB |
| | | | | 320 MHz | 78(82) dBµV/m | 47 dBµV/m | 31(35) dB |
| | | | | 640 MHz | 84(83) dBµV/m | 39 dBµV/m | 45(44) dB |
| Analysis 2 | Field trail; Independent EMC Test; | Rohde % Schwarz ESU26; ETS-Lindgern semi-anechoic chamber | *Dish Hardware*: Receiver Module | 40 MHz | 92(113) dBµV/m | 23 dBµV/m* | 69(90) dB |
| | | | | 120 MHz | 102(114) dBµV/m | 19 dBµV/m* | 83(95) dB |
| | | | | 160 MHz | 104(115) dBµV/m | 18 dBµV/m | 86(97) dB |

*below measurement noise floor.

As shown in the table, the UWA's *SKA phase synchronisation system* exceeds the broad- and narrow-band

SKA1-low radio frequency interference requirements by more than three orders-of-magnitude.

The detailed explanation of the experimental methods and results for Analysis 1 and Analysis 2 is given in:

- Section 9 of: S.W. Schediwy and D.G. Gozzard, *Pre-CDR Laboratory Verification of UWA's SKA Synchronisation System*. SADT Report **620** (2017) 1-79.
  Appendix 7.3.2, Reference [18].

#### 3.3.4.3 Space Requirement

The space requirement for SKA1-low, as defined in [AD1], is:

- The Candidate's solution shall meet the following maximum space requirements as defined by the space allocated to SAT.STFR.FRQ by the SADT NWA model *SKA-TEL-SADT-0000522-MOD_NWAModelLow Revision 3.0*. (SADT NWA Model SKA-low Rev. 03)

The SKA Telescope's correlator is effectively a world-class supercomputer, and this machine takes up a large amount of space. On the other hand, the Central Processing Facility that houses the correlator, must be constructed alongside the SKA's antennas, in a very remote location, so the significant construction costs constrain to keep the facility as small as possible. For similar reasons, the SKA dish pedestals that house the majority of the dish's equipment, as well as any intermediate shelters, are made as small as possible to keep costs down. For these reasons, space at these three locations is at a premium, and so all SKA equipment located there must be made as small as possible.

For the remaining equipment at the Central Processing Facility, UWA's *SKA phase-synchronisation system* is able to keep volume down through the novel use of very compact AOMs as a servo-loop actuator for microwave-frequency transfer systems [RD10]. Standard stabilised microwave-frequency transfer techniques, for example [RD8], require group-delay actuation to compensate for the physical length changes of the fibre link. For practical deployments over long links, this usually involves implementing a combination of fibre stretcher (medium actuation speed and very limited range) in a series with a thermal spool (slow actuation speed and physically bulky). In addition, as the Transmitter Modules are present at this location in the highest quantity, the design focuses on minimising their footprint by incorporating all optical, electronic, and mechanical elements onto a single 'Eurocard' form-factor printed circuit board (see Section 3.4.1.5).

The compliance against SKA1-mid Space Requirement is summarised given in Table 12.

**Table 12 – Compliance against SKA1-low Space Requirement**

| Analysis # | Analysis Method | Key Relevant Measurement Parameters | Space Requirement | Evaluated Space | Excess over Requirement |
|---|---|---|---|---|---|
| Analysis 1 | Construction of Mass Manufacture Archetypes; detailed 3D physical models computer | *CPF Hardware*:<br>Rack Cabinet<br>Optical Source<br>Signal Generator<br>Sub Rack<br>Transmitter Module | 18U | 18U | 1× |
| Analysis 2 | Construction of Mass Manufacture Archetypes; detailed 3D physical models computer | *RPF Hardware*:<br>Receiver Module | 2U | 1U | 2× |

As outlined in Section 3.2.4, the elements of UWA's *SKA phase-synchronisation system* that are present at the SKA1-low Central Processing Facility are:

- Rack Cabinet (141-022900)
- Optical Source (141-022400)

- Signal Generator (141-023100)
- Sub Rack (141-022700)
- Transmitter Module (141-022100)

Note; Rack Cabinet is a collection of rack cabinet accessories that include equipment for heat management and cable management; and sixteen Transmitter Modules are mounted inside every Sub Rack.

On the 'CPF' tab of the SADT NWA Model SKA-low Rev. 03 [RD26], the STFR.FRQ (UWA) equipment is shown to occupy 18 rack units (U); a total of 12U for Sub Rack, 2U for Rack Cabinet, 2U for Optical Source, and 2U for Signal Generator. This is exactly the situation for UWA's *SKA phase-synchronisation system*. Figure 10 shows a render taken from the 3D physical model of the SKA1-low equipment located at Central Processing Facility.

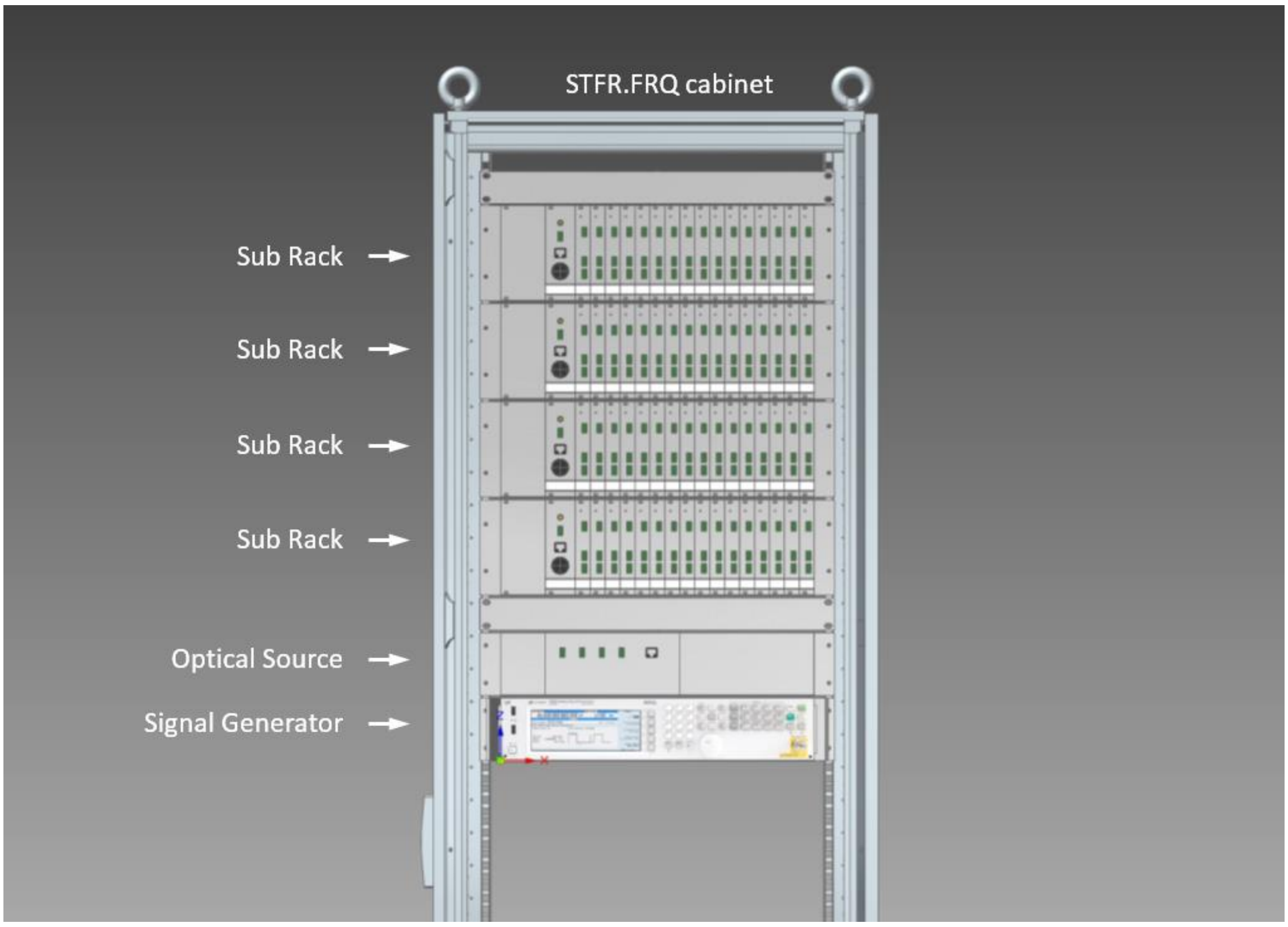

**Figure 10 – SKA1-low equipment located at Central Processing Facility.**

As outlined in Section 3.2.4, the elements of UWA's *SKA phase-synchronisation system* that are present at the SKA1-low Remote Processing Facility are:

- Receiver Module (141-022300)

On the 'RPF' tab of the SADT NWA Model SKA-low Rev. 03 [RD26], the THU equipment is shown to occupy 2 rack units (U) in each of the 36 RFP locations. UWA's *SKA phase-synchronisation system*[6] occupies only 1U. Figure 11 shows a photo of a prototype SKA-low Receiver Module highlighting its 1U dimensions.

[6] Since the release of the SADT NWA Model SKA-low Rev. 03, the detailed design of UWA's *SKA phase-synchronisation system* has been updated to remove the need for any optical amplifiers.

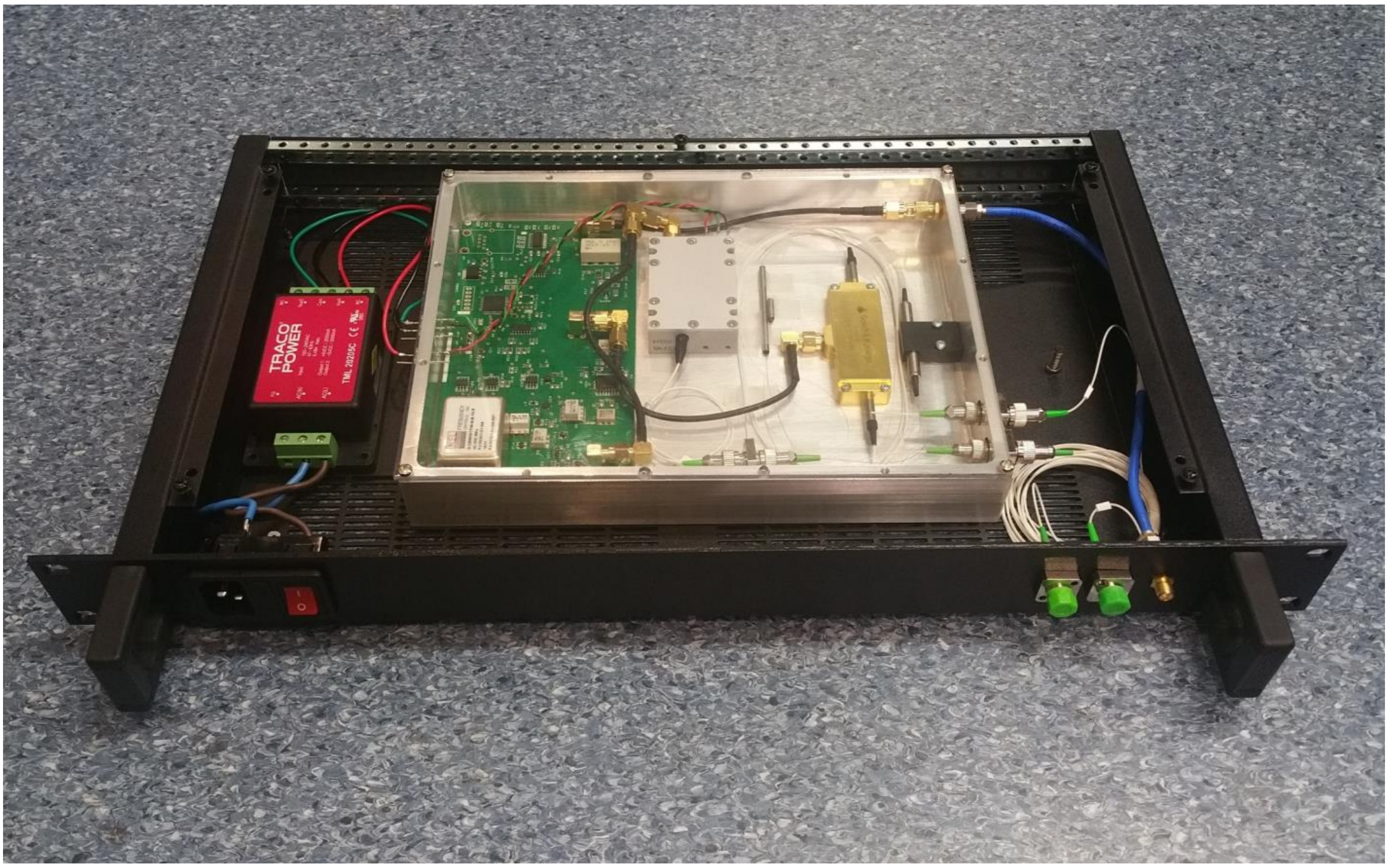

**Figure 11 – SKA1-low equipment located at Remote Processing Facility.**

Renders and/or photos of the individual hardware elements for both the CPF and RPF locations are provided in the relevant sub-sections of the Detailed Design Overview (Section 3.4.1), and further details are provided in Mechanical Detailed Design (Section 3.4.2).

The *SolidEdge* Mechanical Detailed Design Files for are included as a zipped file pack in Appendix 7.5.2.

Further information of the design process for is provided in 'SADT report – Design-for-Manufacture of the SKA1-Mid Frequency Synchronisation System' [RD22], which is included in Appendix 7.3.7. This report focuses on SKA1-mid, but the majority of the design elements are identical between the variants of the two telescopes.

#### 3.3.4.4 Availability Requirements

The space requirement for SKA1-low, as defined in [AD1], is:

- SAT.STFR.FRQ (end-to-end system excluding fibre) shall have 99.9% "Inherent Availability" (SADT.SAT.STFR.FRQ_REQ-3245).

In order to achieve its scientific objectives, the SKA telescope requires to have a high overall duty cycle. As many of the telescope sub-systems, including phase-synchronisation system, are critical for operation, each subsystem must have an even higher inherent availability. UWA's *SKA phase-synchronisation system* has been budgeted with needing to meet an inherent availability 99.9%.

The SADT Consortium contracted the independent firm Frazer-Nash Consultancy, to conduct a 'Reliability, Availability, Maintainability, and Safety' analysis covering each of the SADT sub-elements, including UWA's *SKA phase-synchronisation system*. Figure 12 shows the summary table of SKA1-mid failure modes severity classification taken from the Frazer-Nash Consultancy input worksheet.

As UWA's *SKA phase-synchronisation system* requires three AOMs per link, the AOMs are the overall dominant cost component. However, as demonstrated in the UWA *Detailed Cost Model* (Section 3.10), the dominant cost component and large quantify in use in the system, the AOMs represent the biggest risk in terms of reliability, availability, and maintainability. However, the Frazer-Nash Consultancy's Reliability, Availability, Maintainability, and Safety analysis concludes that demonstrated that the AOM risk criticality is 'tolerable'.

However, the SADT Consortium have identified errors with the reliability data recorded by Frazer-Nash Consultancy, as well as errors in applying the required formulation to determine an estimate of availability. Therefore, the SADT Consortium have decided to conduct an internal availability analysis, using the correct reliability data provided by each work package as input. However, at the time of writing, this analysis has yet to be completed for any SADT work package, so no independent analysis to produce a quantifiable value of inherent availability can be provided to support this requirement.

**Severity Classification SKA1-LOW**

| Level | Class | Consequence to persons or environment | Count |
|---|---|---|---|
| Catastrophic | I | - Total System Loss<br>- High Precision Time Data Loss<br>- Low Precision Time Data Loss<br>- Science Data Loss<br>- A failure that results in Death | 0 |
| Critical | II | - Major Equipment Failure<br>- Major System Damage<br>- Loss of major science programming<br>- Loss of significant time period of science programming<br>- Serious Injury | 0 |
| Major | III | - Partial loss of science programme<br>- Failure causing minor system damage resulting in degraded output, operation , or availability<br>- Failure resulting in minor injury requiring first aid | 4 |
| Minor | IV | - A failure not serious enough to cause injury<br>- A failure not serious enough to cause system damage, but will result in unscheduled maintenance or repair. | 9 |
| Not Credible | N/A | Failure mode not credible during normal operation. | 0 |

**Figure 12 – SKA1-low failure modes severity classification**

The SADT Project Manager indicated that this report can be submitted without this information. In the absence of this information we can only state the following (and this is backed-up by all three independent assessors (see Section 4.2): All aspects of UWA's *SKA phase-synchronisation system* detailed design is engineered using best practice principles; developed in partnership with (University of Manchester, CSIRO, SKA SA); based on the proven designs from previous interferometer telescope array (ALMA, ASKAP, ATCA) and from the international metrology community; and uses only industry standard high-volume commercial components. All indications, based on our testing and use of the various prototypes, and more importantly, our mass-manufacture archetypes over the last three-and-a-half years, is that UWA's *SKA phase-synchronisation system* will prove to have an inherent availability over the project lifetime significantly exceeding the stated requirement.

#### 3.3.4.5 Power Requirements

The radio frequency interference requirements for SKA1-low, as defined in [AD1], are:

- The total power consumption of combined SAT.STFR.FRQ components located in Central Processing Facility (CPF) (LOW) shall be no more than 1.2 kW. (SADT.SAT.STFR.FRQ_REQ-105-148).
- The total power consumption of combined SAT.STFR.FRQ components located in each Remote Processing Facility (RPF) (LOW) shall be no more than 70 Watts (SADT.SAT.STFR.FRQ_REQ-105-149).

The cost of electrical power is one of the primary total cost of ownership limitations for the SKA telescope; the need for world-class power-hungry supercomputing processing being required at very remote locations leads the need to limit the power use of all other SKA subsystems.

UWA's *SKA phase-synchronisation system* for SKA1-low is designed using a combination 6 line replaceable unit building blocks (see Section3.2.4). The two blocks that are replicated with the highest quantity in our design are the Transmitter Module and Receiver Module; and these are built using a bespoke printed circuit board using a highly efficient surface-mount component design. On these PCB's the most power-hungry component is the high-power amplifiers used to drive the two AOMs in Transmitter Module and one AOM in the Receiver Module. We specifically selected low drive-power AOMs, and this allowed us to keep the Transmitter Module power to within 6 W. Other building blocks are requested in much lower quantities, and so the power use per unit is less critical. The Compliance against SKA1-low Power Requirements is summarised in Table 13.

**Table 13 – Compliance against SKA1-low Power Requirements**

| Analysis # | Analysis Method | Measuring Instrument | Key Relevant Measurement Parameters | LRU | Evaluated Power Use | Power Require-ment | Excess over Requirement |
|---|---|---|---|---|---|---|---|
| Analysis 1 | Laboratory demonstration; equipment spec. sheets | Emona EL-302RD Power supply; | Location; Dish Shielded Cabinet | Total | 5 W | 70 W | 14.0× |
| | | | | *Receiver Module* | *5 W* | | |
| Analysis 2 | Laboratory demonstration; equipment spec. sheets | Emona EL-302RD Power supply; | Location; Central Processing Facility | Total | 696 W | 1.2 kW | 1.7× |
| | | | | *Rack Cabinet* | *2× 24 W* | | |
| | | | | *Optical Source* | *80 W* | | |
| | | | | *Signal Generator* | *250 W* | | |
| | | | | *Sub Rack* | *3× 10 W** | | |
| | | | | *Transmitter Module* | *48× 6 W* | | |

*Note, apart from supplying a small number of active components, the primary task of the Sub Rack power supply is to provide the power to the 16 Transmitter Modules. Therefore, the power listed in this cell only includes the power of the components specific to Sub Rack, and the power supply inefficiency of providing power to the Transmitter Modules. Only 3 of the 4 installed Sub Racks are powered at any one time, hence a total of 48 powered Transmitter Units.

It should be noted that this is recursive requirement, as this requirement was based on the estimate of the power consumption values previously provided by the design teams.

### 3.3.5 **Other Key System Parameters**

#### 3.3.5.1 Reflection Mitigation

UWA's *SKA phase-synchronisation system* also utilises AOMs to generate static frequency shifts at the antenna sites to mitigate against unwanted reflections that are inevitably present on real-world links. This mitigation strategy cannot be implemented with modulated frequency transfer techniques, as the carrier and sidebands would be shifted by same frequency. Therefore, these modulated transfer techniques require the returned signal to be rebroadcast at either a different modulation frequency, optical wavelength, or fibre core, to avoid frequency overlap from unwanted reflections on the link. These reflection mitigating methods then can bring about additional complications, including those resulting from optical polarization and chromatic dispersion, which in turn requires further complexity. However, reflection mitigation is absolutely essential for the *SKA phase-synchronisation system*, as there is no way to guarantee that all links will remain completely free of reflections (even if they are free of reflections at the start of operations) over the lifetime of the project.

UWA's *SKA phase-synchronisation system* has been successfully deployed on UWA-Pawsey 31 km fibre link which contains large optical reflections typical of a link that uses optical connectors. The optical time-domain reflectometry (OTDR) trace for one 15.5 km half of the loop-back is displayed in Figure 13. Several large reflections are evident in the trace.

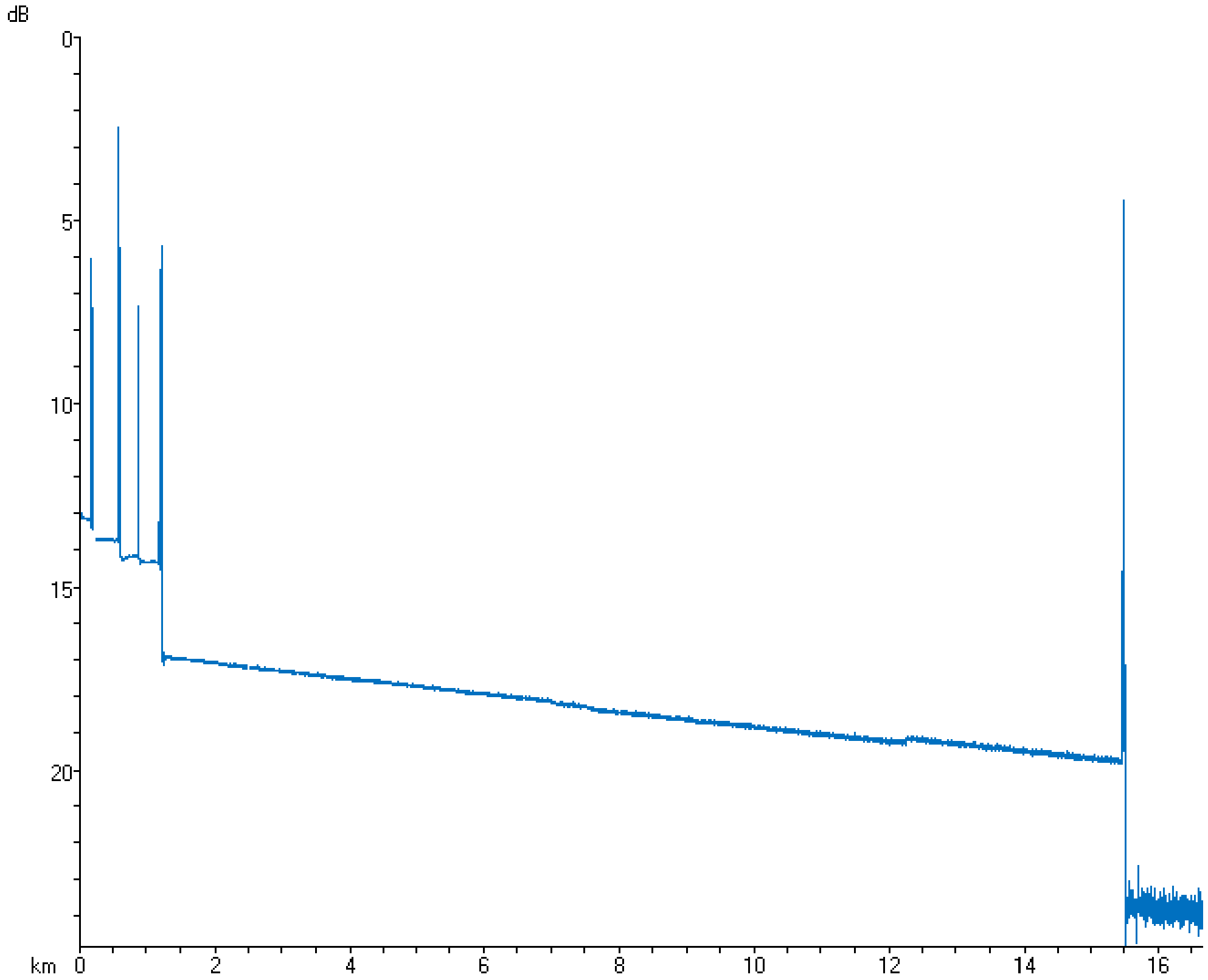

**Figure 13 - Optical time reflectometry trace of the UWA-Pawsey 31 km active fibre link. Figure 5 of [RD12].**

#### 3.3.5.2 Miscellaneous Key System Parameters

UWA's *SKA phase-synchronisation system* use of optical phase sensing allows for the use of Faraday mirrors to give maximum detected signal at the servo photodetector, as is done with stabilised optical transfer. This removes the need for any initial polarisation alignment, or any ongoing polarisation control or polarisation scrambling. The technique can be deployed on standard fibre links and does not require specialty fibre in the fibre link (such as dispersion compensation or polarisation maintaining fibre).

The microwave signal being transmitted on the fibre link arises from only two optical signals, not three as is the case for standard intensity modulation commonly used in RF or MW transfer. Using only two optical signals ensures that that the maximum signal power is available at the antenna site regardless of link length.

Simple and cheap bi-directional optical amplifiers are deployed to extend the range of transmission, and, therefore, potentially complex electronic signal re-generation systems are not required. This also eliminates the potential for a remotely located single point of failure affecting multiple end stations if one re-generation systems are used in a branching system to supply multiple end points. UWA's *SKA phase-synchronisation system* therefore requires only a single laser, reducing system complexity.

## 3.4 Hardware

The SKA1-low Hardware section is split into the following sub-sections:

- Overview of the Detailed Design (3.4.1)
- Mechanical Detailed Design (3.4.2)
- Optical Detailed Design (3.4.3)
- Electronic Detailed Design (3.4.4)

### 3.4.1 Overview of the Detailed Design

The SKA1-low equipment is split into the following six line replaceable units (LRUs):

- Rack Cabinet (141-022900)
- Optical Source (141-022400)
- Signal Generator (141-023100)
- Sub Rack (141-022700)
- Transmitter Module (141-022100)
- Receiver Module (141-022300)

The interdependencies of the STFR.FRQ LRUs are shown in Figure 14.

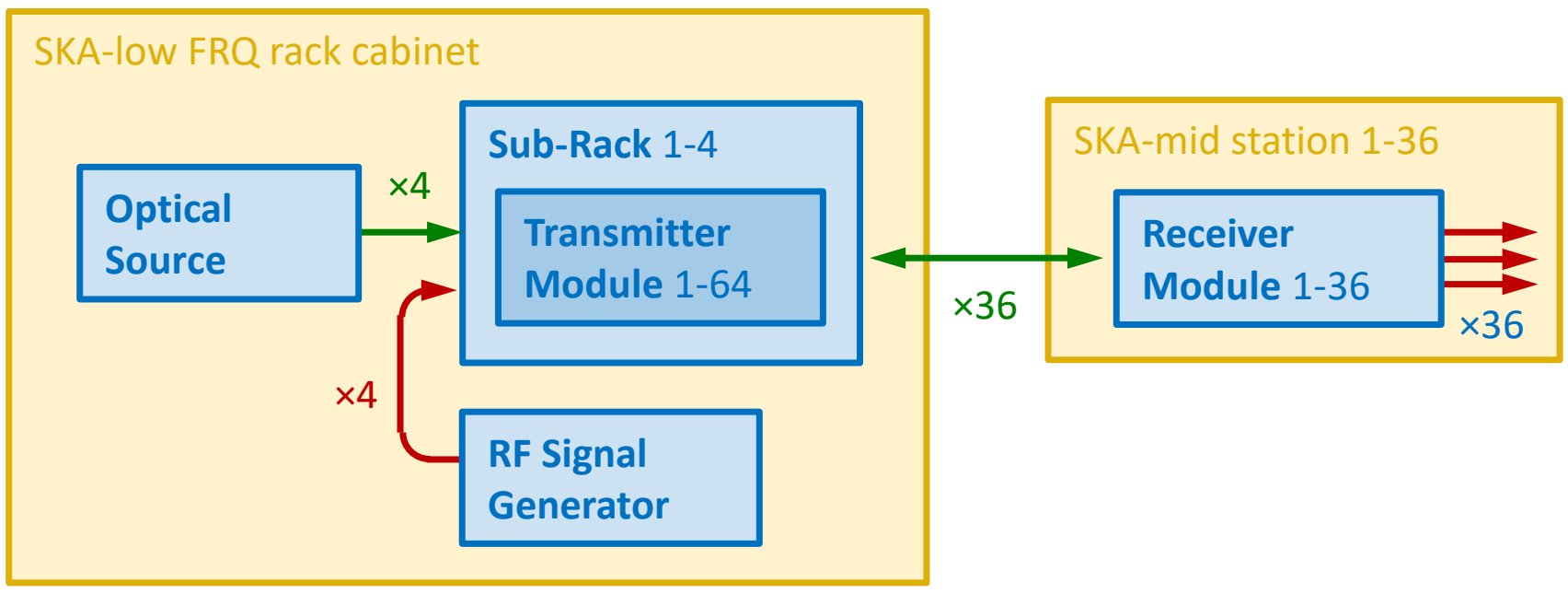

**Figure 14 – SKA1-low overview simplified schematic layout.**

Note, the optical amplifiers are not required for SKA1-low as the longest Fibre Link is 58 km [RD26].

The visual representation of the STFR.FRQ LRUs that are located in the SKA1-low rack cabinet are shown in Figure 15.

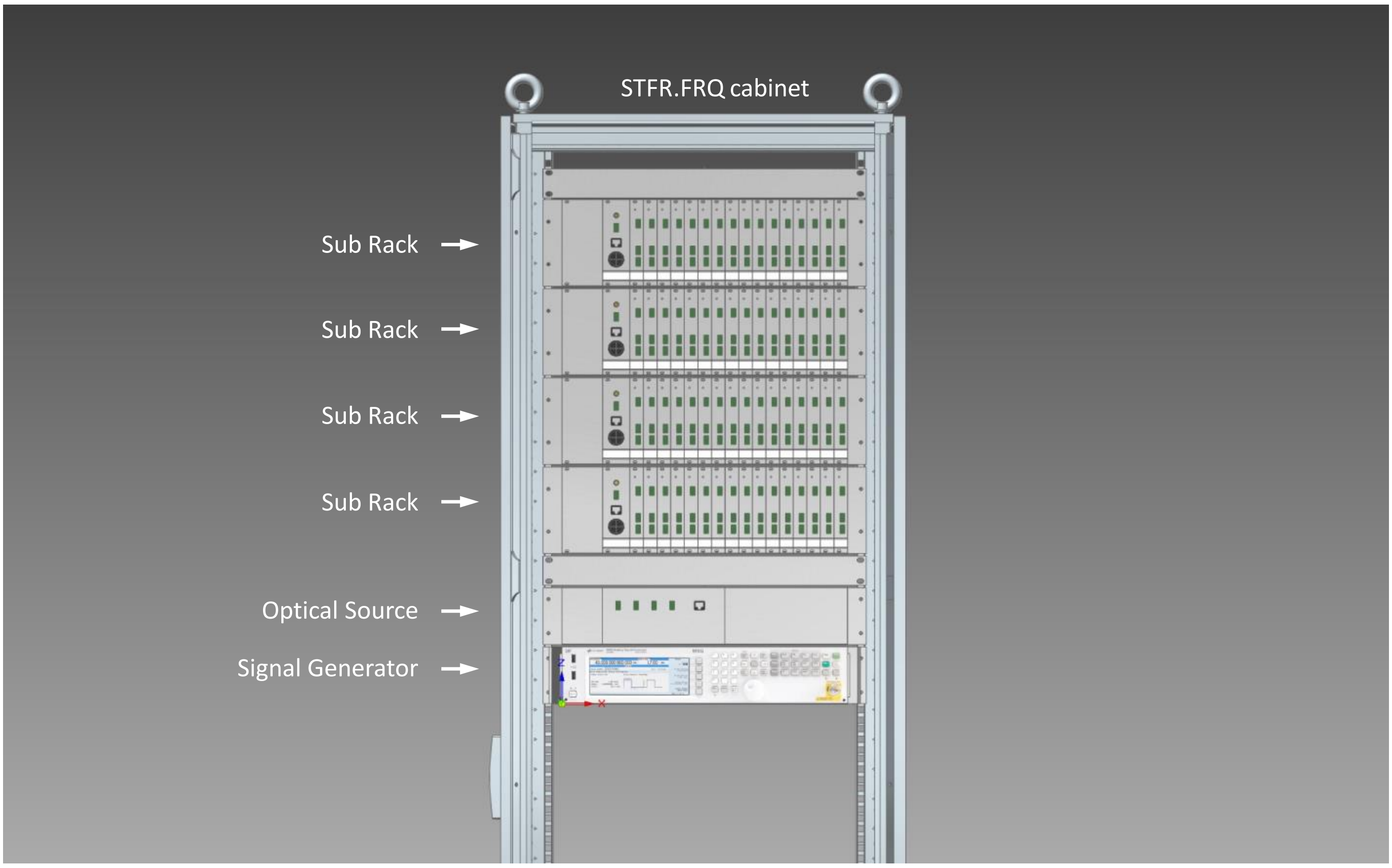

**Figure 15 – SKA1-low overview visual representation.**

#### 3.4.1.1 Rack Cabinet (141-022900)

The LRU is a collection of rack cabinet accessories including equipment for heat management and cable management, and the labour to install this equipment. The visual representation for Rack Cabinet (141-022900) is displayed in Figure 16.

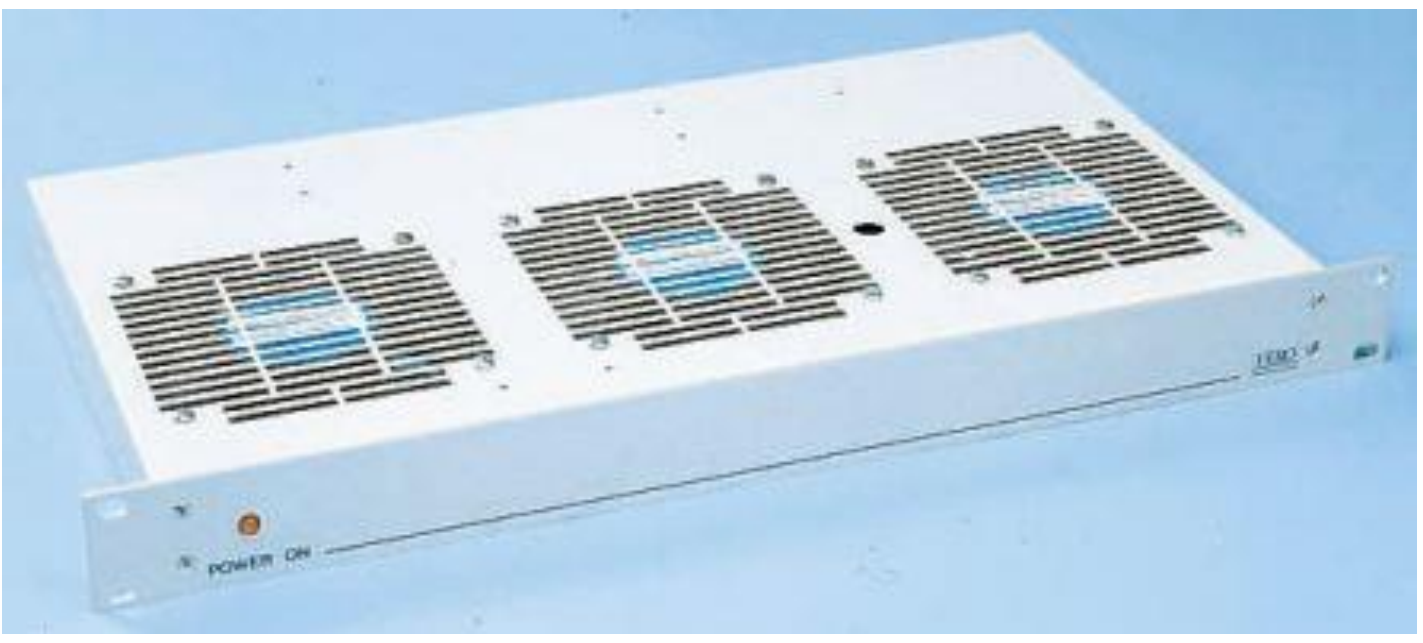

**Figure 16 – SKA1-low Rack Cabinet (141-022900) visual representation.**

#### 3.4.1.2 Optical Source (141-022400)

The simplified schematics layout for Optical Source (141-022400) is displayed in Figure 17.

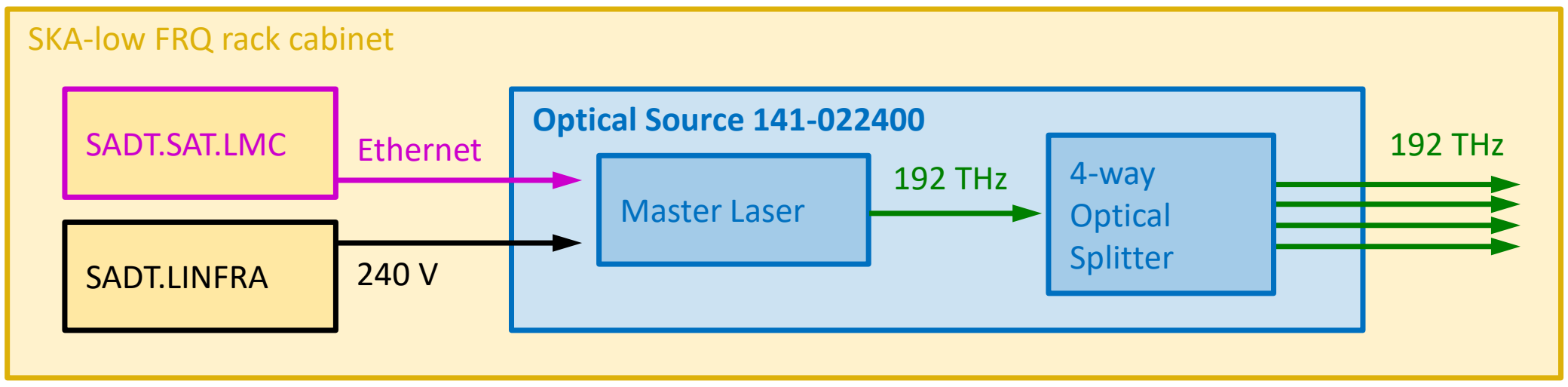

**Figure 17 – SKA1-low Optical Source (141-022400) simplified schematic layout.**

The visual representation for Optical Source (141-022400) external enclosure is displayed in Figure 19.

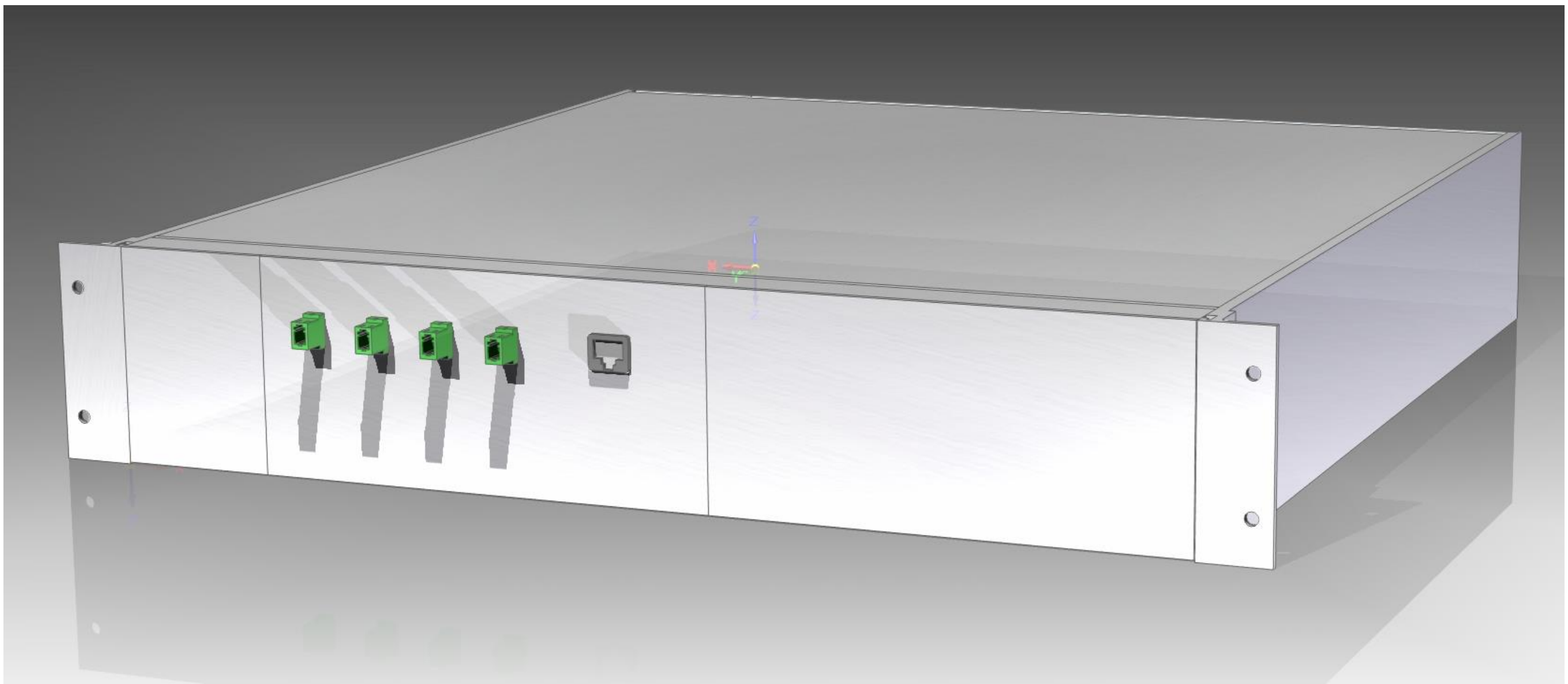

**Figure 18 – SKA1-low Optical Source (141-022400) 3D render (external).**

The visual representation for Optical Source (141-022400) internal equipment is displayed in Figure 19.

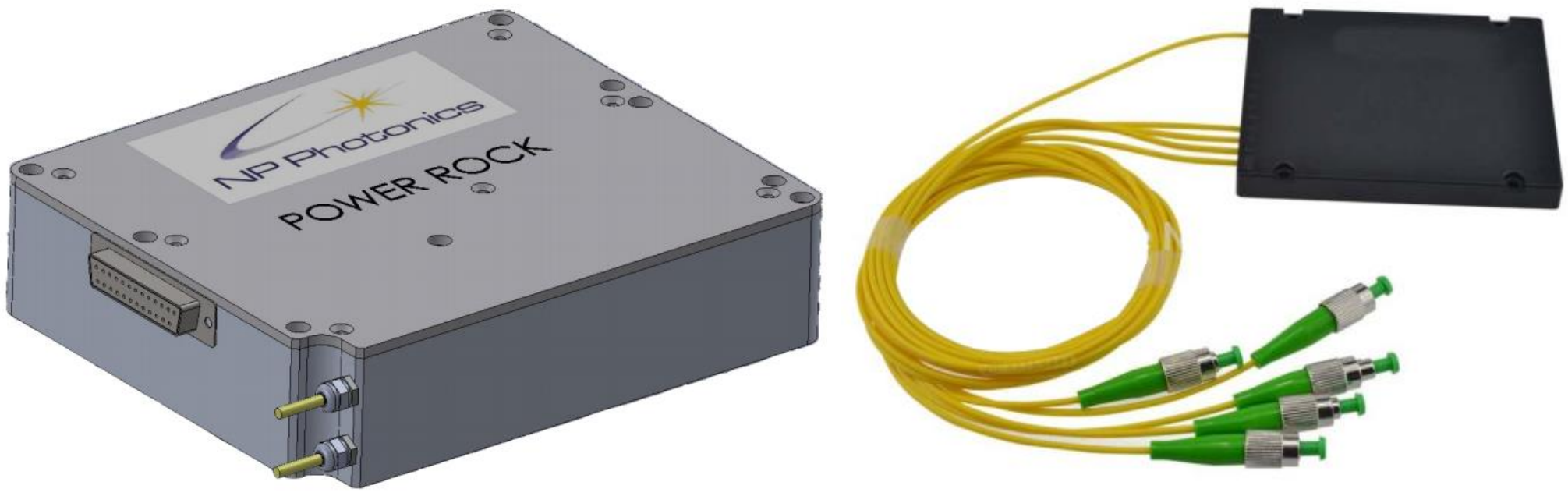

**Figure 19 – SKA1-low Optical Source (141-022400) visual collage.**

#### 3.4.1.3 Signal Generator (141-023100)

The simplified schematics layout for Signal Generator (141-023100) is displayed in Figure 20.

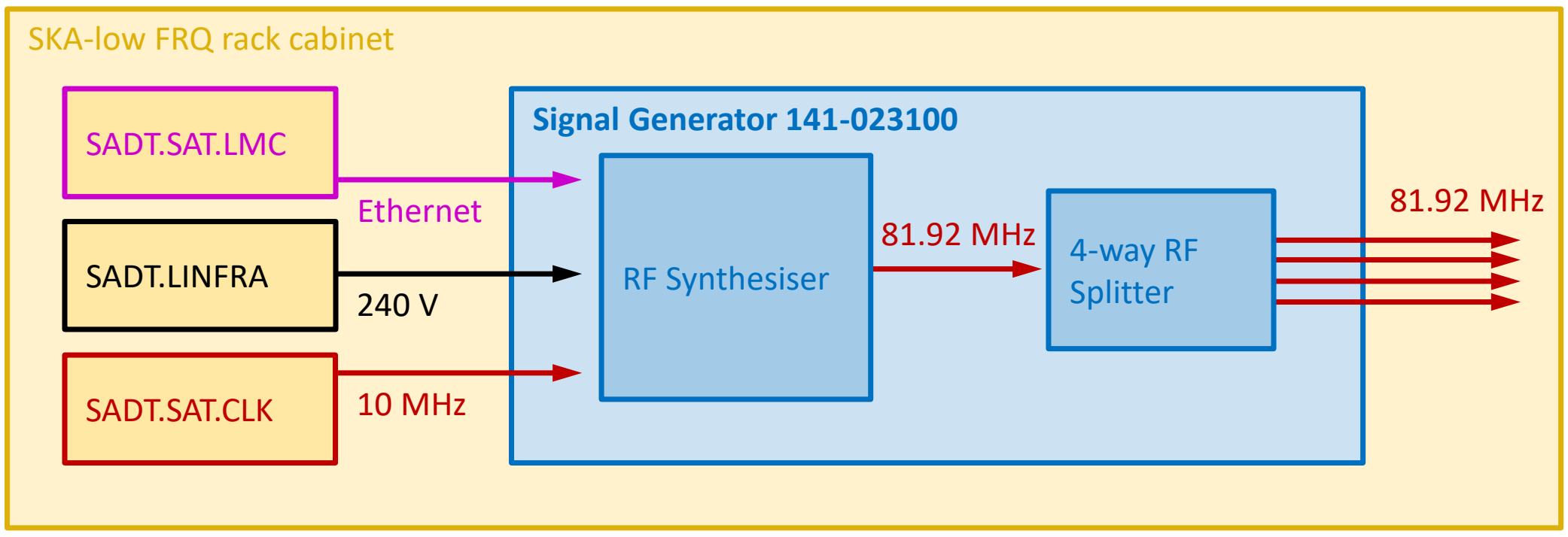

**Figure 20 – SKA1-low Signal Generator (141-023100) simplified schematic layout.**

The visual representation for Signal Generator (141-023100) is displayed in Figure 21.

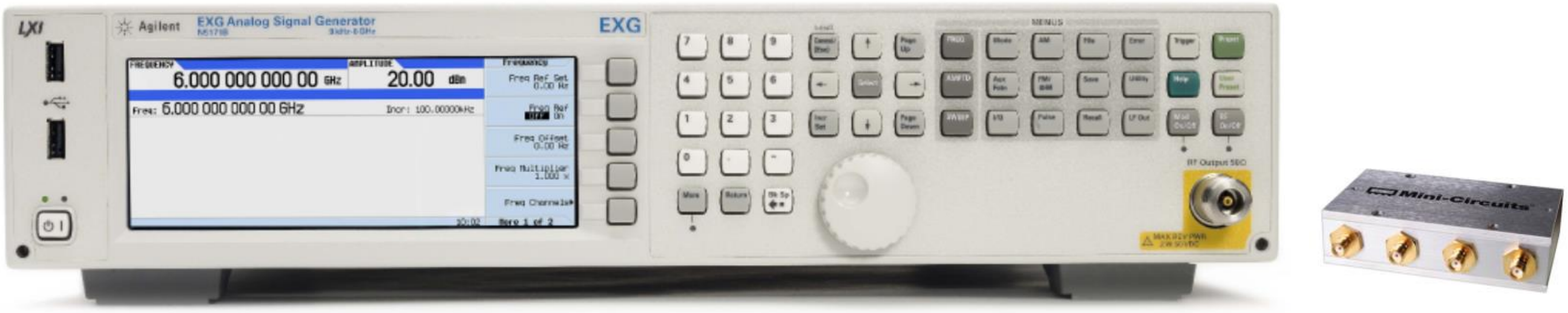

**Figure 21 – SKA1-low Signal Generator (141-023100) photo collage.**

### 3.4.1.4 Sub Rack (141-022700)

The simplified schematics layout for Sub Rack (141-022700) is displayed in Figure 22.

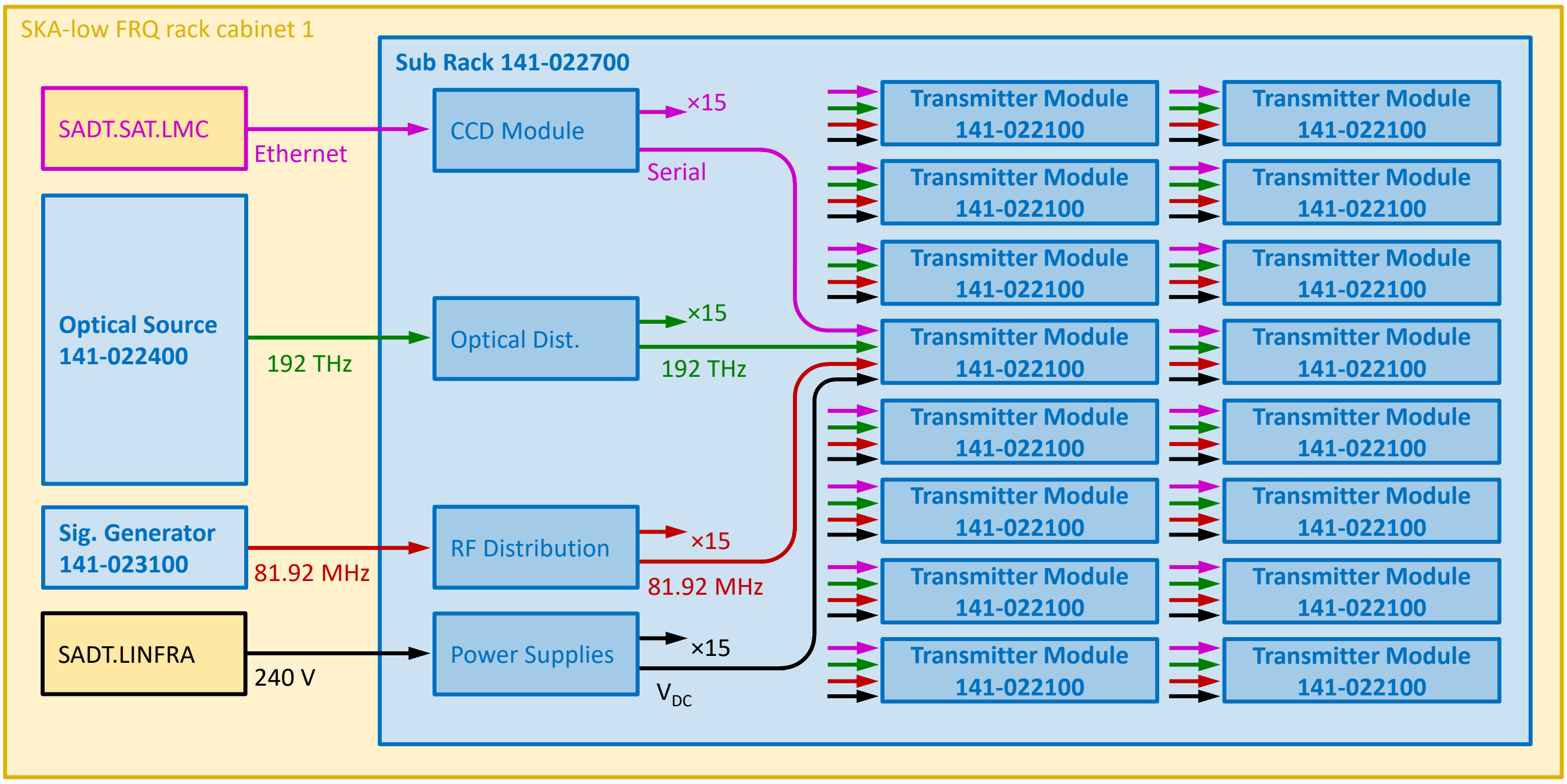

**Figure 22 – SKA1-low Sub Rack (141-022700) simplified schematic layout.**

The visual representation for Sub Rack (141-022700) is displayed in Figure 23.

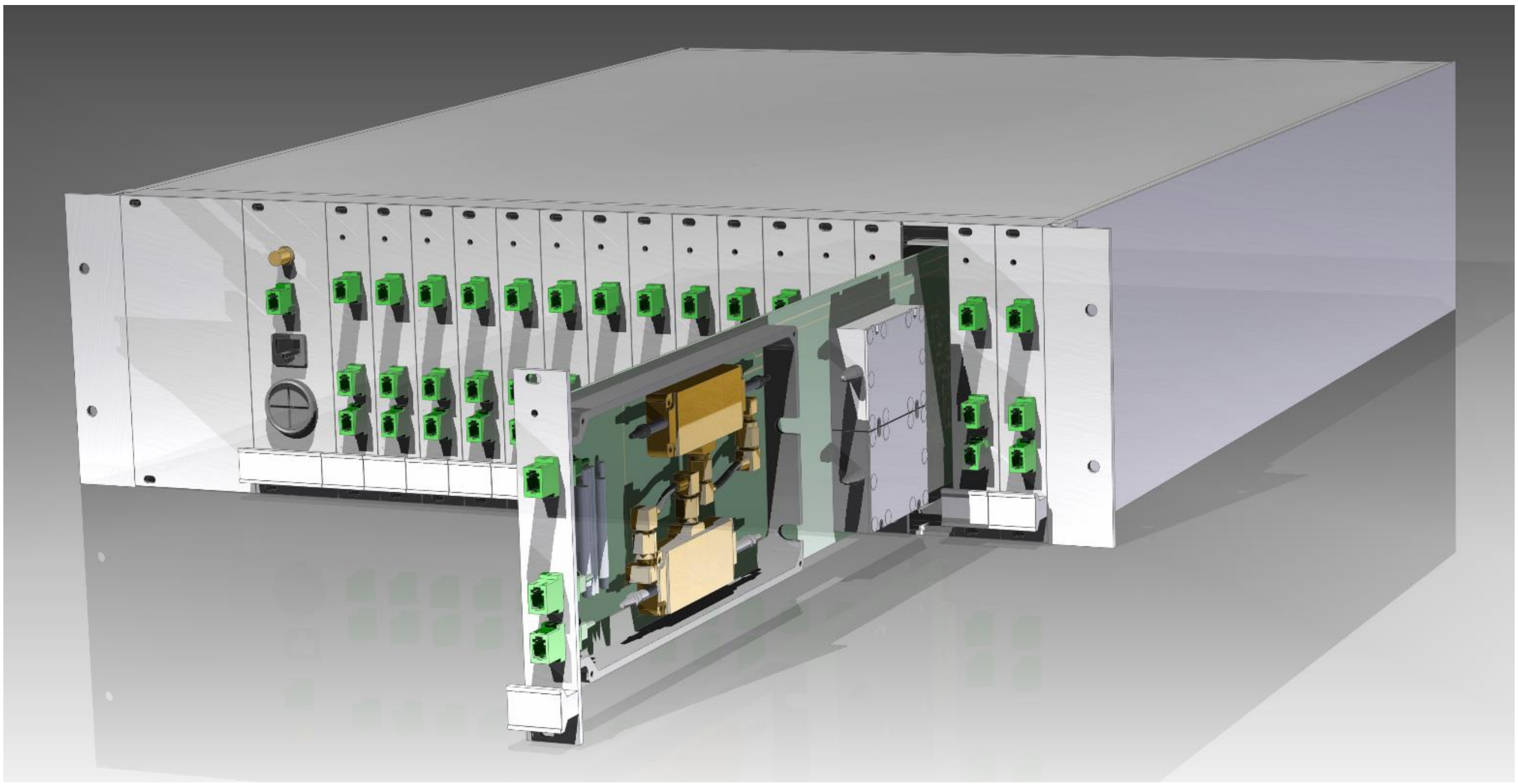

**Figure 23 – SKA1-low Sub Rack (141-022700) 3D render.**

The visual representation for Sub Rack (141-022700) is displayed in Figure 24.

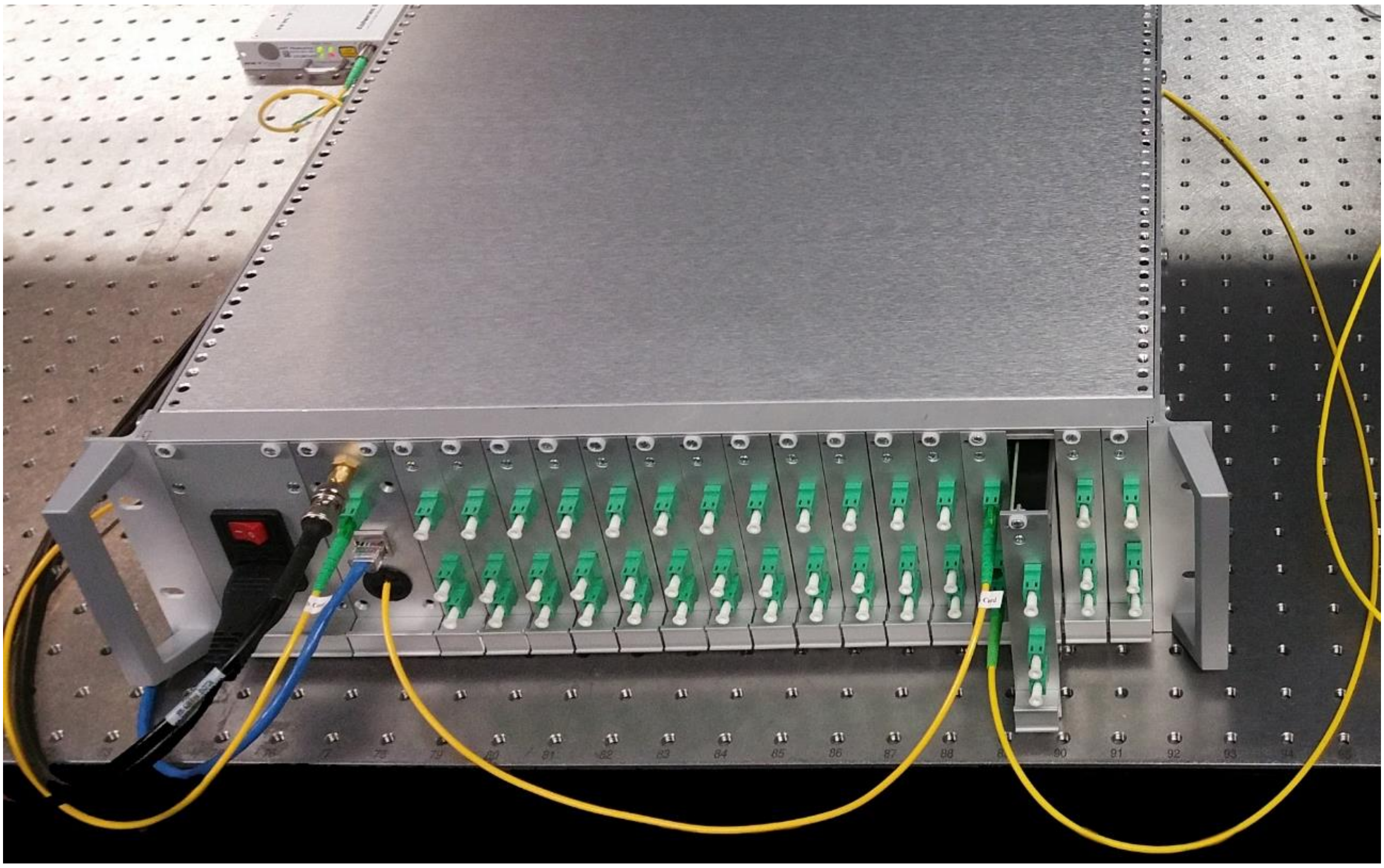

**Figure 24 – SKA1-low Sub Rack (141-022700) photo of prototype.**

### 3.4.1.5 Transmitter Module (141-022100)

The simplified schematics layout for Transmitter Module (141-022100) is displayed in Figure 25.

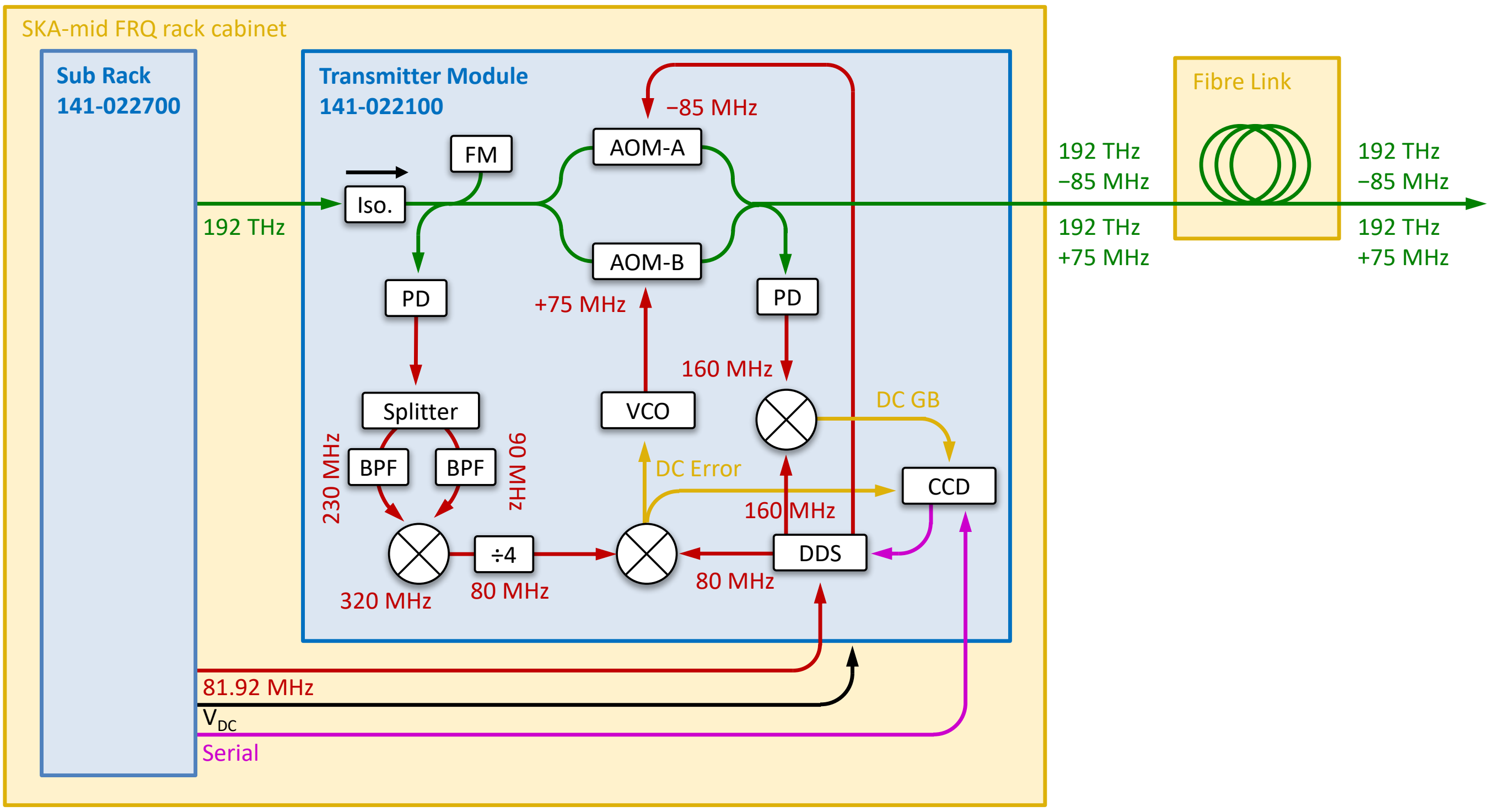

**Figure 25 – SKA1-low Transmitter Module (141-022100) simplified schematic layout.**

The visual representation for Transmitter Module (141-022100) is displayed in Figure 26.

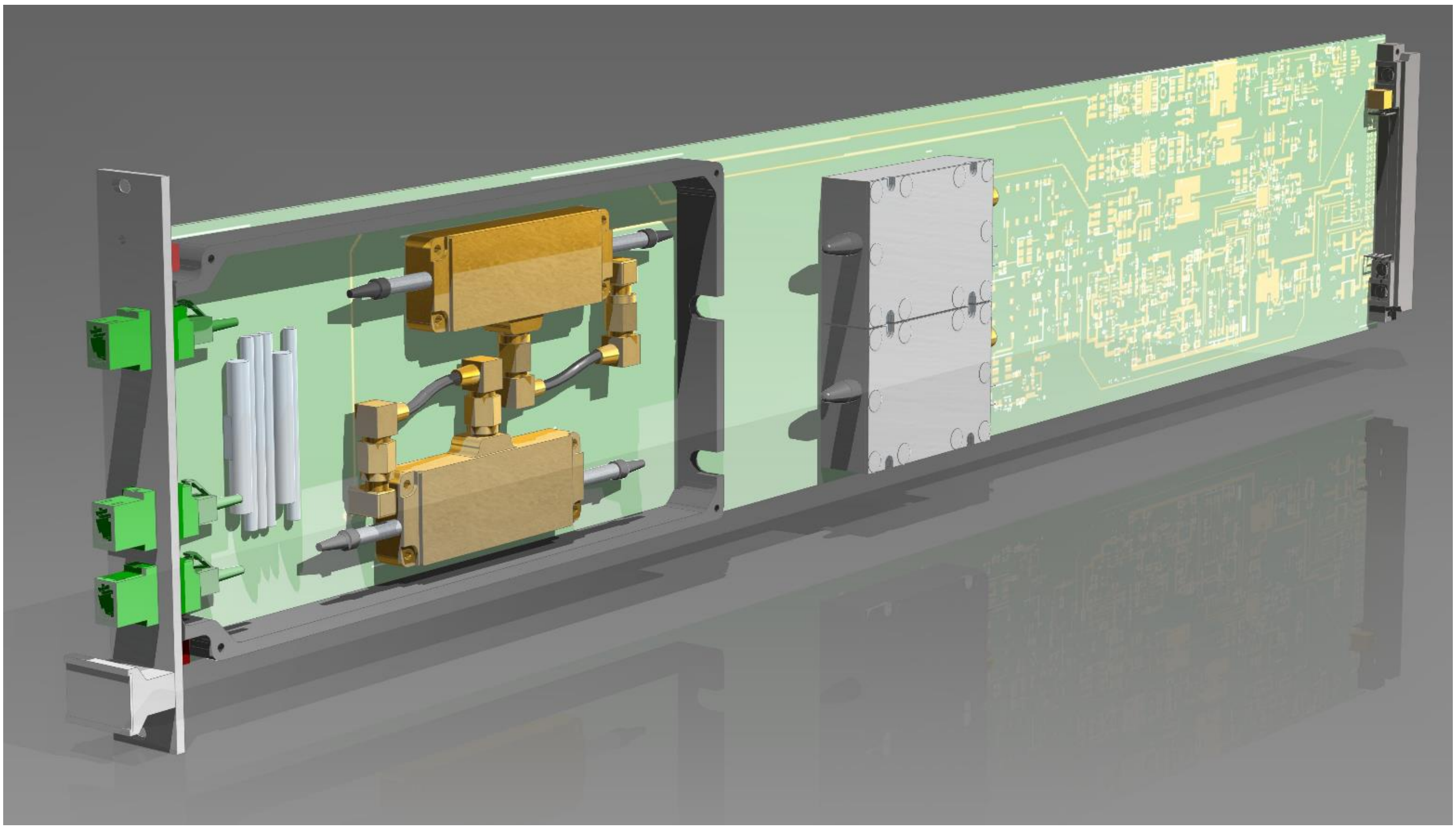

**Figure 26 – SKA1-low Transmitter Module (341-022100) 3D render.**

A photo of the Transmitter Module (141-022100) is displayed in Figure 27.

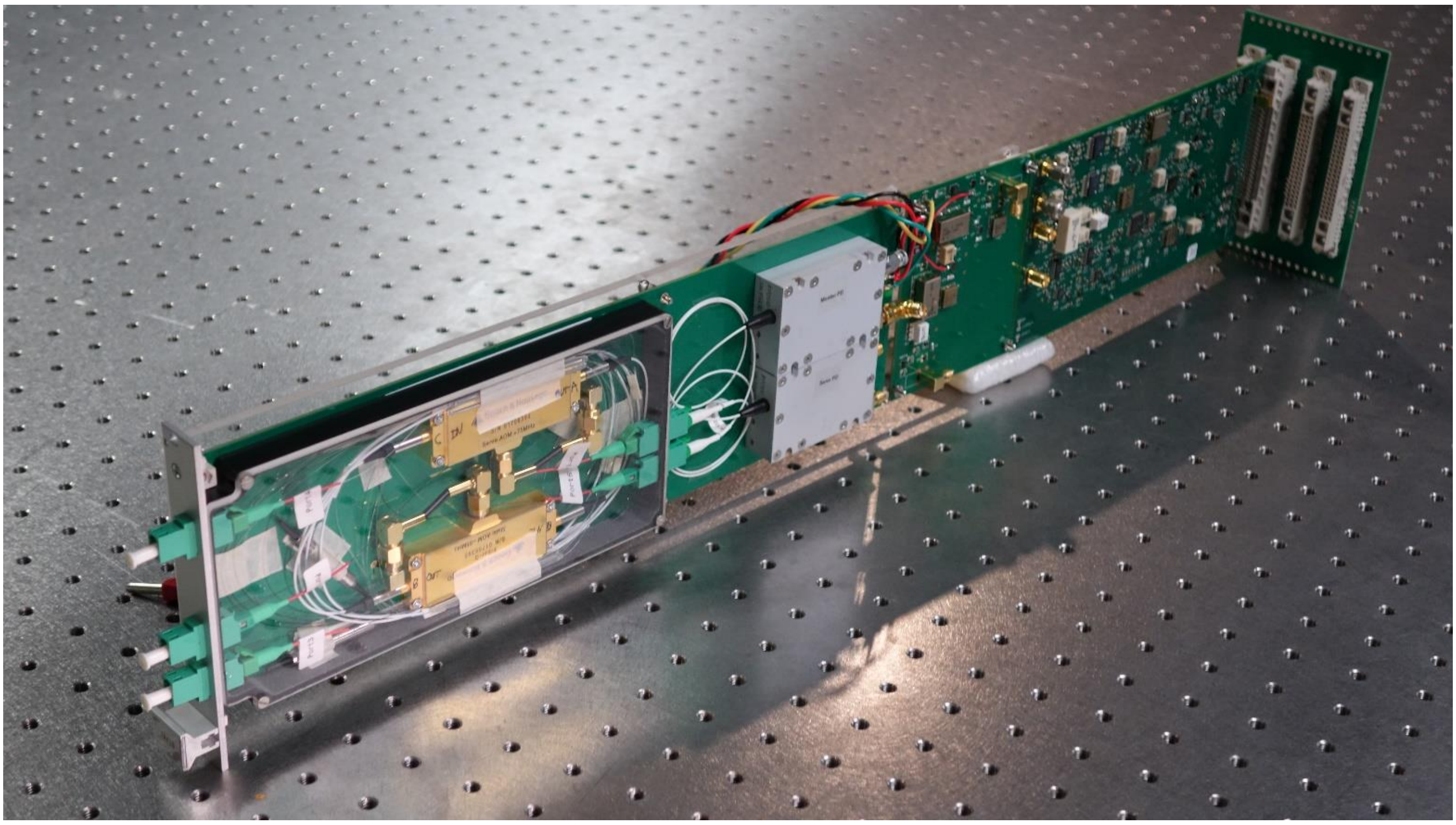

**Figure 27 – SKA1-low Transmitter Module (341-022100) photo.**

### 3.4.1.6 Receiver Module (141-022300)

The simplified schematic layout for Receiver Module (141-022300) is displayed in Figure 28.

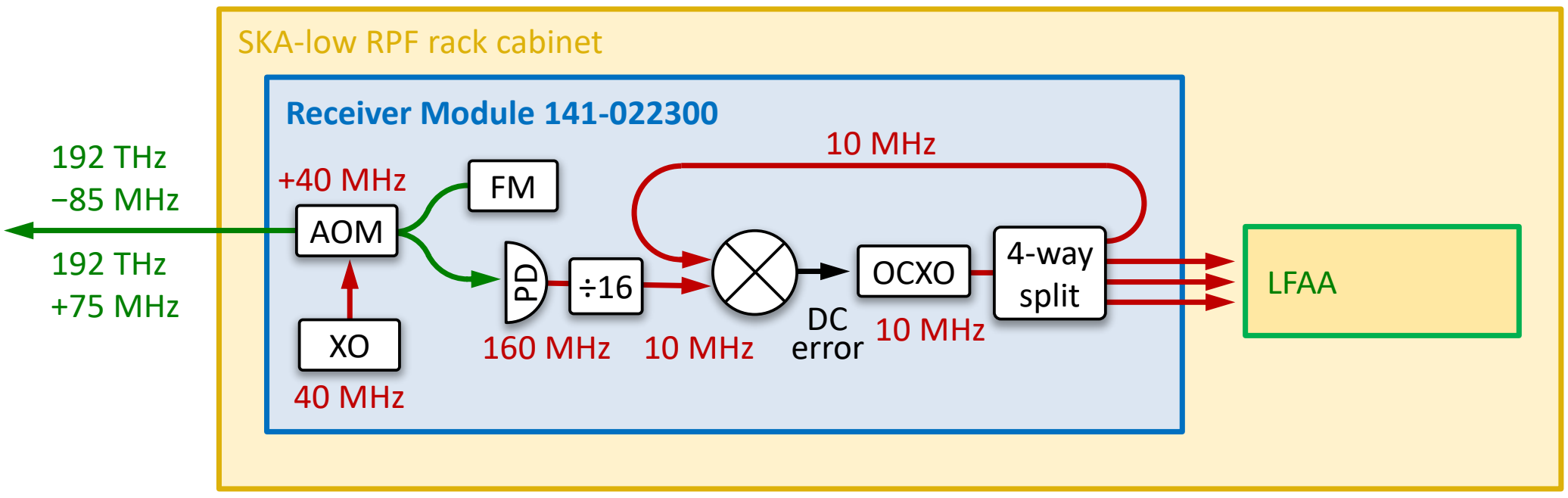

**Figure 28 – SKA1-low Receiver Module (141-022300) simplified schematic layout.**

The visual representation for Receiver Module (141-022300) is displayed in Figure 29.

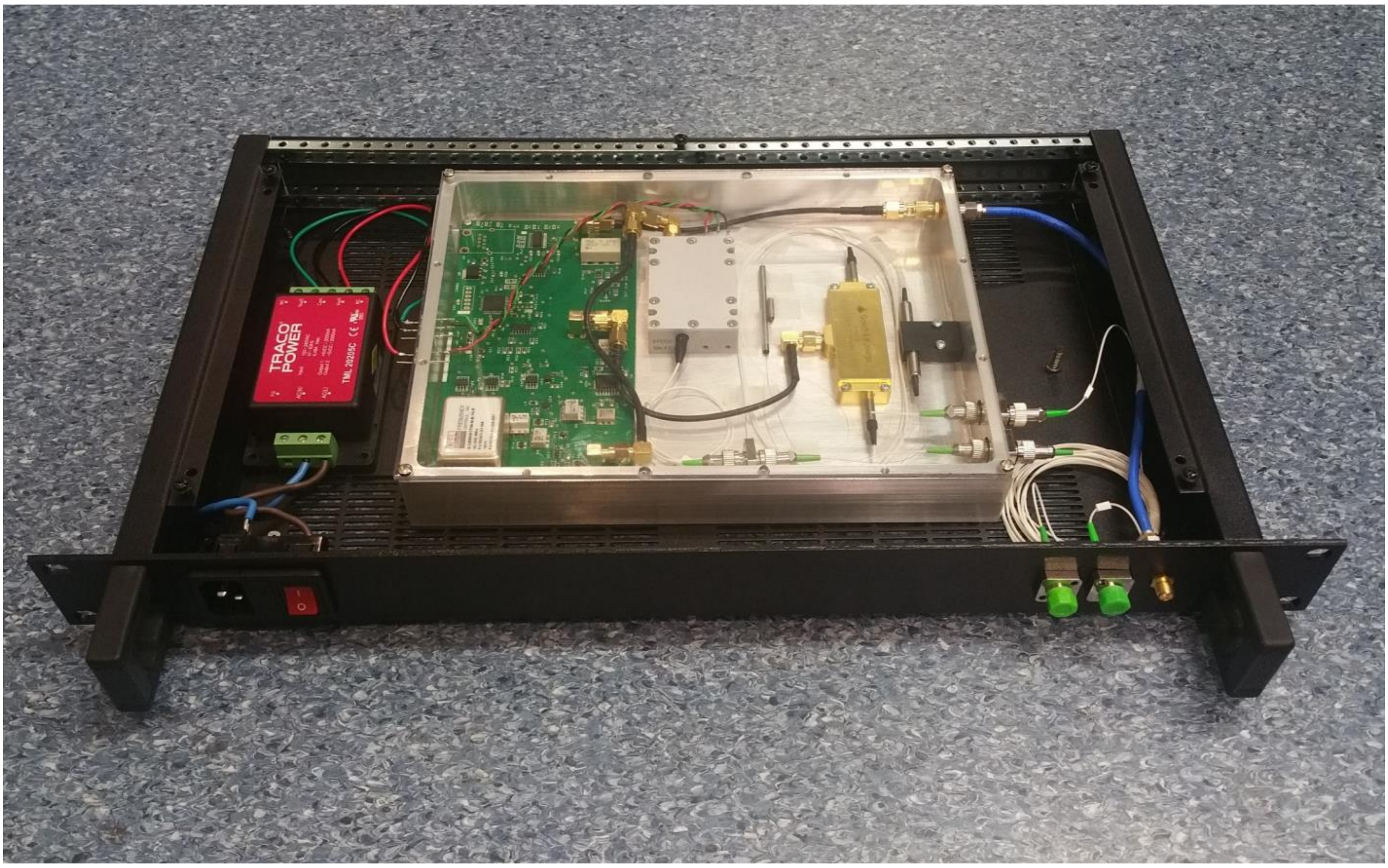

**Figure 29 – SKA1-low Receiver Module (141-022300) photo of prototype.**

## 3.4.2 **Mechanical Detailed Design**

### 3.4.2.1 Rack Cabinet (141-022900)

The mechanical detailed design for Rack Cabinet (141-022900) is given in Table 14.

**Table 14 – SKA1-low Rack Cabinet (141-022900) mechanical detailed design table.**

| Type | Fan Tray |
|---|---|
| Mounting | Rack |
| Rack Units | 1U |
| Number of Fans | 6 |
| Air Flow | 167m$^3$/h |
| Height | 1U |
| Width | 350mm |

The related *Solid Edge* part is included in the Mechanical Detailed Design File pack in Appendix 7.5.2.

### 3.4.2.2 Optical Source (141-022400)

The mechanical detailed design for Optical Source (141-022400) is displayed in Figure 31.

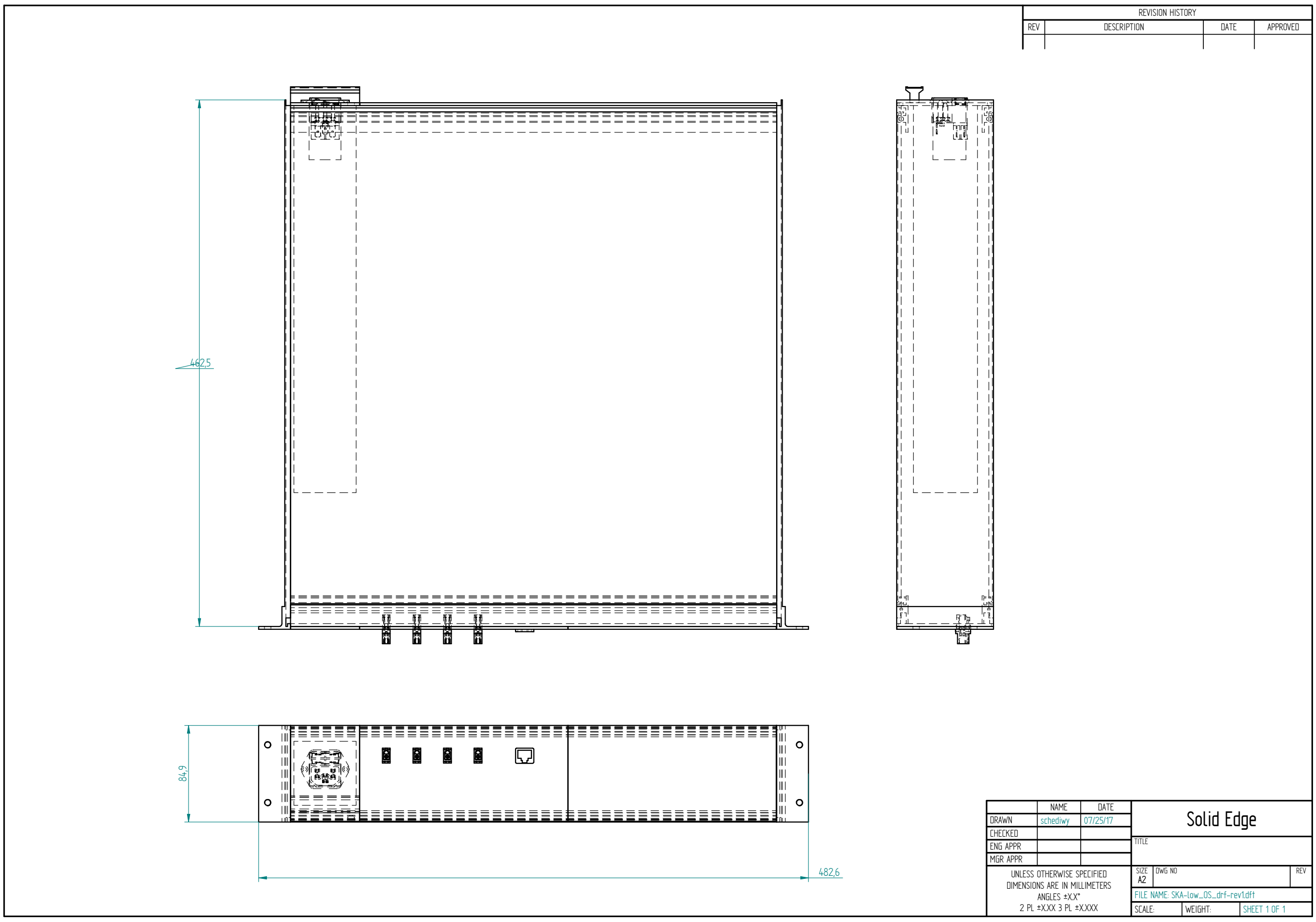

**Figure 30 – SKA1-low Optical Source (141-022400) external enclosure mechanical detailed design figure.**

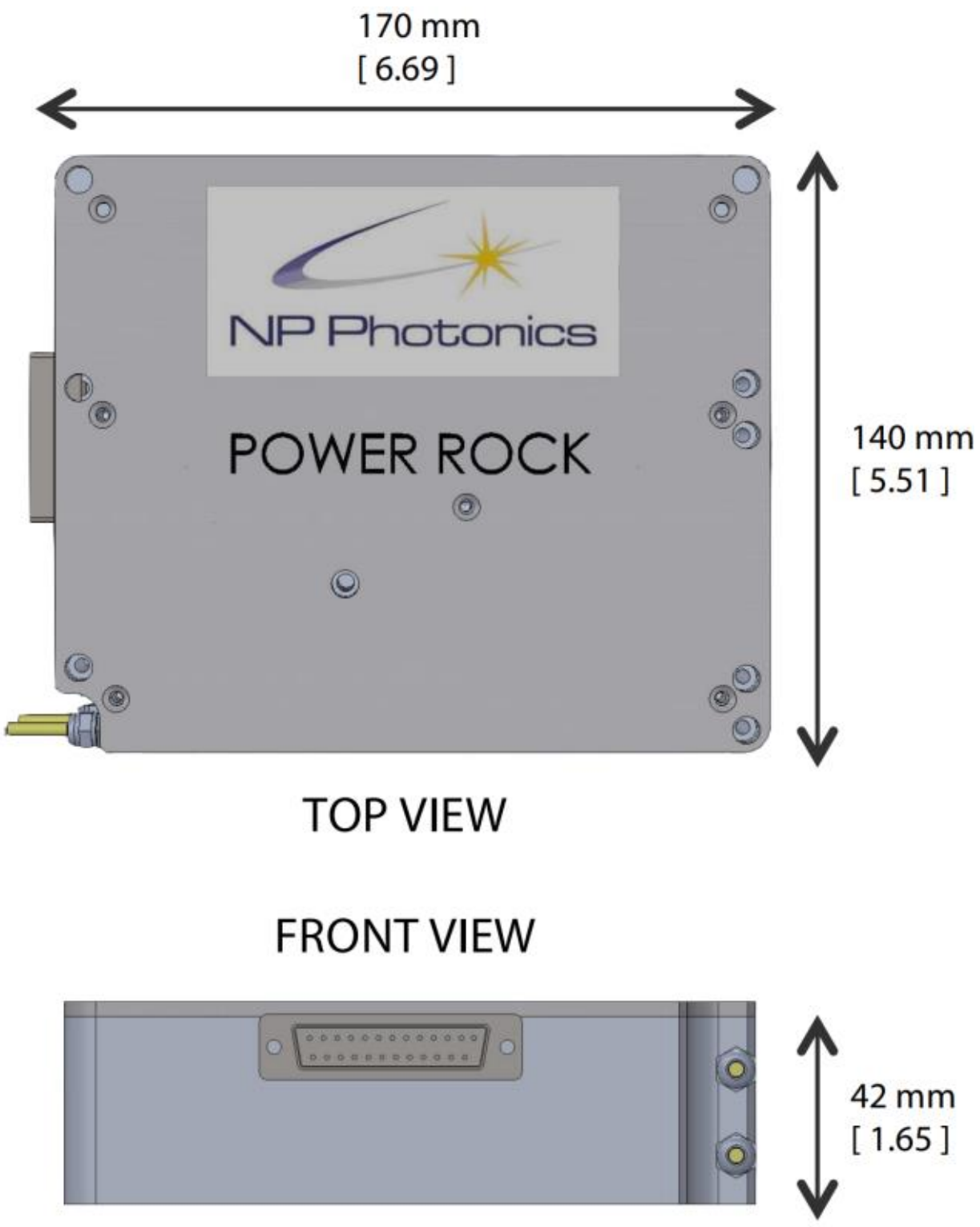

**Figure 31 – SKA1-low Optical Source (141-022400) internal laser mechanical detailed design figure.**

The laser unit shown above, and the 4-way optical splitter, will be housed inside a 2U, 19” rackmount enclosure. The power socket is at the rear.

The related *Solid Edge* part is included in the Mechanical Detailed Design File pack in Appendix 7.5.2.

#### 3.4.2.3 Signal Generator (141-023100)

The mechanical detailed design for Signal Generator (141-023100) is a 3U, 19” rackmount enclosure, with a depth of 489 mm. The power socket is at the rear.

The related *Solid Edge* part is included in the Mechanical Detailed Design File pack in Appendix 7.5.2.

#### 3.4.2.4 Sub Rack (141-022700)

The mechanical detailed design for Sub Rack (141-022700) is displayed in

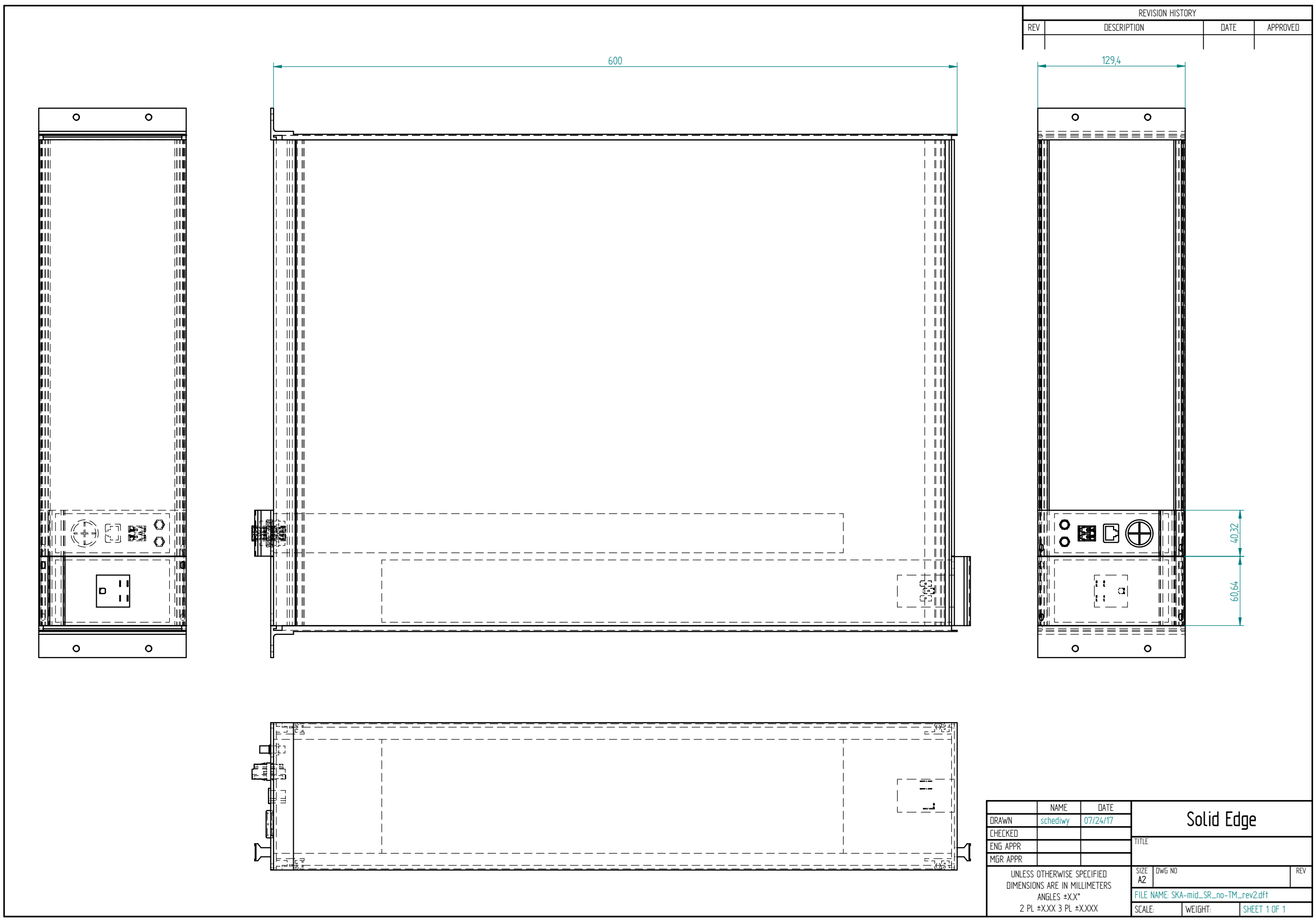

Figure 32.

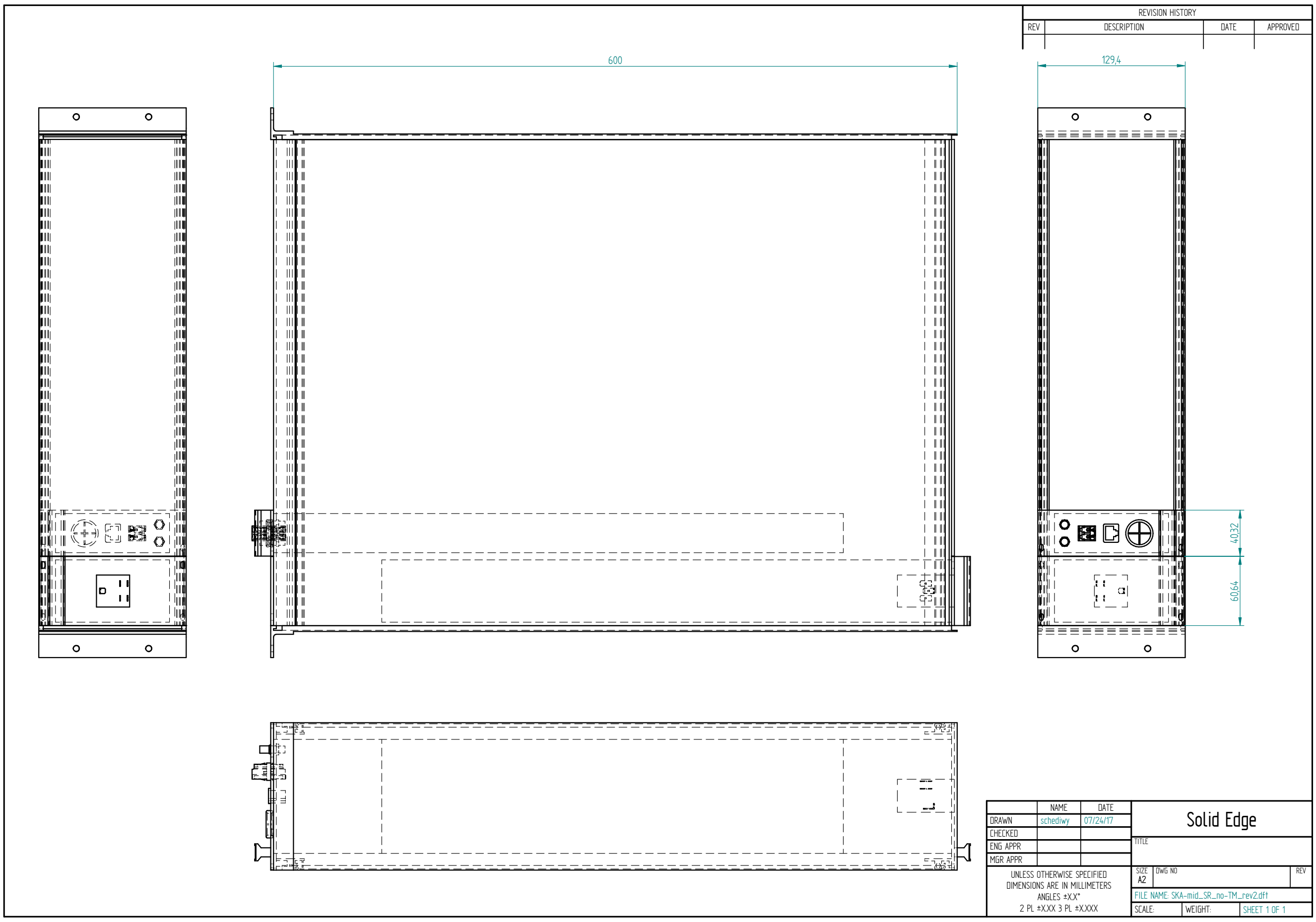

**Figure 32 – SKA1-low Sub Rack (141-022700) mechanical detailed design figure.**

The Sub Rack is a 3U, 19” rackmount enclosure, with a depth of 600 mm (the maximum specified in [RD26]). The TM-power module is 12HP, and the TM-distribution module is 8HP. The power socket is at the rear.

The related *Solid Edge* part is included in the Mechanical Detailed Design File pack in Appendix 7.5.2.

### 3.4.2.5 Transmitter Module (141-022100)

The mechanical detailed design for the Transmitter Module (141-022100) PCB is displayed in.

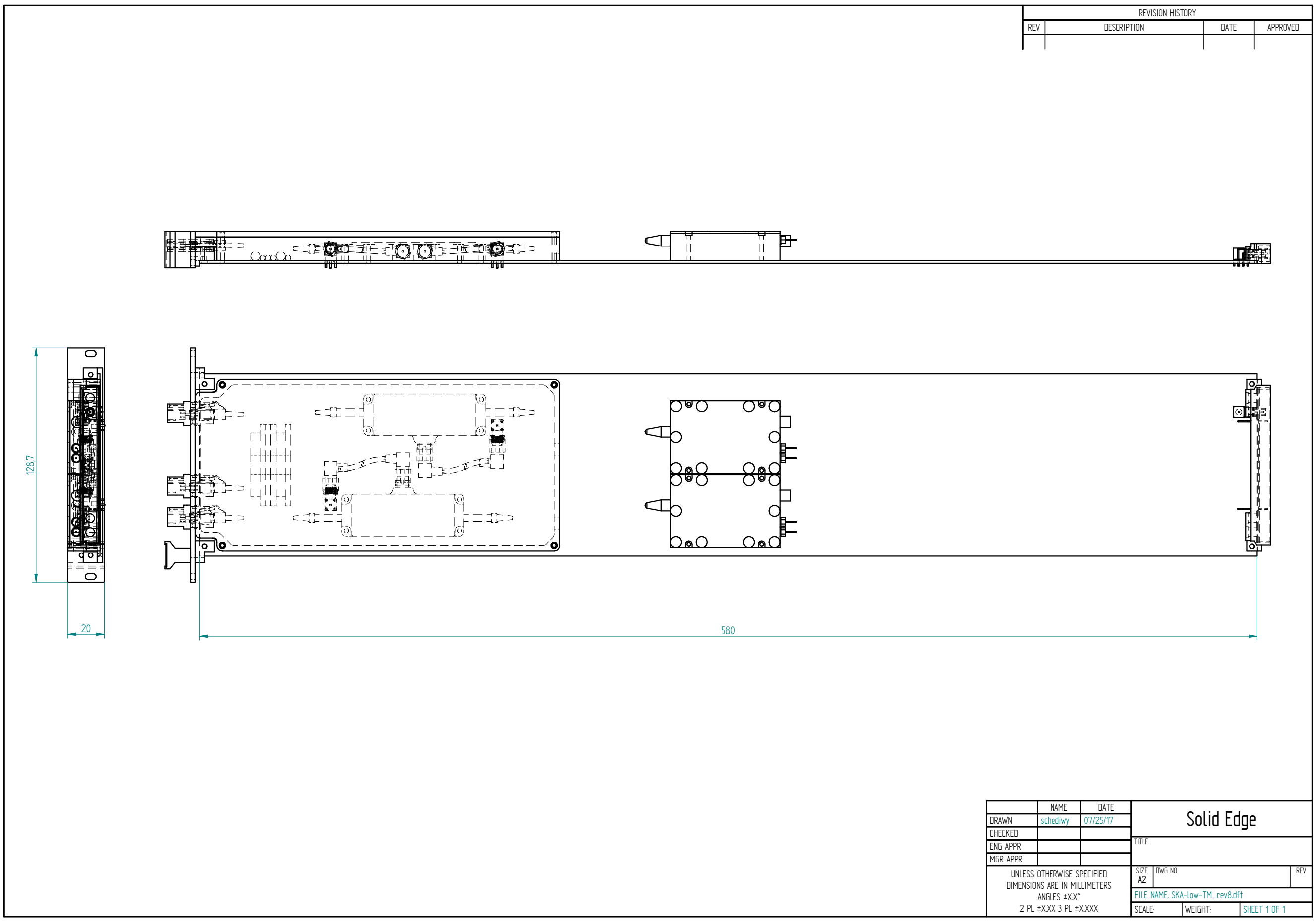

**Figure 33 – SKA1-low Transmitter Module (141-022100) mechanical detailed design figure**

The mechanical detailed design for the Transmitter Module (141-022100) Optics Enclosure is displayed in Figure 34.

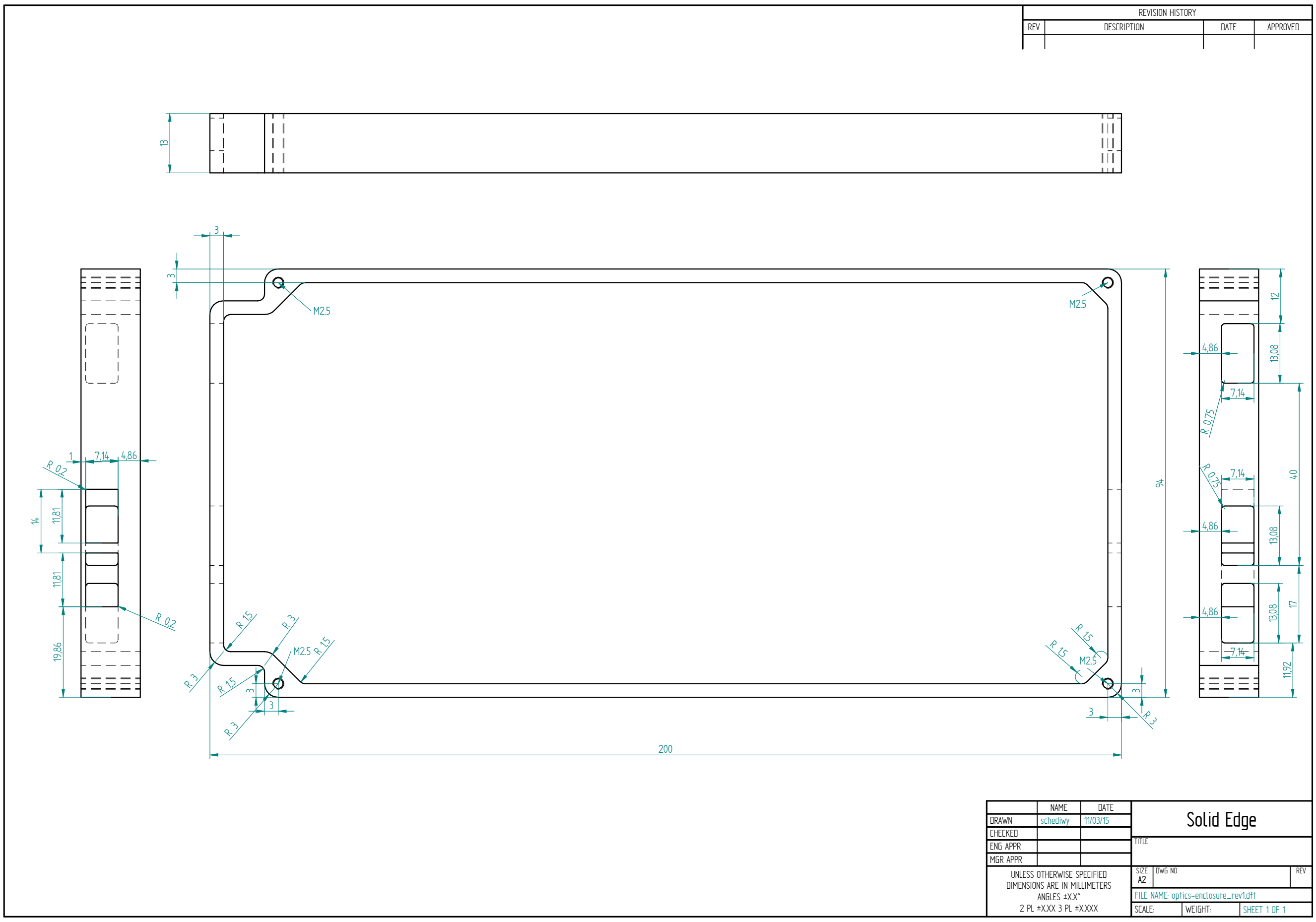

**Figure 34 – SKA1-low Transmitter Module (141-022100) Optics Enclosure mechanical detailed design figure.**

The related *Solid Edge* part is included in the Mechanical Detailed Design File pack in Appendix 7.5.2.

The mechanical detailed design for the Transmitter Module (341-022100) Optics Enclosure Lid is displayed in Figure 35.

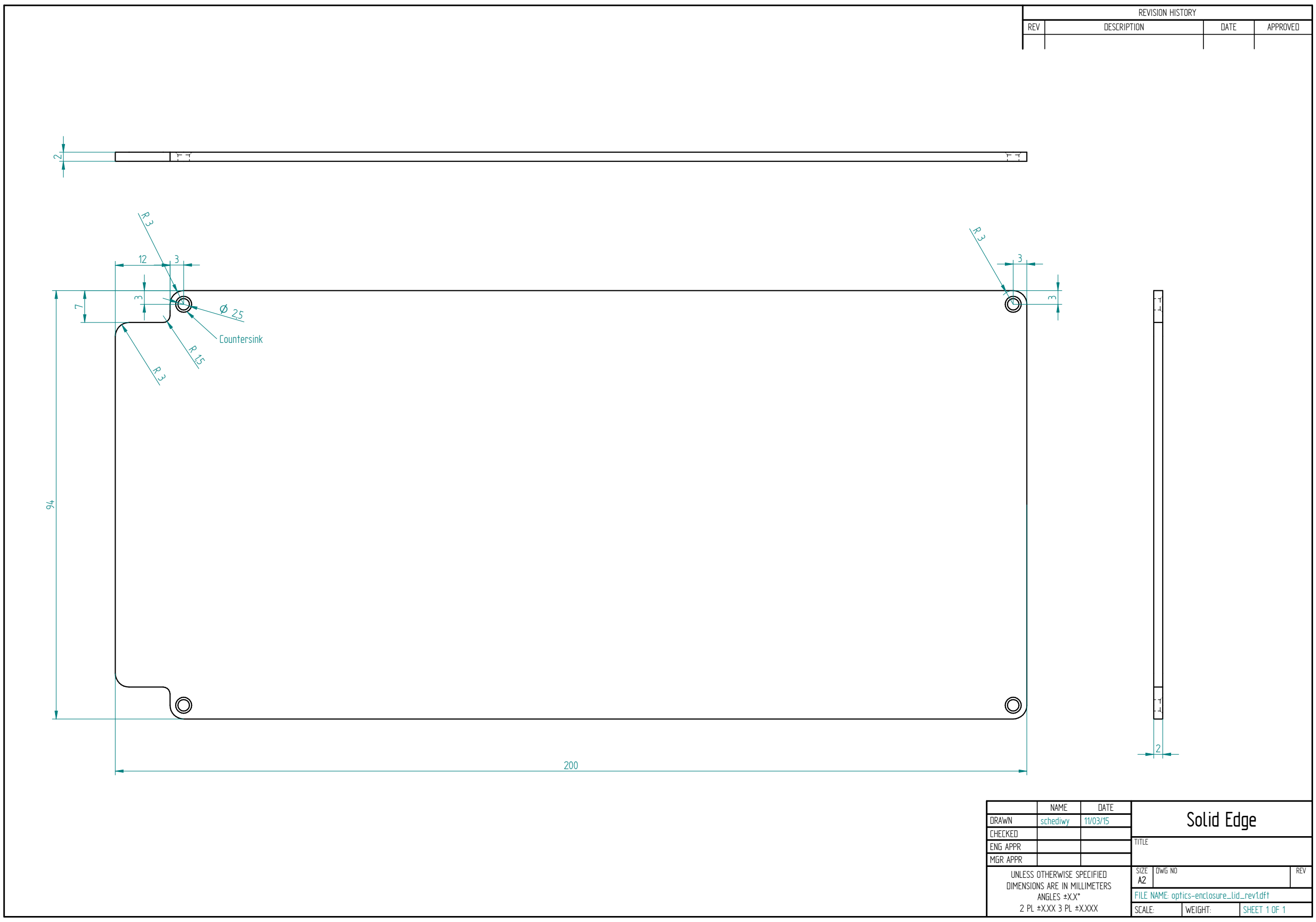

**Figure 35 – SKA1-low Transmitter Module (341-022100) Optics Enclosure Lid mechanical detailed design figure.**

The related *Solid Edge* part is included in the Mechanical Detailed Design File pack in Appendix 7.5.2.

The mechanical detailed design for the Transmitter Module (141-022100) PCB Front Panel is displayed in Figure 36.

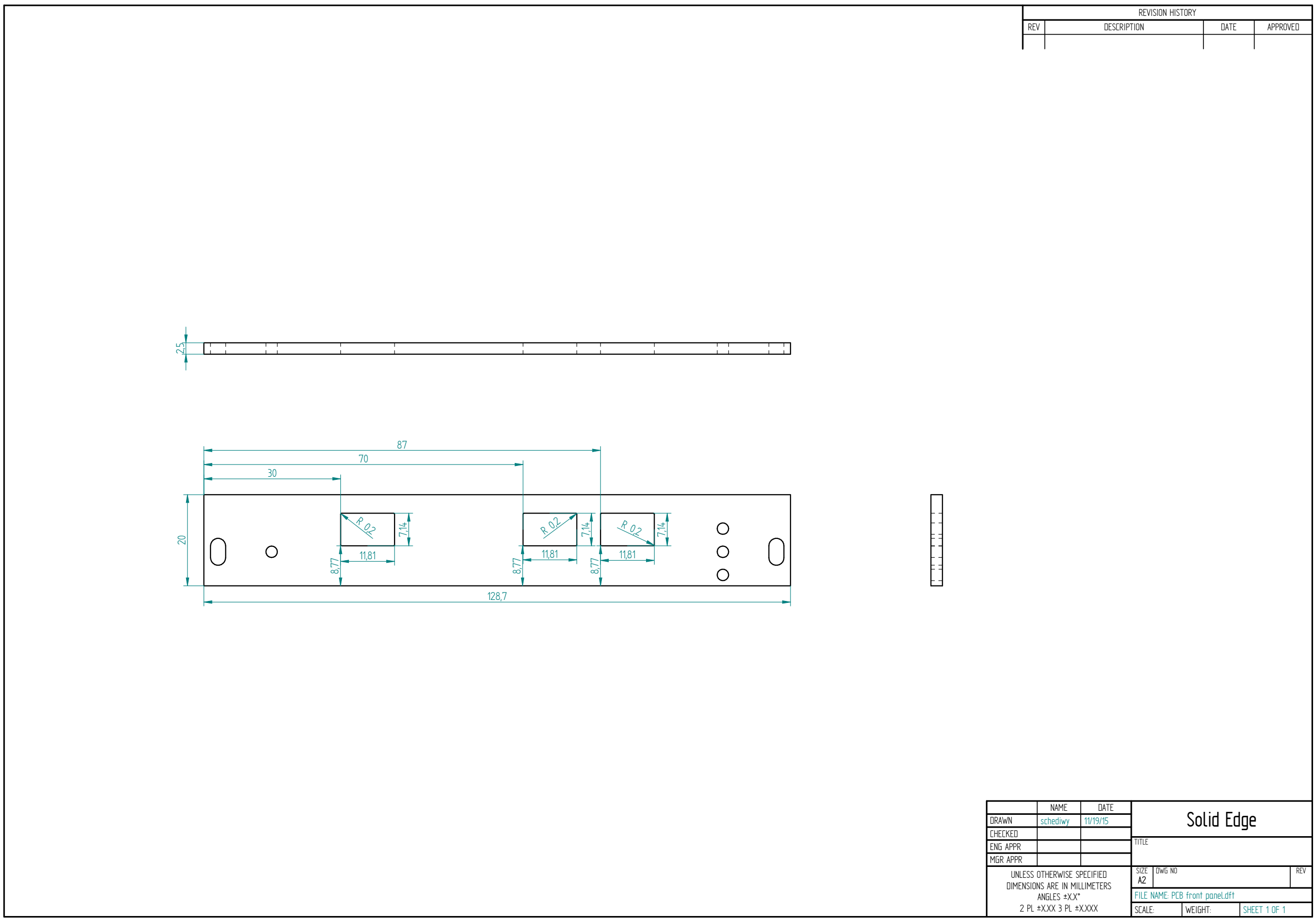

**Figure 36 – SKA1-low Transmitter Module (341-022100) PCB Front Panel mechanical detailed design figure.**

The related *Solid Edge* part is included in the Mechanical Detailed Design File pack in Appendix 7.5.2.

### 3.4.2.6 Receiver Module (141-022300)

The mechanical detailed design for Receiver Module (141-022300) Enclosure Body is displayed in Figure 37.

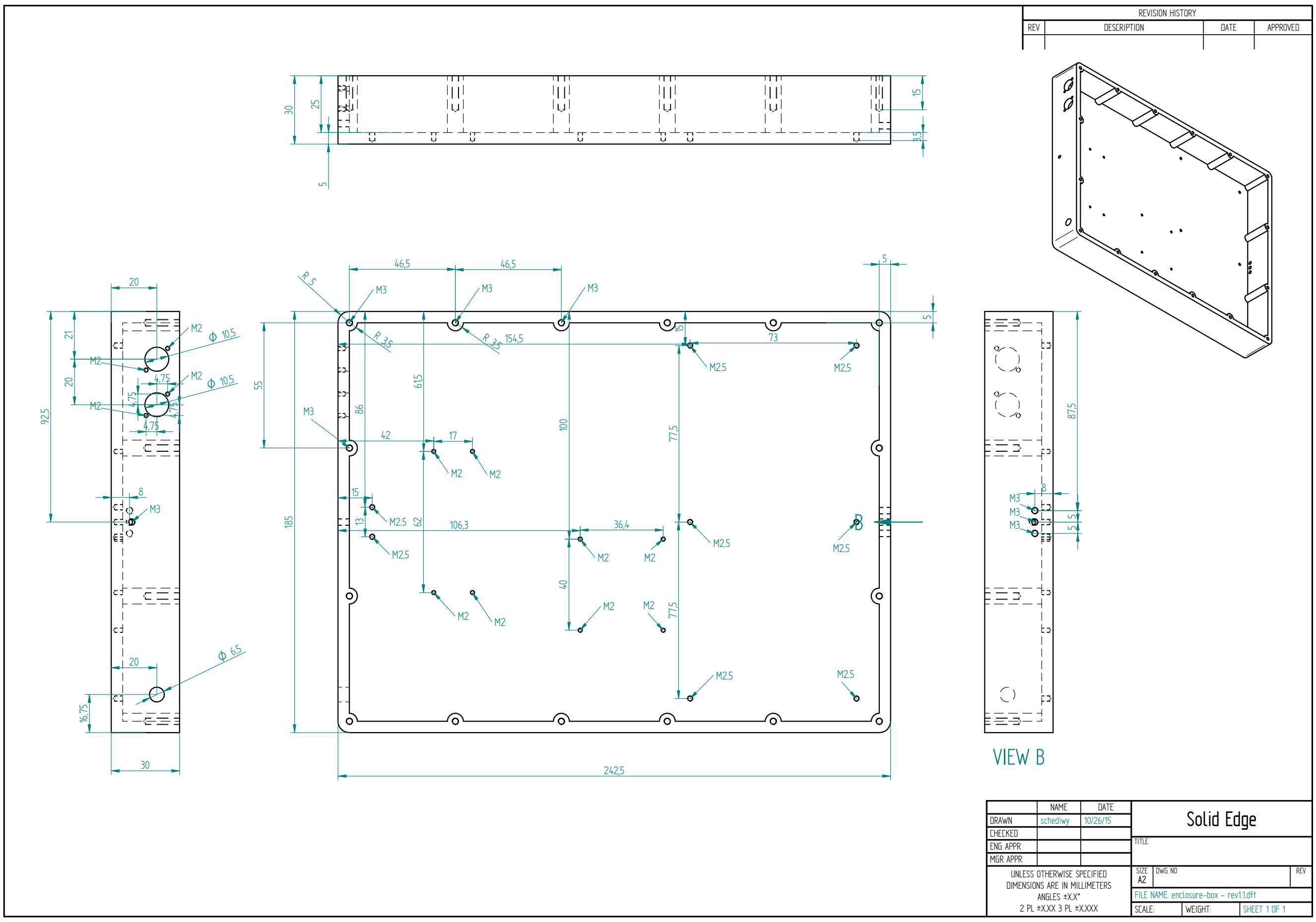

**Figure 37 – SKA1-low Receiver Module (141-022300) Enclosure Body mechanical detailed design figure.**

The related *Solid Edge* part is included in the Mechanical Detailed Design File pack in Appendix 7.5.2.

The Receiver Module, and a DC power supply, will be housed inside a 1U, 19” rackmount enclosure (as shown in Figure 29). The power socket is at the front as is required by [RD26] for equipment located at the RPF.

The mechanical detailed design for Receiver Module (141-022300) Enclosure Lid is displayed in Figure 38.

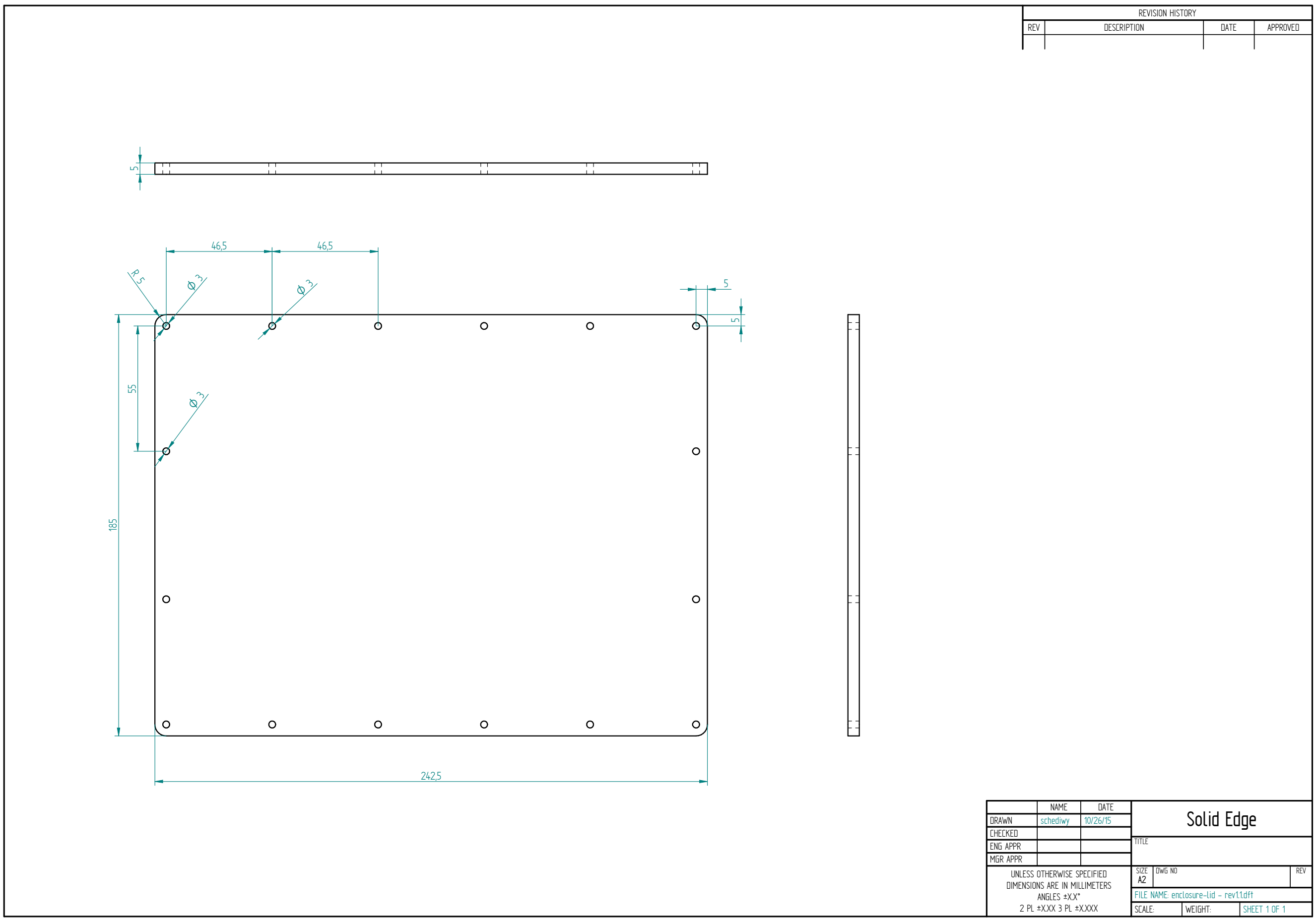

**Figure 38 – SKA1-low Receiver Module (141-022300) Enclosure Lid mechanical detailed design figure.**

The related *Solid Edge* part is included in the Mechanical Detailed Design File pack in Appendix 7.5.2.

### 3.4.3 **Optical Detailed Design**

#### 3.4.3.1 Rack Cabinet (141-022900)

Not applicable.

#### 3.4.3.2 Optical Source (141-022400)

The optical detailed design for Optical Source (141-022400) is displayed in Figure 39.

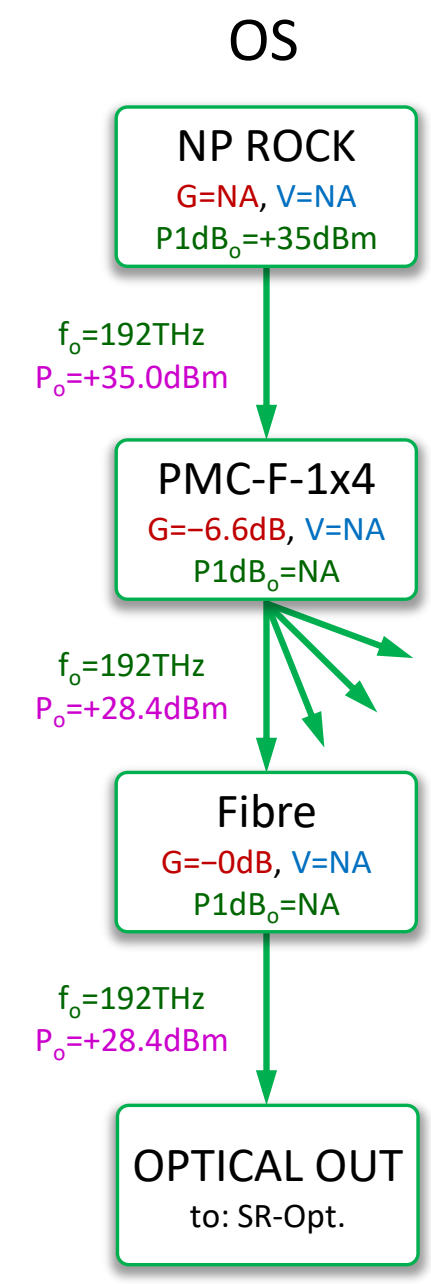

**Figure 39 – SKA1-low Optical Source (141-022400) optical detailed design figure.**

The related *PowerPoint* design is included in the Optical Detailed Design File Pack in Appendix 7.6.2.

#### 3.4.3.3 Signal Generator (141-023100)

Not applicable.

### 3.4.3.4 Sub Rack (141-022700)

The optical detailed design for Sub Rack (141-022700) is displayed in Figure 40.

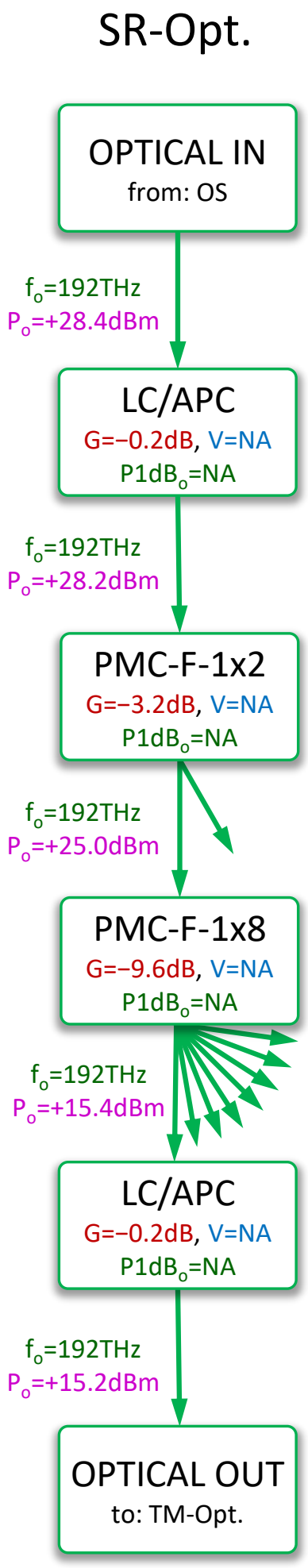

**Figure 40 – SKA1-low Sub Rack (141-022700) optical detailed design figure.**

The related *PowerPoint* design is included in the Optical Detailed Design File Pack in Appendix 7.6.2.

### 3.4.3.5 Transmitter Module (141-022100)

The optical detailed design for Transmitter Module (141-022100) is displayed in Figure 41.

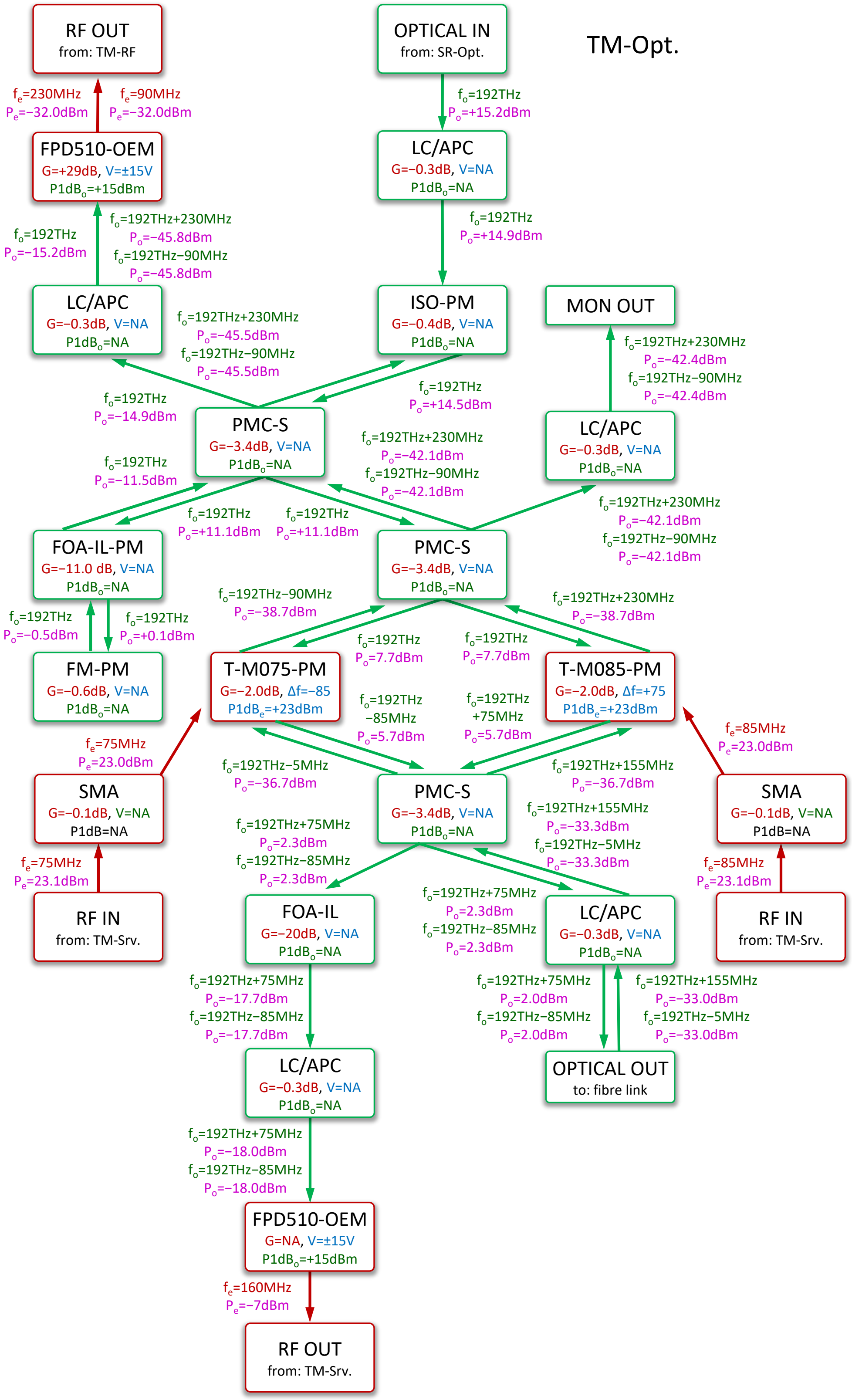

**Figure 41 – SKA1-low Transmitter Module (141-022100) optical detailed design figure.**

The related *PowerPoint* design is included in the Optical Detailed Design File Pack in Appendix 7.6.2.

The optical detailed design for the Fibre Link is displayed in Figure 42.

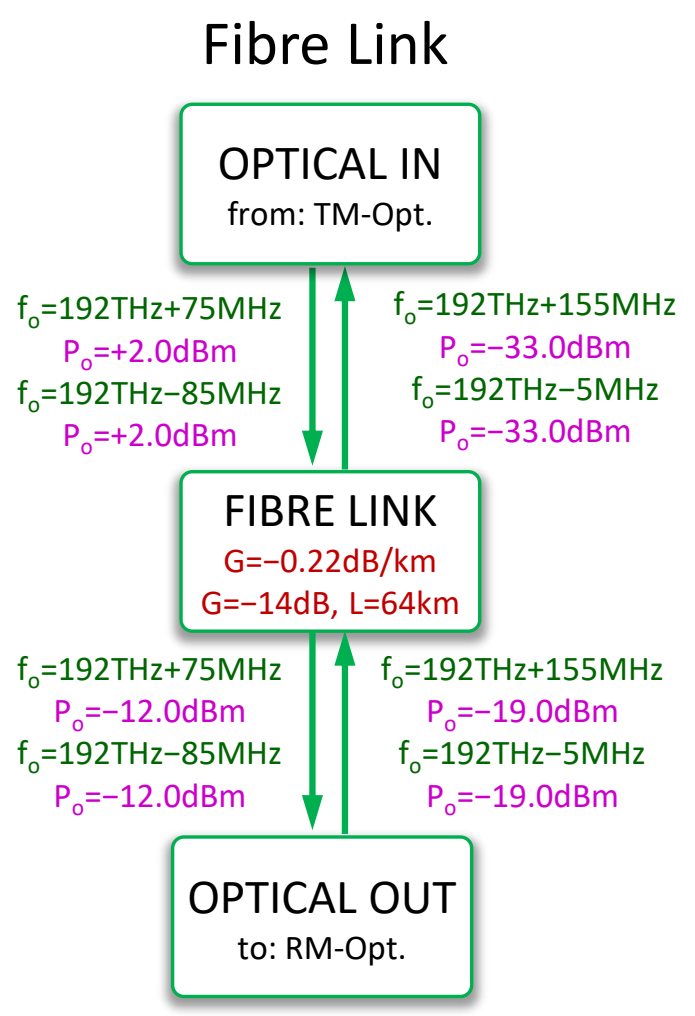

**Figure 42 – SKA1-low Fibre Link optical detailed design figure**

The related *PowerPoint* design is included in the Optical Detailed Design File Pack in Appendix 7.6.2.

The maximum Fibre Link length for the SKA1-low design for UWA's *SKA phase synchronisation system* is 64 km (assuming a conservative loss of 0.22dB/km). The maximum SKA1-low Fibre Link length as per [RD26] is 58 km.

### 3.4.3.6 Receiver Module (141-022300)

The optical detailed design for Receiver Module (141-022300) is displayed in Figure 43.

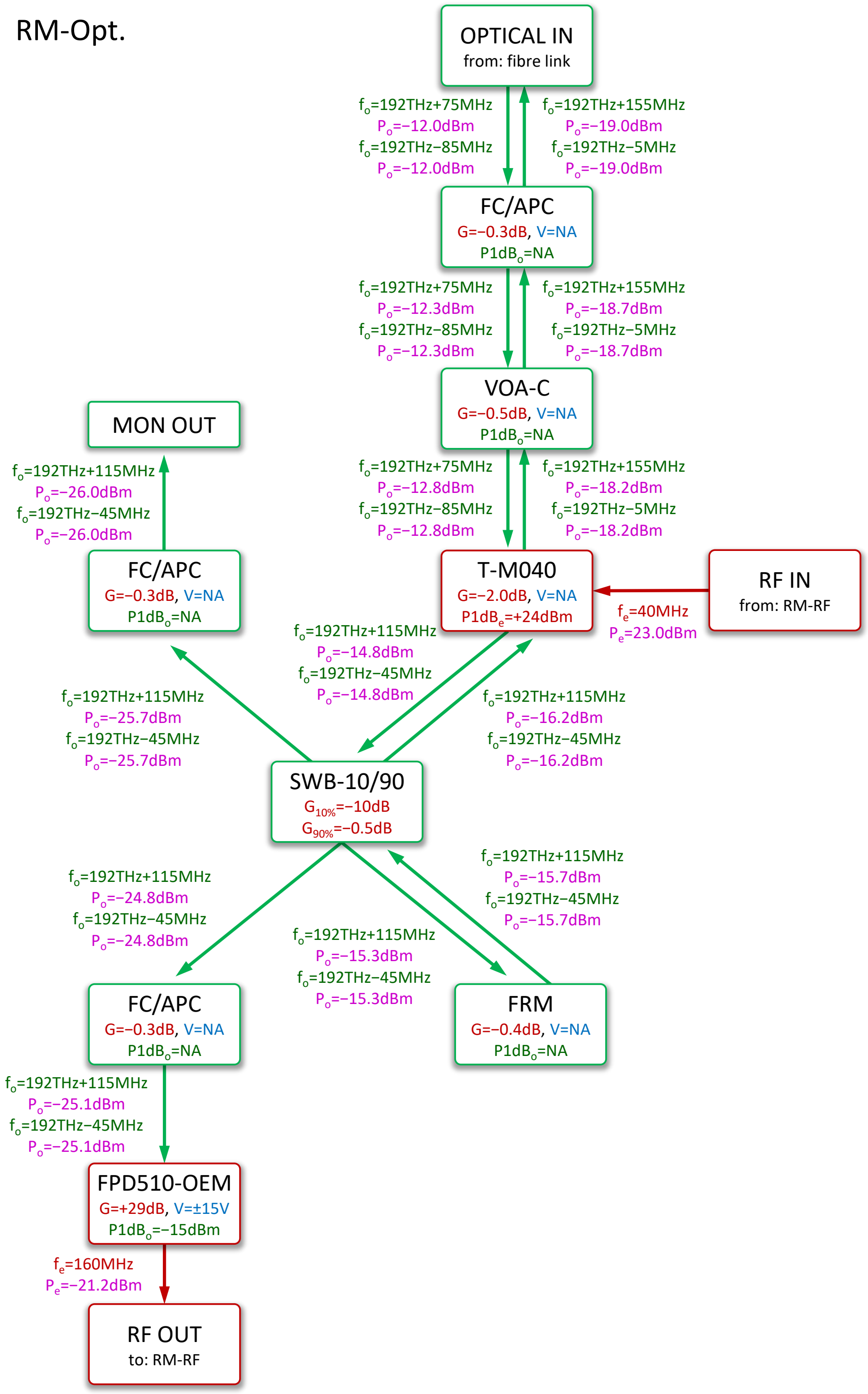

**Figure 43 – SKA1-low Receiver Module (141-022300) optical detailed design figure.**

The related *PowerPoint* design is included in the Optical Detailed Design File Pack in Appendix 7.6.2.

### 3.4.4 **Electronic Detailed Design**

#### 3.4.4.1 Rack Cabinet (141-022900)

Not applicable.

#### 3.4.4.2 Optical Source (141-022400)

Not applicable.

#### 3.4.4.3 Signal Generator (341-023100)

The electronic detailed design for Signal Generator (141-023100) is displayed in Figure 44.

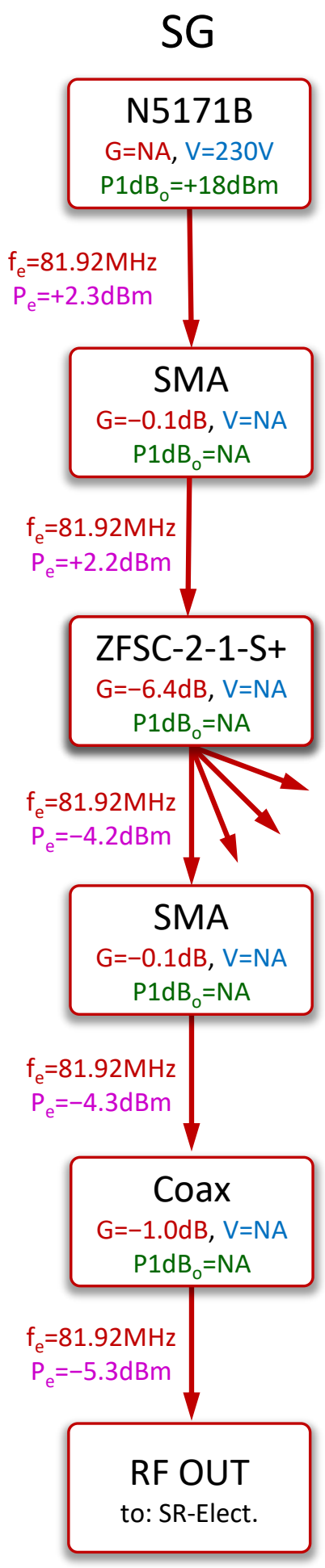

**Figure 44 – SKA1-low Signal Generator (141-023100) electronic detailed design figure.**

The related *PowerPoint* design is included in the Electronic Detailed Design File Pack in Appendix 7.7.2.

### 3.4.4.4 Sub Rack (141-022700)

The electronic detailed design for Sub Rack (141-022700) is displayed in Figure 45.

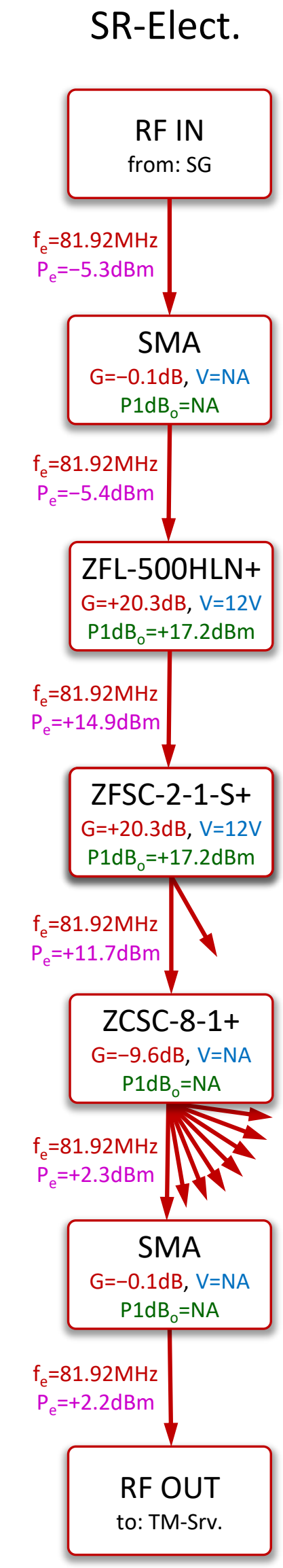

**Figure 45 – SKA1-low Sub Rack (141-022700) electronic detailed design figure.**

The related *PowerPoint* design is included in the Electronic Detailed Design File Pack in Appendix 7.7.2.

### 3.4.4.5 Transmitter Module (141-022100)

The electronic detailed design for Transmitter Module (141-022100) (TM-RF) is displayed in Figure 46.

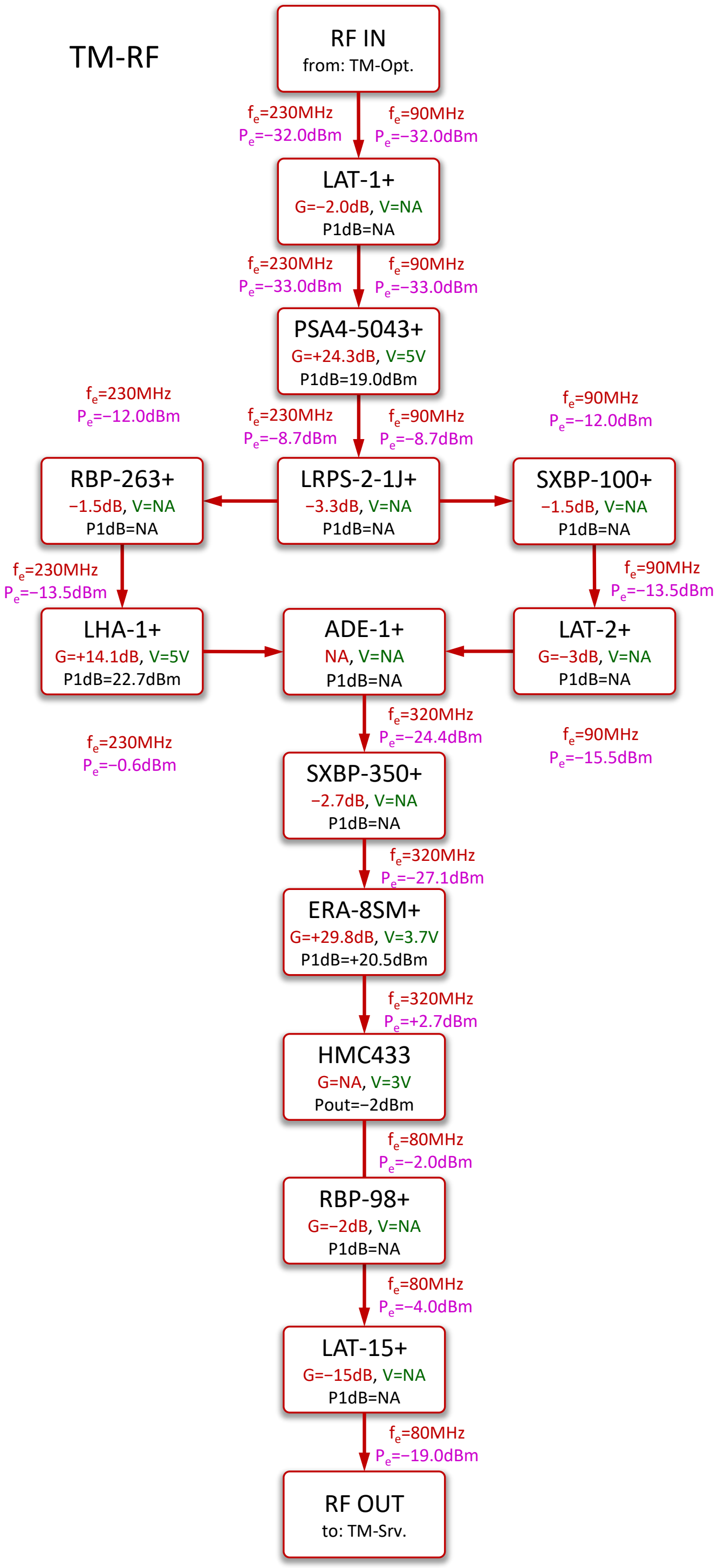

**Figure 46 – SKA1-low Transmitter Module (141-022100) (TM-RF) electronic detailed design figure.**

The related *PowerPoint* design is included in the Electronic Detailed Design File Pack in Appendix 7.7.2.

The electronic circuit schematic for Transmitter Module (141-022100) (TM-RF PCB) is displayed in Figure 47.

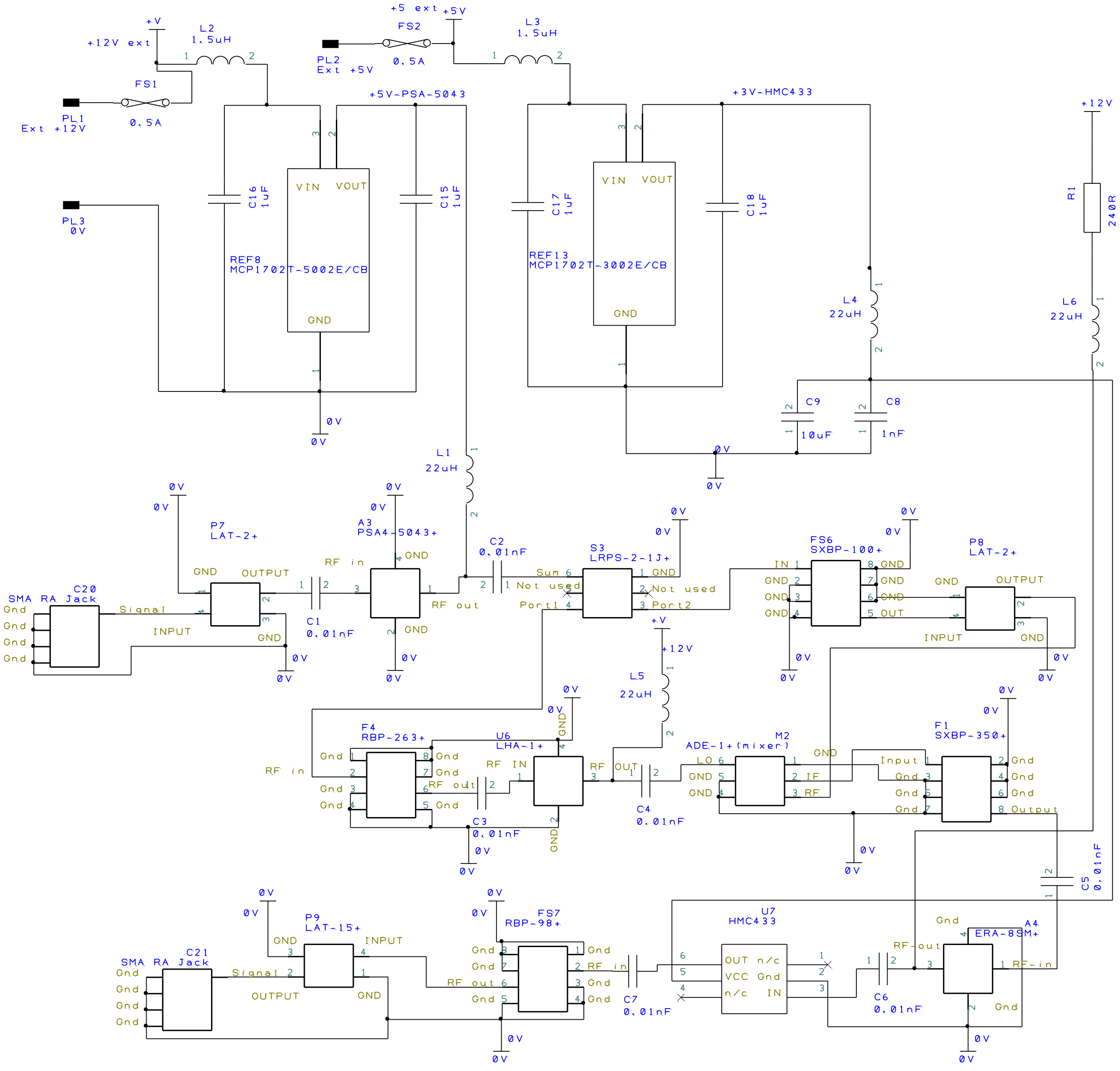

**Figure 47 – SKA1-low Transmitter Module (141-022100) (TM-RF PCB) electronic circuit schematic.**

The related *DesignSpark* circuit schematic is included in the Electronic Detailed Design File Pack in Appendix 7.7.4.

The printed circuit board layout for Transmitter Module (141-022100) (TM-RF PCB) is displayed in Figure 48.

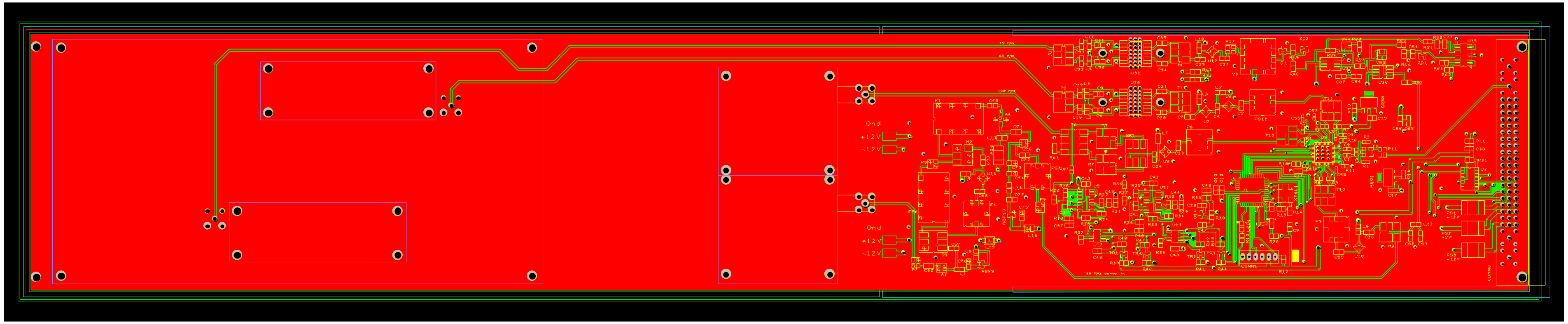

**Figure 48 – SKA1-low Transmitter Module (141-022100) (TM-RF PCB) printed circuit board layout.**

The related *DesignSpark* PCB layout is included in the Electronic Detailed Design File Pack in Appendix 7.7.4.

The electronic detailed design for Transmitter Module (141-022100) (TM-Srv.) is displayed in Figure 49.

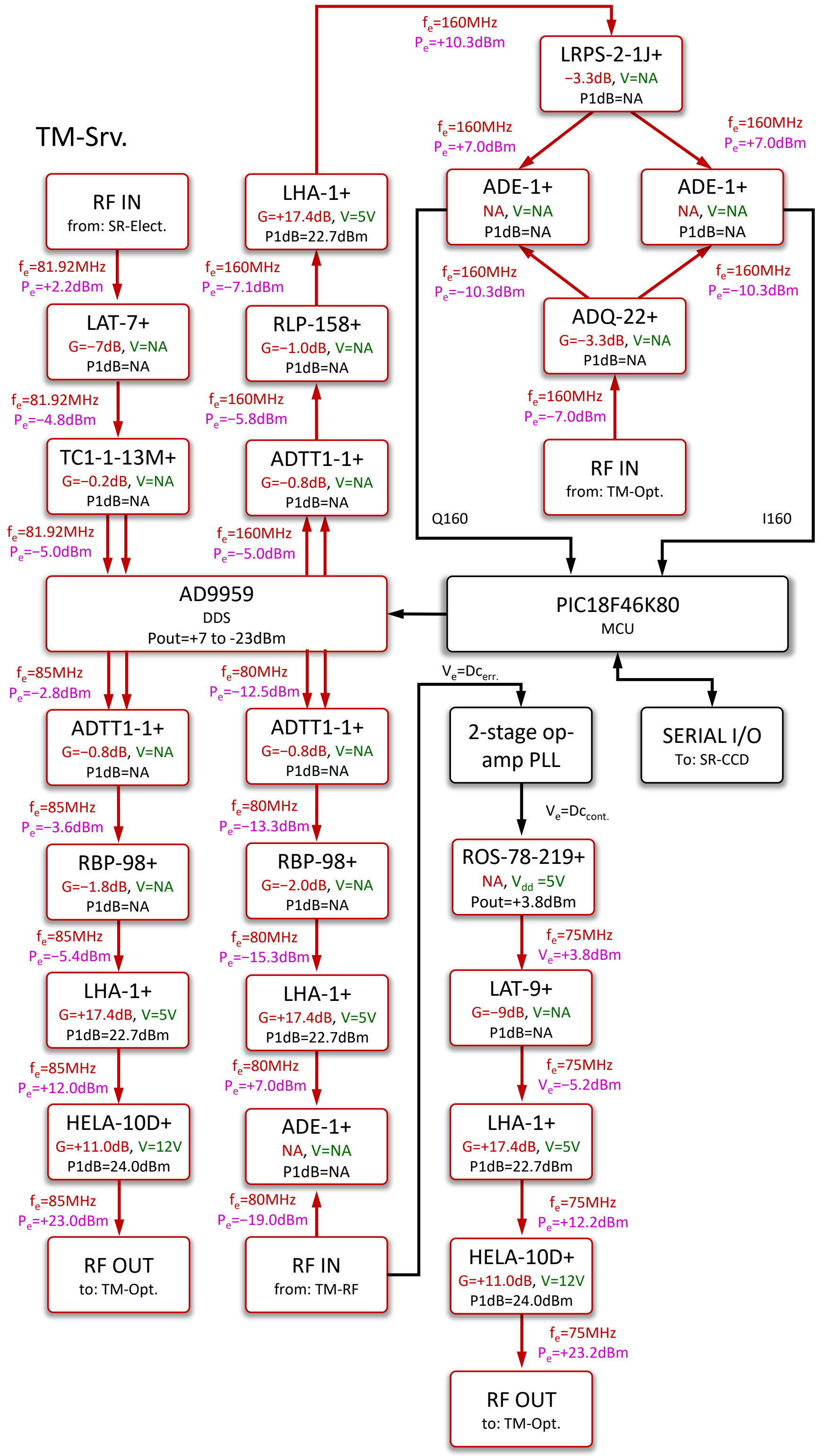

**Figure 49 – SKA1-low Transmitter Module (141-022100) (TM-Srv.) electronic detailed design figure.**

The related *PowerPoint* design is included in the Electronic Detailed Design File Pack in Appendix 7.7.2.

The electronic circuit schematic for Transmitter Module (141-022100) (TM-Srv.) is displayed in Figure 50.

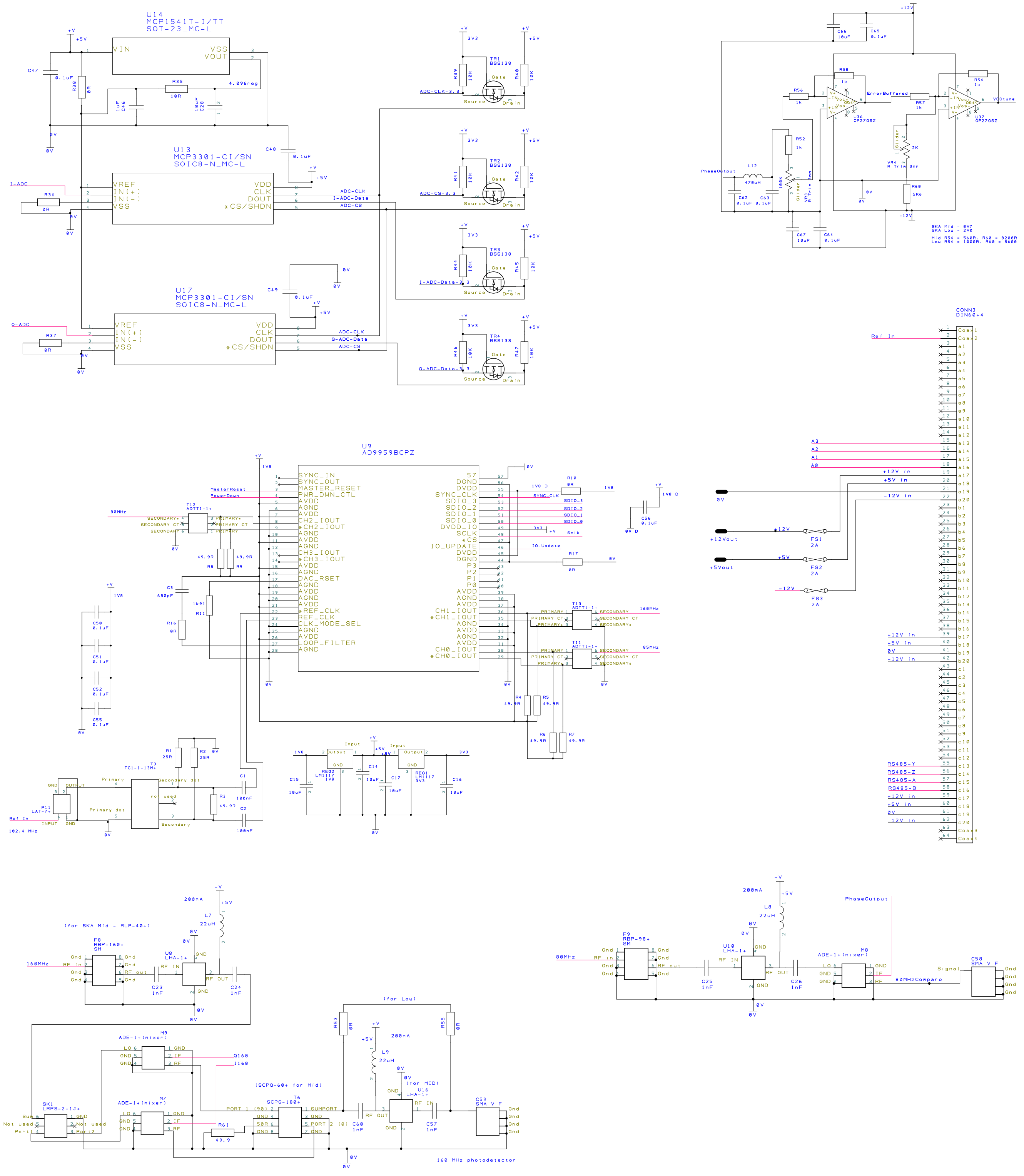

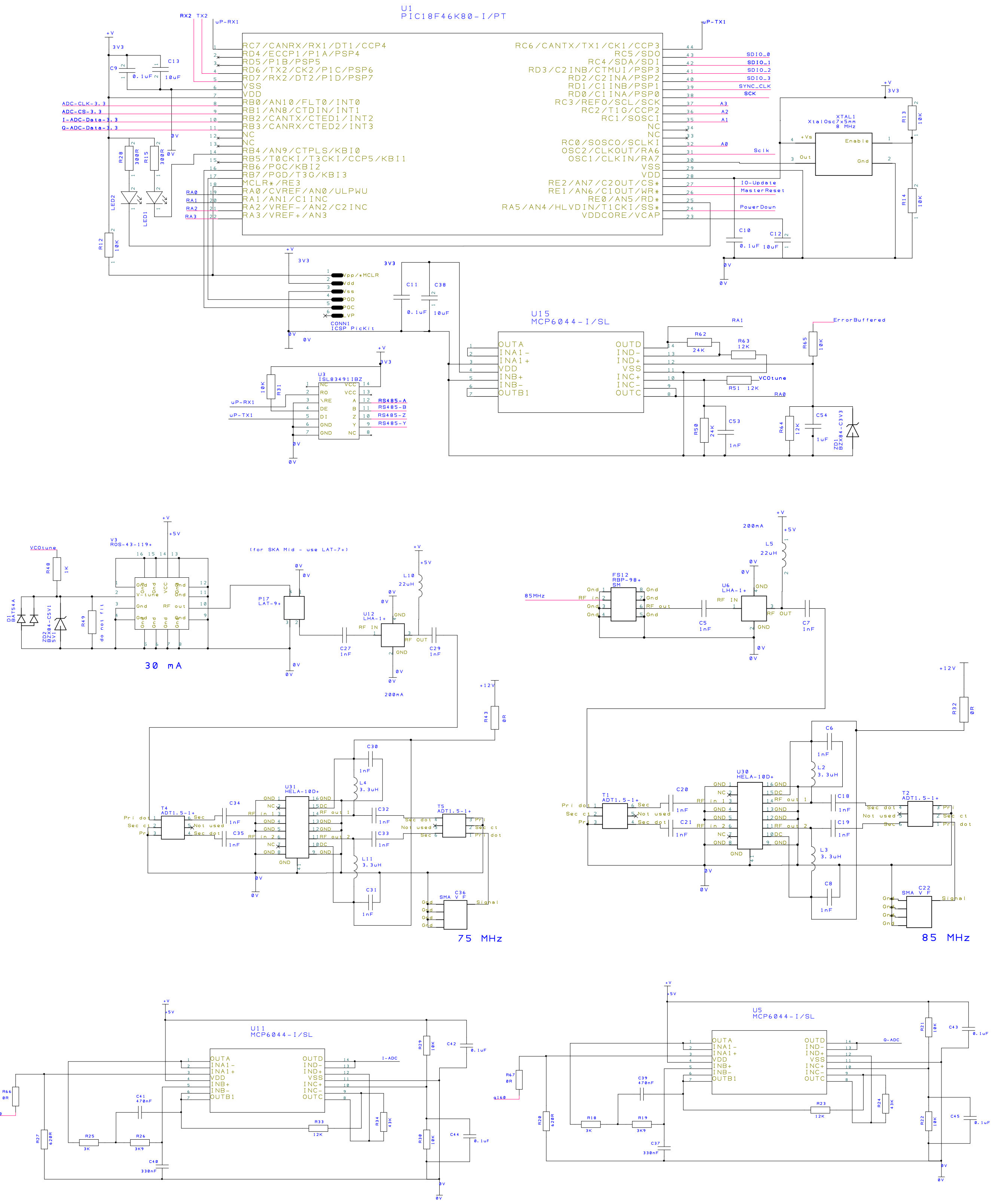

**Figure 50 – SKA1-low Transmitter Module (141-022100) (TM-Srv.) electronic circuit schematic.**

The related *DesignSpark* circuit schematic is included in the Electronic Detailed Design File Pack in Appendix 7.7.4.

### 3.4.4.6 Receiver Module (141-022300)

The electronic detailed design for Receiver Module (141-022300) is displayed in Figure 51.

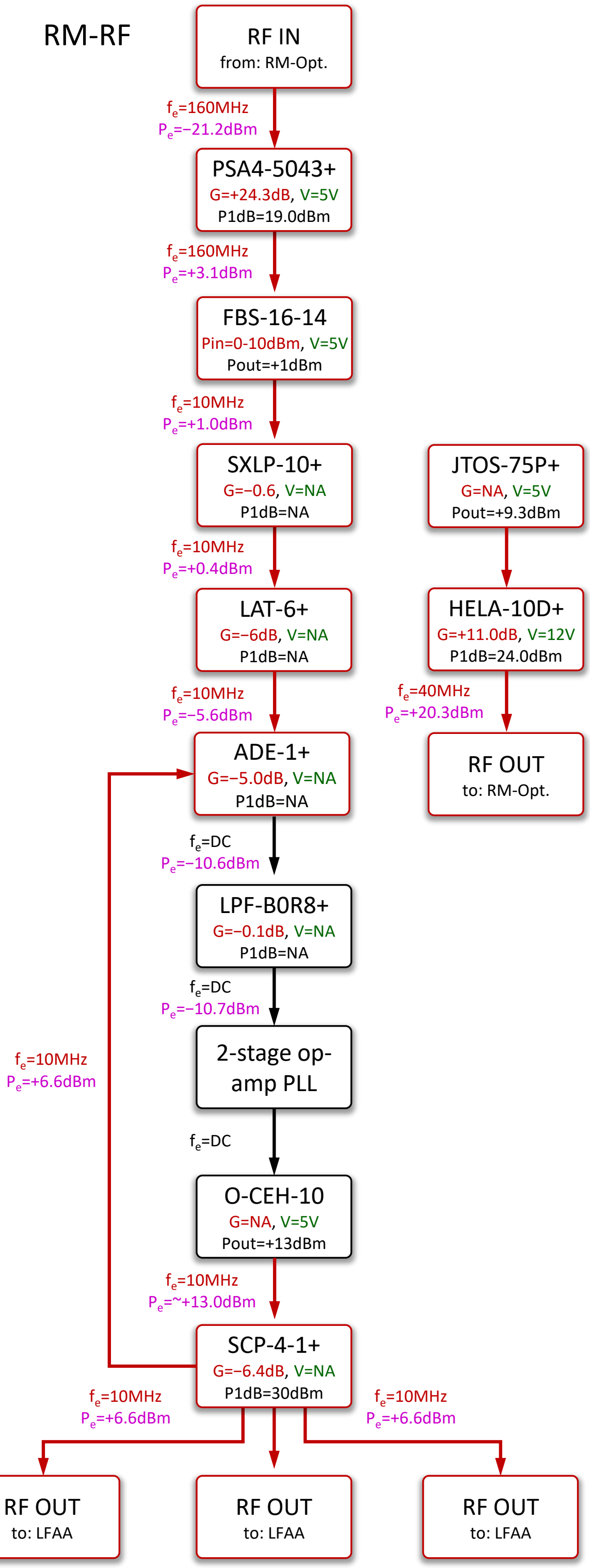

**Figure 51 – SKA1-low Receiver Module (141-022300) electronic detailed design figure.**

The related *PowerPoint* design is included in the Electronic Detailed Design File Pack in Appendix 7.7.2.

The electronic circuit schematic and printed circuit board layout for Receiver Module (141-022300) are a work-in-progress.

While prototype printed circuit boards have already been designed and manufactured as part of the Design for Mass Manufacture work described in Section 5.2.2, the design details has since changed enough that the aforementioned electronic circuit schematic and printed circuit board layout no longer accurately reflect the current design.

However, the related *DesignSpark* circuit schematic and *DesignSpark* PCB layout are still included for the record in the Electronic Detailed Design File Pack in Appendix 7.7.4.

## 3.5 Software

### 3.5.1 System Start-up

The following is the typical start-up procedure for the SADT.SAT.STFR.FRQ (UWA) system.

SAT.LMC commands a power-up of the equipment below in the following order:

- Optical Source (141-022400)
- Signal Generator (141-023100)
- Rack Distribution (141-022800)
- Sub Racks (141-022700); this will also power-up the Transmitter Modules (141-022100)
- Receiver Modules (141-022300)

The frequency stabilisation servo system automatically engages lock. Then run a SAT.LMC system start up diagnostic programme that:

- Confirms the RMS voltage of the servo-loop 'Lock signal' (see Section 3.3.4.1 for an explanation of this signal) for all Transmitter Modules is below the specified threshold voltage.
- Confirms the accumulated phase change as measured by the 'Phase measurement' signal (see Section 3.3.4.1 for an explanation of this signal) is within a set typical range.

If both these conditions are met, and no other faults are reported, then SAT.LMC reports to Telescope Manager that the SADT.SAT.STFR.FRQ (UWA) system is now operational.

### 3.5.2 Normal Operations

No action is required. SAT.LMC monitors health and status.

### 3.5.3 Off-Normal Operations

If only antenna site power is cut:

- Result – The SADT.SAT.STFR.FRQ (UWA) system no longer functions, no radio-frequency reference signals available at remote site.
- Reason – The reflected signal is no longer available (the remote AOM does not transmit light if no RF drive signal is provided), the frequency stabilisation servo-loop cannot lock.
- Diagnostic – The fault is detected by SAT.LMC monitor of Lock signal
  - The fault is trigger by SAT.LMC when the RMS voltage of servo-loop error signal the moves above the specified threshold voltage.
- Resolution – After the power is restored, the frequency stabilisation servo system automatically re-engages.

If transmission though fibre link is cut (break in fibre):

- Result – The SADT.SAT.STFR.FRQ (UWA) system functions with limited long-term coherence. The clean-up oscillators in the Receiver Module continue to function and provide short-term coherence for the array.
- Reason – The reflected signal is no longer available, the frequency stabilisation servo-loop cannot lock.

- Diagnostic – The fault is detected by LMC monitor of Lock signal (see Section 3.3.4.1):
  - The fault is trigger by SAT.LMC when the RMS voltage of servo-loop error signal the moves above the specified threshold voltage.
- Resolution – After the fibre continuity is restored, the frequency stabilisation servo system automatically re-engages.

If only the Central Processing Facility power is cut:

- Result – SADT.SAT.STFR.FRQ (UWA) system functions with limited long-term coherence. The clean-up oscillators in the Receiver Module continue to function and provide short-term coherence for the array.
- Reason – There is no power to transmit the reference signals, or frequency stabilisation servo system.
- Diagnostic – Fault is detected by SAT.LMC when.
  - Status communication ceases from any equipment located within Central Processing Facility.
- Resolution – After the power is restored, a full system start-up is required (as described in Section 3.5.1).

### 3.5.4 System Shutdown

SAT.LMC commands a power-down of the equipment below in the following order:

- Receiver Modules (141-022300)
- Sub Racks (141-022700); this will also power-down Transmitter Modules (141-022100)
- Signal Generator (141-023100)
- Optical Source (141-022400)

## 3.6 Security

The SADT Consortium has stated that the full analysis of Security will only be conducted after the down-select process. This section of the report will be updated at this time.

## 3.7 Safety

### 3.7.1 Hazardous Items

#### 3.7.1.1 Optical Source (141-022400)

The 1552 nm infrared fibre laser that comprises Optical Source has output power of +28.4 dBm (0.7 W). According to the definitions of the IEC 60825-1 standard, this laser is rated as Class 1, given that the laser light is contained within optical fibre. However, should a fibre be accidently disconnected or broken the laser becomes Class 3B (the limit of which is 0.5 W).

This light is then transferred directly to Sub Rack, the outputs of which are +15.2 dBm (33 mW), which is well below the 0.5 W, Class 3B limit and therefore considered eye safe (as Class 3R).

Therefore certified laser training is required for any personnel working or interacting with the optical fibre coming out of the Optical Source.

The optical power being sent down the fibre links is below 10 mW and is therefore Class 1

### 3.7.2 Training

#### 3.7.2.1 Receiver Module (141-022300)

UWA's *SKA phase synchronisation system* has been designed such that the commissioning process requires the optimisation of only one free parameter per link; no other adjustments or optimisations are required across the entire system. The optimisation involves making the total power loss of all optical fibre links (regardless of length) equal; this ensures that the optical power levels all Transmitter Modules and Receiver Modules are equal across the array, which means that resulting electronic signal levels are like-wise optimised and equal between all system. The commissioning personnel will have to be trained to conduct this optimisation process.

The process is very simple, takes less than a minute to execute, and involves only a simple optical power meter. After installation of all necessary equipment a particular link, the optical power meter is temporarily connected to the 'monitor port' of the Receiver Module (see Figure 43). A screwdriver is then used to adjust the trim of a variable optical attenuator which is accessed via recess on the front of the Receiver Module (see Figure 37) until the power meter indicates the nominal optical power value.

## 3.8 Integration

### 3.8.1 Component and System Integration

UWA researchers in partnership with MeerKAT and UoM electronic engineers, progressed the detailed designs described in this document, into a set of mass-manufacture archetypes, effectively getting a head-start at addressing manufacturing issues that may be encounter by contractors during SKA Construction. This includes construction-ready sets of optical, electronic, and mechanical design files, and a bill-of materials based on components purchased from vendors to construct these archetypes – and this forms the basis of the UWA *Detailed Cost Model* (Section 3.10). The first set of mass-manufacture archetypes for SKA1-low were completed in Q2, 2016 [RD21]; and for SKA1-mid in Q1, 2017 [RD22]. The assembly and testing of the completed archetypes was done at UWA. To get an independent evaluation of labour costs associated with assembly and testing, a local optical manufacturing technology consultancy company was employed to provide an independent review of the costings (see Appendix 7.9.5)

### 3.8.2 WSOI Integration

UWA's *phase synchronisation system* has interfaces with several other SKA elements (wider system-of-interest), as outlined in Section 3.8.4 and Section 3.8.5; however, it is fundamentally self-contained.

Described in Section 4.1.1 Test and Verification by Demonstration.

### 3.8.3 Precursor Integration

Previous astronomical verification work with ASKAP [RD19] and ATCA [RD18, 20] is described in Section 4.1.1. We have €14.4k of funding to continue SKA1-low integration tests with AAVS1. UWA and Curtin University have arranged joint MRO field trial. Our plan calls to integrate UWA's SKA *phase synchronisation system* with the SKA1-low AAVS1.

### 3.8.4 SADT Interdependencies

As shown in detail in Section 3.4.1, the SKA1-mid variant of UWA's *phase synchronisation system* has interfaces with the following SADT elements:

- SADT.SAT.CLK
- SADT.SAT.LMC
- SADT.LINFRA
- SADT.NSDN

Further information on the interfaces with these elements is provided in the relevant internal interface control document.

### 3.8.5 Non-SADT (External) Dependencies

As shown in detail in Section 3.4.1, UWA's *phase synchronisation system* has interfaces with only the DISH Consortium hardware. Further information on this interface is provided in the relevant external interface control document.

## 3.9 Interoperability

### 3.9.1 SADT Interoperability

UWA's *phase synchronisation system* has been extensive tested with elements that simulate or replicate the various SADT interfaces outlined in Section 3.8.4. Specifically, the software interfaces with SAT.LMC (and hardware interfaces with SAT.NSDN) was replicated using a mock-up implementation of both the Lock signal and Phase measurement monitoring, and this was demonstrated to the SADT Consortium during the face-to-face meeting in Perth 2016 (see Section 3.3.4.1). The SAT.CLOCK interface (receiving a 10 MHz reference signal from a hydrogen maser) is daily practice; all of the evaluation measurements outlined in Section 4.1.1 were conducted with this in place. Our design for mass manufacture archetypes have been installed in rack mount enclosures for interfacing with the SADT.LINFRA rack cabinet.

### 3.9.2 Non-SADT (External) Interoperability

UWA's *phase synchronisation system* has been extensive tested with an interface that is closely resembling of the LFAA interface outlined in Section 3.8.5 as is possible. During our astronomical verification test with ASKAP [RD19], the output of our Receiver Modules were used to provide the reference clock for the ASKAP receivers. As described in Section 3.3.2, the outcome of these tests were used to verify the Functional Performance Requirements of our system.

## 3.10 Costing

### 3.10.1 Development Costs

The development costs for both the SKA1-mid and SKA1-low variants of UWA's *SKA phase synchronisation system* has been €482.0k. These expenses were incurred between the period start-February 2014 and end-June 2017. This cost is divided as follows:

- Salaries and on-costs: €354.4k
- Laboratory equipment: €64.8k
- Consumables: €17.1k
- Collaborative travel: €25.1k
- Field work: €20.6k

### 3.10.2 Capital Expenditure Equipment Costs

The summary of capital expenditure cost for SKA1-low equipment as determined by the Detailed Cost Model (see Appendix 7.9.2) is given in Figure 52.

**SUMMARY**
**Capex Equipment Cost**

| **SKA-LOW** | **Per Unit** | | **Totals** | | |
|---|---|---|---|---|---|
| **UWA STFR.FRQ LRUs** (-) | **Equipment** (€) | | **Quantity** (#) | **Cost** (€) | |
| Rack Cabinet (141-022900) | € | 3,067 | 1 | € | 3,067 |
| Optical Source (141-022400) | € | 39,869 | 1 | € | 39,869 |
| Signal Generator (141-023100) | € | 6,111 | 1 | € | 6,111 |
| Sub Rack (141-022700) | € | 4,238 | 3 | € | 12,715 |
| Transmitter Module (141-022100) | € | 10,619 | 37 | € | 392,888 |
| Receiver Module (141-022300) | € | 4,979 | 37 | € | 184,216 |
| **Total Cost** | | | | **€** | **638,865** |
| **Cost Per Link** | | | | **€** | **17,267** |

**Figure 52 – SKA1-low summary of the capex equipment detailed cost model database.**

The ancillary parameters used in capital expenditure cost for SKA1-low equipment Detailed Cost Model database are given in Figure 53.

| **Exchange Rate** | **AUD** | **EUR** | **GBP** | **USD** |
|---|---|---|---|---|
| **1 EUR** | 1.46000 | 1.00000 | 0.77483 | 1.12870 |
| Inverse | 0.68493 | 1.00000 | 1.29061 | 0.88598 |
| (2016/04/26) | | | | |

| | | |
|---|---|---|
| **Volume Discount Rate** | 0.0025 | per unit |

**Figure 53 – SKA1-low ancillary parameters used in the capex equipment detailed cost model database.**

#### 3.10.2.1 Rack Cabinet (141-022900)

The detailed cost model database for Rack Cabinet (141-022900) is displayed in Figure 54.

| Rack Cabinet (341-022900) | | |
|---|---|---|
| Equipment Cost | € | 834.36 |
| Total Power | | 60.00 |

| Identifier | Company | Part Number | Part Description | Case Style | CAD/PCE | URL | Q.Each | Min.Q. | | Unit Price | Currency | Exg | | Total |
|---|---|---|---|---|---|---|---|---|---|---|---|---|---|---|
| RC-Equipment | RS | 118-3285 | Heat management | 19", 1U rack | Y | http://au | 8 | 1 | $ | 152.27 | AUD | 0.685 | € | 834.36 |

**Figure 54 – SKA1-low Rack Cabinet (141-022900) capex equipment detailed cost model database.**

#### 3.10.2.2 Optical Source (141-022400)

The detailed cost model database for Optical Source (141-022400) is displayed in Figure 55.

| Optical Source (141-022400) | | |
|---|---|---|
| Equipment Cost | € | 39,868.88 |
| Total Power | | 100.00 |

| Identifier | Company | Part Number | Part Description | Case Style | Model | URL | Q.Each | Min.Q. | | Unit Price | Currency | Exg | | Total |
|---|---|---|---|---|---|---|---|---|---|---|---|---|---|---|
| OS-Laser | NP Photonics | Power Rock | High Power Laser Source | --- | N | http://v | 1 | 1 | $ | 45,000.00 | USD | 0.886 | € | 39,868.88 |

**Figure 55 – SKA1-low Optical Source (141-022400) capex equipment detailed cost model database.**

#### 3.10.2.3 Signal Generator (141-023100)

The detailed cost model database for Signal Generator (141-023100) is displayed in Figure 56.

| Signal Generator (141-023100) | | |
|---|---|---|
| Equipment Cost | € | 6,154.05 |
| Total Power | | 75.00 |

| Identifier | Company | Part Number | Part Description | Case Style | Model | URL | Q.Each | Min.Q. | | Unit Price | Currency | Exg | | Total |
|---|---|---|---|---|---|---|---|---|---|---|---|---|---|---|
| SG-Equipment | Keysight | N5171B-501 | EXG Signal Generator,9 kHz to 1 GHz | 19", 1U rack | Y | http://v | 1 | 1 | $ | 8,921.60 | AUD | 0.685 | € | 6,110.68 |
| SG-Equipment | Minicircuits | ZFSC-2-1-S+ | PWR SPLTR CMBD / BNC / RoHS | K18 | N | https:// | 1 | 1 | $ | 48.95 | USD | 0.886 | € | 43.37 |

**Figure 56 – SKA1-low Signal Generator (141-023100) capex equipment detailed cost model database.**

### 3.10.2.4 Sub Rack (141-022700)

The detailed cost model database for Sub Rack (141-022700) is displayed in Figure 57

| Sub Rack (141-022700) | | |
|---|---|---|
| Equipment Cost | € | 4,259.51 |
| Total Power | | 0.00 |

| Identifier | Company | Part Number | Part Description | Case Style | Model | URL | Q.Each | Min.Q. | | Unit Price | Currency | Exg | | Total |
|---|---|---|---|---|---|---|---|---|---|---|---|---|---|---|
| SR-Eurocard frame | Lektronics | 3684.026 | Rittal Ripac Vario 3Ux84HPx465 sub-rack | --- | Y | | 1 | 1 | $ | 190.00 | AUD | 0.685 | € | 130.14 |
| SR-Eurocard frame | Lektronics | 3636.010 | Rittal Pkt2 3U handles | --- | Y | | 1 | | $ | 18.00 | AUD | 0.685 | € | 12.33 |
| SR-Eurocard frame | Lektronics | 3652.050 | Rittal 3Ux8HP standard front panel kit (with pl | --- | Y | | 1 | | $ | 19.45 | AUD | 0.685 | € | 13.32 |
| SR-Eurocard frame | Lektronics | 3652.070 | Rittal 3Ux12HP standard front panel kit (with | --- | Y | | 1 | | $ | 22.85 | AUD | 0.685 | € | 15.65 |
| SR-Eurocard frame | Lektronics | 3684.656 | Rittal 280mm Plastic snap in card guide | --- | N | | 32 | | $ | 0.90 | AUD | 0.685 | € | 19.73 |
| SR-Eurocard frame | Lektronics | 3684.691 | Rittal 84HPx465 solid unvented lid | --- | Y | | 2 | | $ | 39.80 | AUD | 0.685 | € | 54.52 |
| SR-Eurocard frame | Lektronics | 3684.234 | Rittal mounting blocks (Pkt10) | --- | N | | 2 | | $ | 19.45 | AUD | 0.685 | € | 26.64 |
| SR-Eurocard frame | Lektronics | 3684.233 | Rittal screws (Pkt100) | --- | N | | 1 | | $ | 12.00 | AUD | 0.685 | € | 8.22 |
| SR-Eurocard frame | Lektronics | --- | Shipping | --- | N/A | | 1 | | $ | 25.00 | AUD | 0.685 | € | 17.12 |
| | | | | | | | | | | | | | | |
| SR-CCD Module PCB | Advanced Technolog | --- | FR4 PCB manufacture and assembly | | | | 1 | 1 | $ | 200.00 | AUD | 0.685 | € | 136.99 |
| SR-CCD Module PCB | *various* | *various* | CCD components and Ethernet controller | | | | 1 | 1 | $ | 100.00 | EUR | 1.000 | € | 100.00 |
| SR-Backplane PCB | Advanced Technolog | --- | FR4 PCB manufacture and assembly | | | | 1 | 1 | $ | 200.00 | AUD | 0.685 | € | 136.99 |
| SR-Backplane PCB | *various* | *various* | Passive RF splits and connectors | | | | 1 | 1 | $ | 50.00 | EUR | 1.000 | € | 50.00 |
| | | | | | | | | | | | | | | |
| SR-Optical Distribution | oeMarket | PMC-F-1550-1x2-50/50 | fiberised PM optical coupler (50/50) | --- | N | http://v | 1 | | $ | 299.00 | USD | 0.886 | € | 264.91 |
| SR-Optical Distribution | oeMarket | PMCM-1550-1x8-1-FAF | Secondary PM fibre 1/8 splitters | --- | N | http://v | 2 | | $ | 1,620.00 | USD | 0.886 | € | 2,870.56 |
| | | | | | | | | | | | | | | |
| SR-Power Supplies | Lektronics | 3686.682 | CPCI Power Supply Unit | 3U, 12HP | Y | http://im | 1 | 1 | $ | 550.00 | AUD | 0.685 | € | 376.71 |
| SR-Power Supplies | Lektronics | 3685.33 | CPCI Power Supply Faceplate | 3U, 12HP | Y | http://im | 1 | 1 | $ | 10.00 | AUD | 0.685 | € | 6.85 |
| | | | | | | | | | | | | | | |
| SR-Connectors | RS | | Ethernet bulkhead | | Y | | 1 | 1 | $ | 10.20 | AUD | 0.685 | € | 6.99 |
| SR-Connectors | RS | | SMA bulkhead | | Y | | 2 | 1 | $ | 6.80 | AUD | 0.685 | € | 9.32 |
| SR-Connectors | oeMarket | | SC/APC bulkhead | | Y | | 1 | 1 | $ | 3.70 | AUD | 0.685 | € | 2.53 |

**Figure 57 – SKA1-low Sub Rack (141-022700) capex equipment detailed cost model database.**

### 3.10.2.5 Transmitter Module (141-022100)

The detailed cost model database for Transmitter Module (141-022100) is displayed in Figure 58.

| Transmitter Module (141-022100) | | |
|---|---|---|
| Equipment Cost | € | 11,668.80 |
| Total Power | | 0.00 |

| Identifier | Company | Part Number | Part Description | Case Style | Model | URL | Q.Each | Min.Q. | Unit Price | Currency | Exg | Total |
|---|---|---|---|---|---|---|---|---|---|---|---|---|
| TM-Opt. | Gooch&Housego | T-M080-0.5C8J-3-F2P | +75MHz PM AOM | --- | Y | http://go | 1 | 1 | $ 2,255.00 | GBP | 1.291 | € 2,910.32 |
| TM-Opt. | Gooch&Housego | T-M080-0.5C8J-3-F2P | -85MHz PM AOM | --- | Y | http://go | 1 | | $ 2,255.00 | GBP | 1.291 | € 2,910.32 |
| TM-Opt. | Photonics Solutions | FPD510-OEM | Photodetector | --- | Y | http://ww | 2 | | $ 890.00 | GBP | 1.291 | € 2,297.28 |
| TM-Opt. | OEMarket | FM-PM-1550-0-N | PM Faraday Mirror | --- | | http://ww | 1 | | $ 185.00 | USD | 0.886 | € 163.91 |
| TM-Opt. | OEMarket | PMC-S-1550-2x2-50/5( | 50/50 PM Splitter | --- | | http://ww | 3 | | $ 525.00 | USD | 0.886 | € 1,395.41 |
| TM-Opt. | OEMarket | ISO-PM-1550-S-P-0-N | PM Optical Isolator | --- | | http://ww | 1 | | $ 294.00 | USD | 0.886 | € 260.48 |
| TM-Opt. | OEMarket | PMC-S-1550-1x2-10/9( | PM Fixed Optical Attenuator | --- | | http://ww | 1 | | $ 425.00 | USD | 0.886 | € 376.54 |
| TM-Opt. | OEMarket | FOA-IL-1550-20-0-N | Fixed Optical Attenuator | --- | | http://ww | 1 | | $ 17.99 | USD | 0.886 | € 15.94 |
| TM-Opt. | OEMarket | ADP-MF-FC-PMAF | Optical Connector | --- | | http://ww | 5 | | $ 14.90 | USD | 0.886 | € 66.01 |
| TM-Opt. | Benchmark | | Bespoke plastic enclosure | --- | Y | http://ww | 1 | | $ 315.70 | AUD | 0.685 | € 216.23 |
| TM-Opt. | Element 14 | 124 8989 | SMA Right-angle female | SMA Throug | Y | http://au | 4 | | $ 2.56 | GBP | 1.291 | € 13.22 |
| TM-Opt. | Newbury Electronics | | FR4 4-layer PCB, 1.6mm, Eurocard 100x340mm | --- | Y | | 1 | | $ 55.43 | GBP | 1.291 | € 71.53 |
| TM-Opt. | Lektronics | 3652.010 | Rittal 3Ux4HP standard front panel kit (with pl | --- | Y | | 16 | | $ 18.65 | AUD | 0.685 | € 204.38 |
| TM-RF | Minicircuits | LAT-2+ | Attenuator | MMM168 / | N | http://ww | 1 | | $ 1.95 | USD | 0.886 | € 1.73 |
| TM-RF | Minicircuits | PSA4-5043+ | Amplifier | MMM1362 , | N | http://v | 1 | | $ 2.58 | USD | 0.886 | € 2.29 |
| TM-RF | Minicircuits | LRPS-2-1J+ | Splitter | QQQ569 | N | http://ww | 1 | | $ 10.00 | USD | 0.886 | € 8.86 |
| TM-RF | Minicircuits | RBP-263+ | BP Filter 230MHz | GP731 | N | http://v | 1 | | $ 13.70 | USD | 0.886 | € 12.14 |
| TM-RF | Minicircuits | SXBP-100+ | BP Filter 90MHz | HF1139 | N | http://ww | 1 | | $ 15.95 | USD | 0.886 | € 14.13 |
| TM-RF | Minicircuits | LHA-1+ | Amplifier | FG873 / MC | N | http://v | 1 | | $ 1.64 | USD | 0.886 | € 1.45 |
| TM-RF | Minicircuits | ADE-1+ | Mixer | CDC636 | N | http://v | 1 | | $ 15.00 | USD | 0.886 | € 13.29 |
| TM-RF | Minicircuits | SXBP-350+ | BP Filter 320MHz | HF1139 | N | http://v | 1 | | $ 15.95 | USD | 0.886 | € 14.13 |
| TM-RF | Minicircuits | ERA-8SM+ | Amplifier | WW107 | N | http://v | 1 | | $ - | USD | 0.886 | € - |
| TM-RF | Hittite | HMC433 | Divide by 4 | HMC433 | N | https:// | 1 | | $ 17.92 | USD | 0.886 | € 15.88 |
| TM-RF | Minicircuits | RBP-98+ | Filter 100MHz | GP731 | N | http://v | 1 | | $ 13.70 | USD | 0.886 | € 12.14 |
| TM-RF | Minicircuits | LAT-15+ | Attenuator | MMM168 | N | http://v | 1 | | $ 1.95 | USD | 0.886 | € 1.73 |
| TM-RF | Newbury Electronics | | FR4 4-layer PCB | --- | --- | | 1 | | $ 50.00 | GBP | 1.291 | € 64.53 |
| TM-RF | Newbury Electronics | | Assembly | --- | --- | | 1 | | $ 100.00 | GBP | 1.291 | € 129.06 |
| TM-RF | Minicircuits | LAT-1+ | Attenuator | | | https:// | 1 | | $ 2.15 | USD | 0.886 | € 1.90 |
| TM-Srv. | RS | 9030006262 | Female Copper Alloy DIN Connector | | | http://c | 1 | | $ 9.91 | GBP | 1.291 | € 12.79 |
| TM-Srv. | RS | 9031606901 | Right Angle DIN 41612 Connector | | | https:// | 1 | | $ 8.63 | GBP | 1.291 | € 11.14 |
| TM-Srv. | Minicircuits | HELA-10D+ | Power Amplifier | | | http://ww | 1 | | $ 19.95 | USD | 0.886 | € 17.68 |
| TM-Srv. | Minicircuits | ADE-1+ | Mixer | | | http://v | 3 | | $ 15.00 | USD | 0.886 | € 39.87 |
| TM-Srv. | Minicircuits | LHA-1+ | Amplifier | | | http://v | 4 | | $ 1.64 | USD | 0.886 | € 5.81 |
| TM-Srv. | Minicircuits | RBP-98+ | Filter 100 MHz | | | http://ww | 2 | | $ 13.70 | USD | 0.886 | € 24.28 |
| TM-Srv. | Minicircuits | ADTT1-1+ | Transformer | | | http://v | 3 | | $ 7.88 | USD | 0.886 | € 20.94 |
| TM-Srv. | Minicircuits | OP27EP | DC amplifier | | | | 1 | | $ 6.34 | USD | 0.886 | € 5.62 |
| TM-Srv. | RS | OP27EP | Filter DC | | | | 1 | | $ 6.34 | USD | 0.886 | € 5.62 |
| TM-Srv. | RS | AD9959 | DDS | | | http://ww | 1 | | $ 87.36 | USD | 0.886 | € 77.40 |
| TM-Srv. | Analog Devices | PIC18F46K80-I/PT | uProcessor | | | http://v | 1 | | $ 5.80 | USD | 0.886 | € 5.14 |
| TM-Srv. | RS | 73100-0114 | SMA PCB launcher | | | http://a | 5 | | $ 9.08 | AUD | 0.685 | € 31.10 |
| TM-Srv. | Element 14 | | FR4 4-layer PCB | | | | 1 | | $ 50.00 | GBP | 1.291 | € 64.53 |
| TM-Srv. | Newbury Electronics | | Assembly | | | 1 | $ | 100.00 | GBP | 1.291 | € | 129.06 |
| TM-Srv. | Newbury Electronics | LAT-7+ | Attenuator | | https:// | 1 | $ | 2.15 | USD | 0.886 | € | 1.90 |
| TM-Srv. | Minicircuits | TC1-1-13M+ | RF Transformer | | https:// | 1 | $ | 2.50 | USD | 0.886 | € | 2.21 |
| TM-Srv. | Minicircuits | RLP-158+ | Low-Pass Filter | | https:// | 1 | $ | 7.95 | USD | 0.886 | € | 7.04 |
| TM-Srv. | Minicircuits | LRPS-2-1J+ | Power Splitter | | https:// | 1 | $ | 8.95 | USD | 0.886 | € | 7.93 |
| TM-Srv. | Minicircuits | ADQ-22+ | Power Splitter | | https:// | 1 | $ | 9.35 | USD | 0.886 | € | 8.28 |
| TM-Srv. | Minicircuits | ROS-78-219+ | VCO | | https:// | 1 | $ | 19.95 | USD | 0.886 | € | 17.68 |
| TM-Srv. | Minicircuits | LAT-9+ | Attenuator | | https:// | 1 | $ | 2.15 | USD | 0.886 | € | 1.90 |

**Figure 58 – SKA1-low Transmitter Module (141-022100) capex equipment detailed cost model database.**

### 3.10.2.6 Receiver Module (141-022300)

The detailed cost model database for Receiver Module (141-022300) is displayed in Figure 59.

| Receiver Module (141-022300) | | |
|---|---|---|
| Equipment Cost | € | 5,471.22 |
| Total Power | | 0.00 |

| Identifier | Company | Part Number | Part Description | Case Style | Model | URL | Q.Each | Min.Q. | Unit Price | Currency | Exg | Total |
|---|---|---|---|---|---|---|---|---|---|---|---|---|
| RM-Opt. | Gooch&Housego | T-M040-0.5C8J-3-F2S | 40MHz AOM | | | https://g | 1 | 1 | $ 1,850.00 | GBP | 1.291 | € 2,387.62 |
| RM-Opt. | OEMarket | SMC-1550-2x2-P-10/9( | 10/90 SM Splitter | --- | Y | http://ww | 1 | 1 | $ 17.90 | USD | 0.886 | € 15.86 |
| RM-Opt. | oeMarket | VOA-C-1550-9-0-N | Variable Optical Attenuator | | | http://ww | 1 | 1 | $ 179.95 | USD | 0.886 | € 159.43 |
| RM-Opt. | oeMarket | ADP-MF-FC-PMAF | Optical connector | | | http://ww | 3 | 1 | $ 18.90 | USD | 0.886 | € 50.23 |
| RM-Opt. | oeMarket | FRM-1550-0-N | SM Faraday Mirror | | | http://ww | 1 | 1 | $ 69.95 | USD | 0.886 | € 61.97 |
| RM-Opt. | Photonics Solutions | FPD510-OEM | Photodetector | | | | 1 | 1 | $ 890.00 | GBP | 1.291 | € 1,148.64 |
| RM-RF | Minicircuits | ADE-1+ | Mixer | | Y | http://v | 1 | 1 | $ 15.00 | USD | 0.886 | € 13.29 |
| RM-RF | RS | OP27EP | DC amplifier | | Y | | 1 | 1 | $ 6.34 | USD | 0.886 | € 5.62 |
| RM-RF | RS | OP27EP | Filter DC | | Y | | 1 | 1 | $ 6.34 | USD | 0.886 | € 5.62 |
| RM-RF | Newbury Electronics | | FR4 2-layer PCB | | Y | | 1 | 1 | $ 50.00 | GBP | 1.291 | € 64.53 |
| RM-RF | Newbury Electronics | | Assembly | | Y | | 1 | 1 | $ 100.00 | GBP | 1.291 | € 129.06 |
| RM-RF | Minicircuits | PSA4-5043+ | Amplifier | | | https:// | 1 | 1 | $ 2.58 | USD | 0.886 | € 2.29 |
| RM-RF | RF Bay | FBS-16-14 | Divide by 16 | | | http://r | 1 | 1 | $ 540.00 | USD | 0.886 | € 478.43 |
| RM-RF | Minicircuits | SXLP-10+ | Low Pass Filter | | | https:// | 1 | 1 | $ 8.95 | USD | 0.886 | € 7.93 |
| RM-RF | Minicircuits | LAT-6+ | Attenuator | | | https:// | 1 | 1 | $ 2.15 | USD | 0.886 | € 1.90 |
| RM-RF | Minicircuits | LPF-B0R8+ | Low Pass Filter | | | https:// | 1 | 1 | $ 9.35 | USD | 0.886 | € 8.28 |
| RM-RF | NEL FC Inc | O-CEH-10 | OCXO | | | http://v | 1 | 1 | $ 1,200.00 | USD | 0.886 | € 1,063.17 |
| RM-RF | Minicircuits | SCP-4-1+ | Power Splitter | | | https:// | 1 | 1 | $ 24.95 | USD | 0.886 | € 22.11 |
| RM-RF | Minicircuits | JTOS-75P+ | VCO | | | https:// | 1 | 1 | $ 14.95 | USD | 0.886 | € 13.25 |
| RM-RF | Minicircuits | HELA-10D+ | Amplifier | | | https:// | 1 | 1 | $ 19.95 | USD | 0.886 | € 17.68 |
| RM-Enclosure | Manchester | | Bespoke aluminium box and lid | --- | Y | http://v | 1 | 1 | $ 40.00 | GBP | 1.291 | € 51.62 |
| RM-Enclosure | RS | 885-9960 | SMA feedthrough | --- | Y | http://au | 1 | 1 | $ 19.93 | AUD | 0.685 | € 13.65 |
| RM-Enclosure | oeMarket | | FC/APC bulkhead adaptor | --- | Y | http://v | 3 | 1 | $ 2.19 | USD | 0.886 | € 5.82 |
| RM-Enclosure | Element 14 | 1186435 | Tusonix 4400-093, EMC filter, Type C, Style 1, | --- | Y | http://ww | 3 | 1 | $ 14.95 | AUD | 0.685 | € 30.72 |

**Figure 59 – SKA1-low Receiver Module (141-022300) capex equipment detailed cost model database.**

### 3.10.3 Capital Expenditure Labour Costs

The summary of SKA1-low capital expenditure labour cost as determined by the Detailed Cost Model (Appendix 7.9.4) is given in Figure 60.

| | Per unit | | | | | | For all units | | | | |
|---|---|---|---|---|---|---|---|---|---|---|---|
| Item | Labour time at grade 40 (hours) | Labour time at grade 42 (hours) | Labour time at grade 44 (hours) | Total labour time (hours) | Labour base salary cost, excluding overhead (EUR) | Quantity of item (including 10% spares) | Labour time at grade 40 (hours) | Labour time at grade 42 (hours) | Labour time at grade 44 (hours) | Total labour time (hours) | Labour base salary cost, excluding overhead (EUR) |
| Rack Cabinet (141-022900) | 12 | 24 | 1.75 | 37.75 | € 1,313.03 | 2 | 24 | 48 | 3.5 | 75.5 | € 2,626.07 |
| Optical Source(141-022400) | 2 | 9.41667 | 2.25 | 13.6667 | € 490.50 | 2 | 4 | 18.8333 | 4.5 | 27.333 | € 980.99 |
| Signal Generator (141-023100) | 0.5 | 1.86667 | 4.25 | 6.61667 | € 308.42 | 2 | 1 | 3.73333 | 8.5 | 13.233 | € 616.83 |
| Sub Rack (141-022700) | 1 | 16.5333 | 1.25 | 18.7833 | € 612.47 | 4 | 4 | 66.1333 | 5 | 75.133 | € 2,449.88 |
| Transmitter Module (141-022100) | 1 | 5.68333 | 2.75 | 9.43333 | € 363.39 | 41 | 41 | 233.017 | 112.75 | 386.77 | € 14,898.93 |
| Receiver Module (141-022300) | 1 | 11.7333 | 2.75 | 15.4833 | € 547.79 | 41 | 41 | 481.067 | 112.75 | 634.82 | € 22,459.36 |
| | | | | | | **TOTALS** | **115** | **850.783** | **247** | **1212.8** | **€ 44,032.06** |
| | | | | | | | | | | | **(excluding overheads)** |

**Figure 60 – SKA1-low summary of the capex labour detailed cost model database.**

These values were independently determined by the consultancy firm Light Touch Solutions, and their Capex Labour Costing Analysis report is attached as Appendix 7.9.5.

The ancillary parameters used in the SKA1-low capital expenditure labour cost Detailed Cost Model database are given in Figure 61.

| | | Yearly salary (AUD) | Hourly rate (AUD) | Hourly rate (EUR) |
|---|---|---|---|---|
| **Technician, Grade =** | 40 | $ 80,100 | $ 44.50 | € 30.48 |
| **Assistant engineer, Grade =** | 42 | $ 106,500 | $ 59.17 | € 40.53 |
| **Senior engineer, Grade =** | 44 | $ 143,000 | $ 79.44 | € 54.41 |
| | | | | |
| **Time per splice** | 0.25 | | | |
| **Time per fibre secure** | 0.166666667 | | | |
| **Time per RF connection** | 0.083333333 | | | |
| **Time per optical connection** | 0.033333333 | | | |
| **Time per RF connection** | 0.033333333 | | | |
| **Time per ethernet connection** | 0.033333333 | | | |
| **Time per power connection** | 0.033333333 | | | |
| **Time per cable secure** | 0.083333333 | | | |

**Figure 61 – SKA1-low ancillary parameters used in the capex labour detailed cost model database.**

The colour key used in the SKA1-low capital expenditure labour cost Detailed Cost Model database is presented in Figure 62.

| |
|---|
| * Indicated building of receiver/transmitter/Microwave shift/Rack distribution |
| Requires Check |
| Requires input |
| Do not change |

**Figure 62 – SKA1-low key to shading colours used in the capex labour detailed cost model database.**

### 3.10.3.1 Rack Cabinet (141-022900)

The detailed cost model database for Rack Cabinet (141-022900) is displayed in Figure 63.

| Activity | Quantity | Time (hours) | Grade | Hourly rate for grade (EUR) | Cost (EUR) |
|---|---|---|---|---|---|
| **At sub-contractor's site:** | | | | | |
| Receiving at sub-contractor site | | 2 | 40 | $ 40.53 | $ 81.05 |
| * Preparation for building of RM/TM/MS/RD | | 0 | 42 | $ 30.48 | $ - |
| * Fibre splices | 0 | 0 | 42 | $ 30.48 | $ - |
| * Securing fibre component and fibres | 0 | 0 | 42 | $ 30.48 | $ - |
| * RF or microwave connections | 0 | 0 | 42 | $ 30.48 | $ - |
| * Testing | | 0 | 44 | $ 54.41 | $ - |
| Assembly of rack-mounting | | 3 | 42 | $ 30.48 | $ 91.44 |
| Mounting onto rack | | 1.5 | 42 | $ 30.48 | $ 45.72 |
| Optical connections | 0 | 0 | 42 | $ 30.48 | $ - |
| RF or microwave connections | 0 | 0 | 42 | $ 30.48 | $ - |
| Ethernet connections | 0 | 0 | 42 | $ 30.48 | $ - |
| Power connections | 0 | 0 | 42 | $ 30.48 | $ - |
| Fastening of cables | 0 | 0 | 42 | $ 30.48 | $ - |
| Testing/verification including documentation | | 1.5 | 44 | $ 54.41 | $ 81.62 |
| Outgoing inspection | | 0.25 | 44 | $ 54.41 | $ 13.60 |
| packaging for transport | | 5 | 40 | $ 40.53 | $ 202.63 |
| Transport logistics | | 5 | 40 | $ 40.53 | $ 202.63 |
| **At in-situ site:** | | | | | |
| Travel time to in-situ site | | 7.5 | 42 | $ 30.48 | $ 228.60 |
| Receiving at in-situ site | | 2 | 42 | $ 30.48 | $ 60.96 |
| installation at in-situ site | | 2 | 42 | $ 30.48 | $ 60.96 |
| Testing/verification | | 0.5 | 42 | $ 30.48 | $ 15.24 |
| Travel time return from in-situ site | | 7.5 | 42 | $ 30.48 | $ 228.60 |
| | | | | **Labour base salary cost per module (EUR):** | **$ 1,313.03** |

| | Grade | Hourly rate (EUR) |
|---|---|---|
| **Technician, Grade =** | 40 | $ 30.48 |
| **Assistant engineer, Grade =** | 42 | $ 40.53 |
| **Senior engineer, Grade =** | 44 | $ 54.41 |

| | |
|---|---|
| **Time per splice** | 0.25 |
| **Time per fibre secure** | 0.166667 |
| **Time per RF connection** | 0.083333 |
| **Time per optical connection** | 0.033333 |
| **Time per RF connection** | 0.033333 |
| **Time per ethernet connection** | 0.033333 |
| **Time per power connection** | 0.033333 |
| **Time per cable secure** | 0.083333 |
| **Total Labour time at grade 40:** | **12** |
| **Total Labour time at grade 42:** | **24** |
| **Total Labour time at grade 44:** | **1.75** |
| **Total labour time at all grades:** | **37.75** |

**(Excluding overheads)**

**Figure 63 – SKA1-low Rack Cabinet (141-022900) capex labour detailed cost model database.**

### 3.10.3.2 Optical Source (141-022400)

The detailed cost model database for Optical Source (141-022400) is displayed in Figure 64.

| Activity | Quantity | Time (hours) | Grade | Hourly rate for grade (EUR) | Cost (EUR) |
|---|---|---|---|---|---|
| **At sub-contractor's site:** | | | | | |
| Receiving at sub-contractor site | | 2 | 40 | $ 40.53 | $ 81.05 |
| * Preparation for building of RM/TM/MS/RD | | 0 | 42 | $ 30.48 | $ - |
| * Fibre splices | 0 | 0 | 42 | $ 30.48 | $ - |
| * Securing fibre component and fibres | 0 | 0 | 42 | $ 30.48 | $ - |
| * RF or microwave connections | 0 | 0 | 42 | $ 30.48 | $ - |
| * Testing | | 0 | 44 | $ 54.41 | $ - |
| Assembly of rack-mounting | | 7.5 | 42 | $ 30.48 | $ 228.60 |
| Mounting onto rack | | 1 | 42 | $ 30.48 | $ 30.48 |
| Optical connections | 5 | 0.1666667 | 42 | $ 30.48 | $ 5.08 |
| RF or microwave connections | 0 | 0 | 42 | $ 30.48 | $ - |
| Ethernet connections | 1 | 0.0333333 | 42 | $ 30.48 | $ 1.02 |
| Power connections | 1 | 0.0333333 | 42 | $ 30.48 | $ 1.02 |
| Fastening of cables | 7 | 0.5833333 | 42 | $ 30.48 | $ 17.78 |
| Testing/verification including documentation | | 2 | 44 | $ 54.41 | $ 108.83 |
| Outgoing inspection | | 0.25 | 44 | $ 54.41 | $ 13.60 |
| packaging for transport | | 0 | 40 | $ 40.53 | $ - |
| Transport logistics | | 0 | 40 | $ 40.53 | $ - |
| **At in-situ site:** | | | | | |
| Travel time to in-situ site | | 0 | 42 | $ 30.48 | $ - |
| Receiving at in-situ site | | 0 | 42 | $ 30.48 | $ - |
| installation at in-situ site | | 0 | 42 | $ 30.48 | $ - |
| Testing/verification | | 0.1 | 42 | $ 30.48 | $ 3.05 |
| Travel time return from in-situ site | | 0 | 42 | $ 30.48 | $ - |
| | | | | **Labour base salary cost per module (EUR):** | **$ 490.50** |

| | Grade | Hourly rate (EUR) |
|---|---|---|
| **Technician, Grade =** | 40 | $ 30.48 |
| **Assistant engineer, Grade =** | 42 | $ 40.53 |
| **Senior engineer, Grade =** | 44 | $ 54.41 |

| | |
|---|---|
| **Time per splice** | 0.25 |
| **Time per fibre secure** | 0.166667 |
| **Time per RF connection** | 0.083333 |
| **Time per optical connection** | 0.033333 |
| **Time per RF connection** | 0.033333 |
| **Time per ethernet connection** | 0.033333 |
| **Time per power connection** | 0.033333 |
| **Time per cable secure** | 0.083333 |
| **Total Labour time at grade 40:** | **2** |
| **Total Labour time at grade 42:** | **9.416667** |
| **Total Labour time at grade 44:** | **2.25** |
| **Total labour time at all grades:** | **13.66667** |

**(Excluding overheads)**

**Figure 64 – SKA1-low Optical Source (141-022400) capex labour detailed cost model database.**

### 3.10.3.3 Signal Generator (141-023100)

The detailed cost model database for Signal Generator (141-023100) is displayed in Figure 65.

| Activity | Quantity | Time (hours) | Grade | Hourly rate for grade (EUR) | Cost (EUR) |
|---|---|---|---|---|---|
| **At sub-contractor's site:** | | | | | |
| Receiving at sub-contractor site | | 0.5 | 40 | $ 40.53 | $ 20.26 |
| * Preparation for building of RM/TM/MS/RD | | 0 | 42 | $ 30.48 | $ - |
| * Fibre splices | 0 | 0 | 42 | $ 30.48 | $ - |
| * Securing fibre component and fibres | 0 | 0 | 42 | $ 30.48 | $ - |
| * RF or microwave connections | 0 | 0 | 42 | $ 30.48 | $ - |
| * Testing | | 0 | 44 | $ 54.41 | $ - |
| Assembly of rack-mounting | | 0 | 42 | $ 30.48 | $ - |
| Mounting onto rack | | 1 | 42 | $ 30.48 | $ 30.48 |
| Optical connections | 0 | 0 | 42 | $ 30.48 | $ - |
| RF or microwave connections | 6 | 0.2 | 42 | $ 30.48 | $ 6.10 |
| Ethernet connections | 1 | 0.0333333 | 42 | $ 30.48 | $ 1.02 |
| Power connections | 1 | 0.0333333 | 42 | $ 30.48 | $ 1.02 |
| Fastening of cables | 6 | 0.5 | 42 | $ 30.48 | $ 15.24 |
| Testing/verification including documentation | | 4 | 44 | $ 54.41 | $ 217.66 |
| Outgoing inspection | | 0.25 | 44 | $ 54.41 | $ 13.60 |
| packaging for transport | | 0 | 40 | $ 40.53 | $ - |
| Transport logistics | | 0 | 40 | $ 40.53 | $ - |
| **At in-situ site:** | | | | | |
| Travel time to in-situ site | | 0 | 42 | $ 30.48 | $ - |
| Receiving at in-situ site | | 0 | 42 | $ 30.48 | $ - |
| installation at in-situ site | | 0 | 42 | $ 30.48 | $ - |
| Testing/verification | | 0.1 | 42 | $ 30.48 | $ 3.05 |
| Travel time return from in-situ site | | 0 | 42 | $ 30.48 | $ - |
| | | | | **Labour base salary cost per module (EUR):** | **$ 308.42** |

| | Grade | Hourly rate (EUR) |
|---|---|---|
| **Technician, Grade =** | 40 | $ 30.48 |
| **Assistant engineer, Grade =** | 42 | $ 40.53 |
| **Senior engineer, Grade =** | 44 | $ 54.41 |

| | |
|---|---|
| **Time per splice** | 0.25 |
| **Time per fibre secure** | 0.166667 |
| **Time per RF connection** | 0.083333 |
| **Time per optical connection** | 0.033333 |
| **Time per RF connection** | 0.033333 |
| **Time per ethernet connection** | 0.033333 |
| **Time per power connection** | 0.033333 |
| **Time per cable secure** | 0.083333 |
| **Total Labour time at grade 40:** | **0.5** |
| **Total Labour time at grade 42:** | **1.866667** |
| **Total Labour time at grade 44:** | **4.25** |
| **Total labour time at all grades:** | **6.616667** |

**(Excluding overheads)**

**Figure 65 – SKA1-low Signal Generator (141-023100) capex labour detailed cost model database.**

### 3.10.3.4 Sub Rack (141-022700)

The detailed cost model database for Sub Rack (141-022700) is displayed in Figure 66.

| Activity | Quantity | Time (hours) | Grade | Hourly rate for grade (EUR) | Cost (EUR) |
|---|---|---|---|---|---|
| **At sub-contractor's site:** | | | | | |
| Receiving at sub-contractor site | | 1 | 40 | $ 40.53 | $ 40.53 |
| * Preparation for building of RM/TM/MS/RD | | 0 | 42 | $ 30.48 | $ - |
| * Fibre splices | 0 | 0 | 42 | $ 30.48 | $ - |
| * Securing fibre component and fibres | 0 | 0 | 42 | $ 30.48 | $ - |
| * RF or microwave connections | 0 | 0 | 42 | $ 30.48 | $ - |
| * Testing | | 0 | 44 | $ 54.41 | $ - |
| Assembly of rack-mounting | | 7.5 | 42 | $ 30.48 | $ 228.60 |
| Mounting onto rack | | 1 | 42 | $ 30.48 | $ 30.48 |
| Optical connections | 17 | 0.5666667 | 42 | $ 30.48 | $ 17.27 |
| RF or microwave connections | 17 | 0.5666667 | 42 | $ 30.48 | $ 17.27 |
| Ethernet connections | 17 | 0.5666667 | 42 | $ 30.48 | $ 17.27 |
| Power connections | 17 | 0.5666667 | 42 | $ 30.48 | $ 17.27 |
| Fastening of cables | 68 | 5.6666667 | 42 | $ 30.48 | $ 172.72 |
| Testing/verification including documentation | | 1 | 44 | $ 54.41 | $ 54.41 |
| Outgoing inspection | | 0.25 | 44 | $ 54.41 | $ 13.60 |
| packaging for transport | | 0 | 40 | $ 40.53 | $ - |
| Transport logistics | | 0 | 40 | $ 40.53 | $ - |
| **At in-situ site:** | | | | | |
| Travel time to in-situ site | | 0 | 42 | $ 30.48 | $ - |
| Receiving at in-situ site | | 0 | 42 | $ 30.48 | $ - |
| installation at in-situ site | | 0 | 42 | $ 30.48 | $ - |
| Testing/verification | | 0.1 | 42 | $ 30.48 | $ 3.05 |
| Travel time return from in-situ site | | 0 | 42 | $ 30.48 | $ - |
| | | | | **Labour base salary cost per module (EUR):** | **$ 612.47** |

| | Grade | Hourly rate (EUR) |
|---|---|---|
| **Technician, Grade =** | 40 | $ 30.48 |
| **Assistant engineer, Grade =** | 42 | $ 40.53 |
| **Senior engineer, Grade =** | 44 | $ 54.41 |

| | |
|---|---|
| **Time per splice** | 0.25 |
| **Time per fibre secure** | 0.166667 |
| **Time per RF connection** | 0.083333 |
| **Time per optical connection** | 0.033333 |
| **Time per RF connection** | 0.033333 |
| **Time per ethernet connection** | 0.033333 |
| **Time per power connection** | 0.033333 |
| **Time per cable secure** | 0.083333 |
| **Total Labour time at grade 40:** | **1** |
| **Total Labour time at grade 42:** | **16.53333** |
| **Total Labour time at grade 44:** | **1.25** |
| **Total labour time at all grades:** | **18.78333** |

**(Excluding overheads)**

**Figure 66 – SKA1-low Sub Rack (141-022700) capex labour detailed cost model database.**

### 3.10.3.5 Transmitter Module (141-022100)

The detailed cost model database for Transmitter Module (141-022100) is displayed in Figure 67.

| Activity | Quantity | Time (hours) | Grade | Hourly rate for grade (EUR) | Cost (EUR) |
|---|---|---|---|---|---|
| **At sub-contractor's site:** | | | | | |
| Receiving at sub-contractor site | | 1 | 40 | $ 40.53 | $ 40.53 |
| * Preparation for building of RM/TM/MS/RD | | 2 | 42 | $ 30.48 | $ 60.96 |
| * Fibre splices | 3 | 0.75 | 42 | $ 30.48 | $ 22.86 |
| * Securing fibre component and fibres | 4 | 0.6666667 | 42 | $ 30.48 | $ 20.32 |
| * RF or microwave connections | 7 | 0.5833333 | 42 | $ 30.48 | $ 17.78 |
| * Testing | | 2 | 44 | $ 54.41 | $ 108.83 |
| Assembly of rack-mounting | | 0.5 | 42 | $ 30.48 | $ 15.24 |
| Mounting onto rack | | 0.5 | 42 | $ 30.48 | $ 15.24 |
| Optical connections | 1 | 0.0333333 | 42 | $ 30.48 | $ 1.02 |
| RF or microwave connections | 2 | 0.0666667 | 42 | $ 30.48 | $ 2.03 |
| Ethernet connections | 1 | 0.0333333 | 42 | $ 30.48 | $ 1.02 |
| Power connections | 1 | 0.0333333 | 42 | $ 30.48 | $ 1.02 |
| Fastening of cables | 5 | 0.4166667 | 42 | $ 30.48 | $ 12.70 |
| Testing/verification including documentation | | 0.5 | 44 | $ 54.41 | $ 27.21 |
| Outgoing inspection | | 0.25 | 44 | $ 54.41 | $ 13.60 |
| packaging for transport | | 0 | 40 | $ 40.53 | $ - |
| Transport logistics | | 0 | 40 | $ 40.53 | $ - |
| **At in-situ site:** | | | | | |
| Travel time to in-situ site | | 0 | 42 | $ 30.48 | $ - |
| Receiving at in-situ site | | 0 | 42 | $ 30.48 | $ - |
| installation at in-situ site | | 0 | 42 | $ 30.48 | $ - |
| Testing/verification | | 0.1 | 42 | $ 30.48 | $ 3.05 |
| Travel time return from in-situ site | | 0 | 42 | $ 30.48 | $ - |
| | | | | **Labour base salary cost per module (EUR):** | **$ 363.39** |

| | Grade | Hourly rate (EUR) |
|---|---|---|
| **Technician, Grade =** | 40 | $ 30.48 |
| **Assistant engineer, Grade =** | 42 | $ 40.53 |
| **Senior engineer, Grade =** | 44 | $ 54.41 |

| | |
|---|---|
| **Time per splice** | 0.25 |
| **Time per fibre secure** | 0.166667 |
| **Time per RF connection** | 0.083333 |
| **Time per optical connection** | 0.033333 |
| **Time per RF connection** | 0.033333 |
| **Time per ethernet connection** | 0.033333 |
| **Time per power connection** | 0.033333 |
| **Time per cable secure** | 0.083333 |
| **Total Labour time at grade 40:** | **1** |
| **Total Labour time at grade 42:** | **5.683333** |
| **Total Labour time at grade 44:** | **2.75** |
| **Total labour time at all grades:** | **9.433333** |

**(Excluding overheads)**

**Figure 67 – SKA1-low Transmitter Module (141-022100) capex labour detailed cost model database.**

### 3.10.3.6 Receiver Module (141-022300)

The detailed cost model database for Receiver Module (141-022300) is displayed in Figure 68.

| Activity | Quantity | Time (hours) | Grade | Hourly rate for grade (EUR) | Cost (EUR) |
|---|---|---|---|---|---|
| **At sub-contractor's site:** | | | | | |
| Receiving at sub-contractor site | | 1 | 40 | $ 40.53 | $ 40.53 |
| * Preparation for building of RM/TM/MS/RD | | 2 | 42 | $ 30.48 | $ 60.96 |
| * Fibre splices | 2 | 0.5 | 42 | $ 30.48 | $ 15.24 |
| * Securing fibre component and fibres | 4 | 0.6666667 | 42 | $ 30.48 | $ 20.32 |
| * RF or microwave connections | 4 | 0.3333333 | 42 | $ 30.48 | $ 10.16 |
| * Testing | | 2 | 44 | $ 54.41 | $ 108.83 |
| Assembly of rack-mounting | | 0.5 | 42 | $ 30.48 | $ 15.24 |
| Mounting onto rack | | 0.5 | 42 | $ 30.48 | $ 15.24 |
| Optical connections | 1 | 0.0333333 | 42 | $ 30.48 | $ 1.02 |
| RF or microwave connections | 1 | 0.0333333 | 42 | $ 30.48 | $ 1.02 |
| Ethernet connections | 0 | 0 | 42 | $ 30.48 | $ - |
| Power connections (inc soldering) | 5 | 0.1666667 | 42 | $ 30.48 | $ 5.08 |
| Fastening of cables | 3 | 0.25 | 42 | $ 30.48 | $ 7.62 |
| Testing/verification including documentation | | 0.5 | 44 | $ 54.41 | $ 27.21 |
| Outgoing inspection | | 0.25 | 44 | $ 54.41 | $ 13.60 |
| packaging for transport | | 0 | 40 | $ 40.53 | $ - |
| Transport logistics | | 0 | 40 | $ 40.53 | $ - |
| **At in-situ site:** | | | | | |
| Travel time to in-situ site | | 2.5 | 42 | $ 30.48 | $ 76.20 |
| Receiving at in-situ site | | 0.25 | 42 | $ 30.48 | $ 7.62 |
| installation at in-situ site | | 1 | 42 | $ 30.48 | $ 30.48 |
| Testing/verification | | 0.5 | 42 | $ 30.48 | $ 15.24 |
| Travel time return from in-situ site | | 2.5 | 42 | $ 30.48 | $ 76.20 |
| | | | | **Labour base salary cost per module (EUR):** | **$ 547.79** |

| | Grade | Hourly rate (EUR) |
|---|---|---|
| **Technician, Grade =** | 40 | $ 30.48 |
| **Assistant engineer, Grade =** | 42 | $ 40.53 |
| **Senior engineer, Grade =** | 44 | $ 54.41 |

| | |
|---|---|
| **Time per splice** | 0.25 |
| **Time per fibre secure** | 0.166667 |
| **Time per RF connection** | 0.083333 |
| **Time per optical connection** | 0.033333 |
| **Time per RF connection** | 0.033333 |
| **Time per ethernet connection** | 0.033333 |
| **Time per power connection** | 0.033333 |
| **Time per cable secure** | 0.083333 |
| **Total Labour time at grade 40:** | **1** |
| **Total Labour time at grade 42:** | **11.73333** |
| **Total Labour time at grade 44:** | **2.75** |
| **Total labour time at all grades:** | **15.48333** |

**(Excluding overheads)**

**Figure 68 – SKA1-low Receiver Module (141-022300) capex labour detailed cost model database.**

# 4 EVALUATION

## 4.1 Evaluation

Between February 2014 and July 2017, UWA's *SKA phase synchronisation system* was evaluated using a total of 17 test campaigns, as defined in [RD47], and outlined in Section 4.1.1. These tests successfully managed to verify by demonstration the functionally of the system against all SKA requirements, as well as many other more rigorous practical considerations.

In addition, UWA researchers in partnership with MeerKAT and UoM electronic engineers, progressed the detailed designs described in this document, into a set of mass-manufacture archetypes, effectively getting a head-start at addressing manufacturing issues that may be encounter by contractors during SKA Construction. This includes construction-ready sets of optical, electronic, and mechanical design files, and a bill-of materials based on components purchased from vendors to construct these archetypes – and this forms the basis of the UWA *Parametric Cost Model* (Section 3.10). The first set of mass-manufacture archetypes for SKA1-low were completed in Q2, 2016 [RD21]; and for SKA1-mid in Q1, 2017 [RD22]. The assembly and testing of the completed archetypes was done at UWA. To get an independent evaluation of labour costs associated with assembly and testing, a local optical manufacturing technology consultancy company was employed to provide an independent review of the costings (see Appendix 7.9.5).

The report on the construction of the *SKA phase synchronisation system* mass manufacture archetype [RD22] is summarised in Section 4.1.2. While this report focuses on SKA1-mid, it was actually the SKA1-low archetype that was first developed and completed, so much of the outcomes of this report is applicable to both mass manufacture archetypes.

### 4.1.1 Test and Verification by Demonstration

The following 11 tests are relevant to SKA1-low:

Test(1) – SKA STFR.FRQ (UWA) v1 – Preliminary Laboratory Test

Test(2) – SKA-LOW STFR.FRQ (UWA) v1 – Laboratory Demonstration

Test(4) – SKA-LOW STFR.FRQ (UWA) v1 – Astronomical Verification

Test(5) – SKA STFR.FRQ (UWA) v2 – Transmitter Module Stabilisation

Test(6) – SKA-LOW STFR.FRQ (UWA) v2 – Astronomical Verification

Test(7) – SKA1 Overhead Optical Fibre Characterisation Field Trial

Test(8) – SKA1 STFR.FRQ (UWA) v2 – Long-Haul Overhead Fibre Verification

Test(10) – SKA-LOW STFR.FRQ (UWA) v2 – Laboratory Demonstration

Test(12) – SKA-LOW STFR.FRQ (UWA) v2 – Temperature and Humidity

Test(14) – SKA-LOW STFR.FRQ (UWA) v2 – Seismic Resilience

Test(16) – SKA-LOW STFR.FRQ (UWA) v2 – Electromagnetic Compatibility

#### 4.1.1.1 Test(1) – SKA STFR.FRQ (UWA) v1 – Preliminary Laboratory Test

- SADT technical report: *Pre-PDR Laboratory Verification of UWA's SKA Synchronisation System* [RD12].

This document reports on the laboratory verification of the dual-servo version of UWA's SKA Frequency Transfer System STFR.FRQ (UWA) for stabilising the radio frequency (RF) or microwave (MW) frequency separation between two optical signals. The dual-servo version is indicated by "v1" to distinguish it from the latter version that only uses a single servo; that is, the single-servo version, designated "v2".

The transition from "v1" to "v2" occurred following our first SKA-low 'Astronomical Verification' field trail with the Australian SKA Pathfinder (ASKAP) in November 2014. This period was after the of the preliminary design review (PDR) documentation submission, so the PDR 'SADT.SAT.STFR Design Report' contains information about the v1 system only. By the PDR assessment panel meeting, the transition to v2 was complete. The full explanation of the transition including the motivation and process are described in the UWA 'SKA-low Astronomical Verification' report.

Therefore, the technical details of the system described in this report are strictly not applicable to the current design ("v2") that UWA adopted following PDR. However, the changes in technical details between "v1" and "v2" are minor enough that all of the Key System Characteristics described in the next section are unchanged.

These tests described in this section of the report were performed in the UWA SADT laboratory over 0 km (short patch leads) and a 1 km spool of fibre. A 70 MHz RF signal was transferred using an acousto-optic modulator (AOM) to produce the frequency separation between two optical frequencies.

The results of this series of tests confirmed that the dual-servo system worked as planned and gave us confidence that its performance would also be sufficient to meet the SKA performance requirements at the higher MW frequencies. The results were presented at the SADT face-to-face meeting in June 2014 and at SAT Kick-Off Meeting in Beijing.

The aim of this series of tests was to confirm that phase-stabilising two optical frequencies independently would result in effective phase-stabilisation of the frequency difference between the two optical signals. These tests also provided preliminary data sufficient to confirm that, in principle, the performance of this frequency transfer technique is sufficient to meet SKA performance requirements.

Key STFR.FRQ (UWA) System Characteristics specifically investigated as part of this test:

- Transfer stability
- Transfer accuracy

#### 4.1.1.2 Test(2) – SKA-LOW STFR.FRQ (UWA) v1 – Laboratory Demonstration

- SADT technical report: *Pre-PDR Laboratory Verification of UWA's SKA Synchronisation System* [RD12].

Having proved that the frequency transfer system presented in Section 3 of this report effectively stabilised a transmitted RF signal, the experiment was modified to be more representative of the systems we proposed for use in the SKA.

UWA's SKA Frequency Transfer System comes in two variants: The SKA-LOW STFR.FRQ (UWA) design is optimised for transmission of RF frequencies for SKA-low, where currently the highest required frequency is 800 MHz; used to clock the SKA-low analogue-to-digital converters. The SKA-MID STFR.FRQ (UWA) design is optimised for transmission of MW frequencies for SKA-mid. SKA-mid at the time of measurement required both 4 GHz and 32 GHz for local oscillators for separate receiver bands (with 3.2 GHz and 5 GHz also potential required in the future if other receiver bands are funded). This section describes the construction and testing of the SKA-low system in the laboratory at UWA.

Frequencies of 16 MHz and 20 MHz were transmitted using the same components to demonstrate the system's ability to transmit different frequencies as desired. The system was tested at 20 MHz because this was the frequency produced by the interferometer when the AOMs were operated at their nominal frequency. 16 MHz was tested to provide information about the performance of the system for a subsequent set of tests at the ASKAP/BETA telescope, which uses a 16 MHz photonic reference distributed to each antenna.

The tests conducted with this Laboratory Prototype at 16 MHz and 20 MHz were presented at the SADT face-to-face meeting in June 2014. The performance of the system was shown to be better than the SKA coherence requirements by more than two orders-of-magnitude when transmitted over 31 km of active fibre. A number of key system characteristics were also demonstrated.

The tests outlined here took place between 2nd and 16th May, 2014.

The aim of this series of tests was to demonstrate the performance of the SKA-LOW STFR.FRQ (UWA) at 16 MHz (for later tests at ASKAP/BETA) as well as to experimentally verify several of the key system characteristics presented at the SAT Kick-Off Meeting. Since SKA-LOW STFR.FRQ (UWA) is less complex than the SKA-mid MW transfer system, these tests were also conducted with the aim of establishing a best-case frequency transfer performance of RF frequencies in the laboratory.

Key STFR.FRQ (UWA) System Characteristics specifically investigated as part of this test:

- Transfer stability
- Transfer accuracy
- Fibre bandwidth
- Immunity from reflections
- Polarisation alignment
- Optical amplification
- Frequency offset
- Optical source
- Fibre topology
- Signal fading

#### 4.1.1.3 Test(4) – SKA-LOW STFR.FRQ (UWA) v1 – Astronomical Verification

- SADT technical report: *SKA-low Astronomical Verification* [RD19].

The motivation for this work was the Astronomical Verification of the SKA-low variant of UWA's SKA frequency transfer system. ASKAP was specifically chosen for these tests (over low-frequency arrays such as the MWA), as its higher operating frequency would require a much larger frequency multiplication – from 16 MHz or 10 MHz, up to 4,432 MHz. This allowed us to effectively multiply the intrinsic instability of our transfer system up above the instability of the atmosphere, thereby enabling us to make a direct astronomical measurement of the transfer stability of our system using a radio telescope array. The ultimate aim was to show that this independent astronomical measurement produced the same results as the various metrology techniques we previously used in the laboratory.

The Field Deployable Prototype dual-servo RF transfer system is a ruggedized, portable version of the 16 MHz RF transfer SKA-low Laboratory Prototype described in the UWA 'Pre-PDR Laboratory Verification' report.

This unit was designed to produce a stabilised 16 MHz photonic signal at the remote site in order to be compatible with the ASKAP/BETA antennas at the MRO. The construction and testing of this unit was completed between 23rd September and 30th October and was deployed at the MRO between 1st and 6th November 2014.

The 16 MHz transfer stability was tested over 1 km, 2 km, 5 km, 25 km and 30 km of fibre spool in the laboratory; a 31 km active urban fibre link; and over 2 km and 4 km transmission distances to ASKAP/BETA antennas at the MRO.

During the field trials at the MRO some technical issues were encountered. The Mach-Zehnder interferometer setup used by the UWA system to produce the RF signal from two independently stabilised optical signals was more sensitive to vibrations than expected. The fan cooling systems in the MRO correlator room produced strong acoustic-frequency vibrations on all working surfaces and this affected the operation of the Transmitter Module.

The aim of this series of tests was to construct a Field Deployable Prototype of UWA's SKA Frequency Transfer System that is sufficiently robust to be used for field trials, and for this unit to be compatible with the ASKAP/BETA systems for the purpose of determining engineering compatibility between UWA's SKA Frequency Transfer System and likely SKA systems.

The objective was to replace one of the standard 16 MHz photonic references delivered to all of the ASKAP/BETA antennas with a stabilised 16 MHz reference provided by the UWA Field Deployable Prototype, and to compare the astronomical observations taken by the ASKAP/BETA telescope when using the stabilised 16 MHz signal with similar observations performed using the standard system.

Key STFR.FRQ.UWA System Characteristics specifically investigated as part of this test:

- Transfer stability
- Transfer accuracy
- Optical interface
- Polarisation alignment
- Servo robustness
- Remote site volume
- Environment at remote site
- Environment at local site
- Lifetime before failure

#### 4.1.1.4 Test(5) – SKA STFR.FRQ (UWA) v2 – Transmitter Module Stabilisation

- SADT technical report: *SKA-low Astronomical Verification* [RD19].

A key issue encountered when initially testing the dual-servo version (v1) of UWA's SKA Frequency Transfer System was that acoustic vibrations in the Transmitter Module were significantly degrading the stability of the transmitted 16 MHz signal. This was attributed to the fact that:

a) Approximately 0.5 m of fibre in the beginning of each arm of the Mach-Zehnder interferometer was outside the servo loop, and therefore any relative thermal and/or force gradients between the Mach-Zehnder interferometer arms led to frequency fluctuations which could not be compensated for by the servos

b) The output frequency of the Mach-Zehnder interferometer was especially sensitive to thermal / acoustic fluctuations in the Transmitter Module since each arm was spatially separated.

The sensitivity of this Mach-Zehnder interferometer was characterised using a vibration rig at acoustic frequencies, and analysing the spectrum of an error signal representing the relative phase fluctuations in each arm of the interferometer.

The aims of this test are to eliminate the Transmitter Module's sensitivity to environmental conditions at the Local Site.

Key STFR.FRQ.UWA System Characteristics specifically investigated as part of this test:

- Environment at local site

#### 4.1.1.5 Test(6) – SKA-LOW STFR.FRQ (UWA) v2 – Astronomical Verification

- SADT technical report: *SKA-low Astronomical Verification* [RD19].

Based on the results of the first MRO field campaign, UWA's SKA-low frequency transfer system was redesigned to stabilise the Mach-Zehnder interferometer as well as the optical link to the remote site. The system achieves RF transfer with direct RF sensing, but with optical phase stabilisation utilising one active AOM.

A single servo now actively controls optical-frequency-scale length fluctuations originating in the Mach-Zehnder interferometer in the Transmitter Module, as well as RR-scale length fluctuations in the optical fibre link as sensed by the Michelson interferometer.

This version of SKA-low was designed to transmit a photonic 10 MHz signal which, when demodulated with a photodetector at the remote end, could be used as the 10 MHz reference input to a commercial frequency synthesiser. The synthesiser selected for this task had a cross-over frequency of its internal clean-up oscillator that was configurable between 25 Hz and 650 Hz.

The 10 MHz transfer was tested over 1 km and 5 km fibre spools in the lab, and over a 1 km link to an antenna on the ASKAP/BETA network. This system was deployed to the MRO during a second ASKAP field campaign between 26th January and 5th February 2015.

The aim of this series of tests was to ascertain that the issues of susceptibility to external vibrations of the previous SKA-low Transmitter Module design had been solved and that the new system effectively provided a stable reference signal to the ASKAP antennas.

The interface between the transmitted reference signal and the ASKAP RSL was also changed in order to eliminate the technical timing interface issue. This issue was solved by bypassing the ASKAP digitisers that caused the failures during the previous campaign and supplying an electronic signal for the Lower LO directly.

Key STFR.FRQ (UWA) System Characteristics specifically investigated as part of this test:

- Transfer stability
- Transfer accuracy
- Servo robustness
- Lifetime before failure
- Optical interface
- Polarisation alignment
- Remote site volume
- Environment at remote site
- Environment at local site

#### 4.1.1.6 Test(7) – SKA1 Overhead Optical Fibre Characterisation Field Trial

- Journal paper: *Characterization of Optical Frequency Transfer over 154 km of Aerial Fiber* [RD15].
- SADT technical report: *UWA South African SKA Site Long-Haul Overhead Fibre Field Trial Report* [RD14].

The UWA High-Precision Fibre Characterisation System revealed that overhead optical fibre links are significantly more susceptible to environmental impacts, when compared to non-overhead links of comparable length. In the environmental conditions present during our trials, this resulted in frequency fluctuations over 1,000 times greater (at 1 second of integration) than those previously measured using the same system on conventional links (buried, conduit, spool). To put this into perspective, the 153 km South African overhead fibre link exhibits about 500 times more frequency noise than the 1840 km world-record link across Germany [RD48].This additional noise was measured to be related to the severity of the environmental conditions (specifically, wind speed), and so it can be expected that more severe weather conditions will result in frequency fluctuations greater than the 500 times increase measured.

Two optical fibre reticulation options are being considered for the spiral arm component of SKA-mid's Synchronisation and Timing (SAT) system. The default option is a trenched route outside the core region that has to avoid physical obstructions. This results in a route with a total length of 604 km across the three spiral arms, or an average of around 200 km per spiral arm. This is approximately 50% longer than a route that directly follows the layout of the spiral arms determined to be 393 km, or about 130 km per arm. With the additional length of the trenched fibre in the core, the maximum SKA-mid arm length for this configuration is about 150 km.

The total SADT local infrastructure costs of the trenched route based on the default option is currently calculated to be around 11 million euros. However, if the SAT system can be shown to be able to operate on overhead fibre, then all of SADT fibre cable could simply be suspended on the same overhead poles that distribute power along the spiral arms. This route has a total distance of only 394 km across the three spiral arms. Therefore, the only infrastructure cost specific for SAT are the additional (now shorter) SAT fibre cables, estimated to be less than 300 thousand euros. Detailed cost analysis studies for moving all SADT fibre onto the overhead route are ongoing, but it is expected that not having to trench the SKA-mid spiral arms could save the SKA project several million euros.

To quantify the difference between overhead and conventional fibre links in terms of SAT frequency transfer, several members of the SADT consortium deployed the UWA High-Precision Fibre Characterisation System over various overhead fibre links on the South African SKA site between Sunday 31st of May and Wednesday the 3rd of June, 2015. These were compared to frequency transfer data taken in Australia on conventional links of comparable distance as measured during the previous few weeks. The overhead optical fibre characterisation field trials included deployment over a 153 km overhead fibre link which is an ideal analogue for the longest planned SKA-mid fibre links, if an overhead route can be accommodated.

The aims and objectives of this work was to:

- Conduct a highly-precise comparison measurements of overhead fibre links compared to conventional links (buried, conduit, spool) of similar length, in order to ascertain the feasibility of using overhead fibre frequency transfer for the SKA.
- Test the robustness of AOM-based servo actuation systems in conditions that required a factor of 6,000 greater actuation range and speed than the maximum required for SKA frequency transfer (192 THz frequency transfer compared with the 32 GHz transfer required SKA-mid Band 5).
- Confirm that the coherence length of a moderately priced, commercial, off-the-shelf diode laser (Koheras X15) is suitable for SKA long-haul frequency transfer.
- Test component reliability by exposing them to severe physical punishment during air transport to South Africa (objective applied retroactively after we discovered our shock-proof equipment case failed during transport resulting in external damage to our equipment).

#### 4.1.1.7 Test(8) – SKA1 STFR.FRQ (UWA) v2 – Long-Haul Overhead Fibre Verification

- Journal paper: *Stabilized Radio-Frequency Transfer Over 186 km of Aerial Fiber* [RD16].
- SADT technical report: *UWA South African SKA Site Long-Haul Overhead Fibre Field Trial Report* [RD14].

UWA's SKA Frequency Transfer System was demonstrated to be capable of frequency transfer over the longest SKA1 overhead fibre links.

Two optical fibre reticulation options are being considered for the spiral arm component of SKA-mid's Synchronisation and Timing (SAT) system. The default option is a trenched route outside the core region that has to avoid physical obstructions. This results in a route with a total length of 604 km across the three spiral arms, or an overage of about 200 km per spiral arm. This is approximately 50% longer than a route that directly follows the layout of the spiral arms determined to be 393 km, or about 130 km per arm. With the additional length of the trenched fibre in the core, the maximum SKA-mid arm length for this configuration is about 150 km.

The total SADT local infrastructure costs of the trenched route based on the default option is currently calculated to be around 11 million euros. However, if the SAT system can be shown to be able to operate on overhead fibre, then all of SADT fibre cable could simply be suspended on the same overhead poles that distribute power along the spiral arms. This route has a total distance of only 394 km across the three spiral arms. Therefore, the only infrastructure cost specific for SAT are the additional (now shorter) SAT fibre cables, estimated to be less than 300 thousand euros. Detailed cost estimate studies for moving all SADT fibre onto the overhead route are ongoing, but it is expected that not having to trench the SKA-mid spiral arms could save the SKA project several million euros.

The SAT system comprises two independent systems; frequency transfer and absolute time transfer, with the absolute time transfer system being developed by other members of the SADT consortium. Given its much lower susceptibility to optical link fluctuations than the frequency transfer system, we expect the SKA absolute time transfer system to be able to operate nominally on overhead fibre. Nonetheless, the SADT consortium has put forward plans to also verify this system on overhead fibre in the near future.

UWA's SKA Frequency Transfer System was used to transmit a stabilised radio frequency signal over several different lengths of overhead optical fibre at the South African SKA site. An out-of-loop measurement system was used to independently measure the frequency stability of the frequency transfer. The system was shown to operate reliably for continuous periods up to 48 hours, over links up to 186 km, and during average wind speeds up to 56 km/h.

The aims and objectives of this work was to:

- Confirm that frequency signals transferred using UWA's SKA Frequency Transfer System meets the SKA phase stability requirements when deployed on overhead fibre links, over the longest planned SKA fibre links, and during extreme environmental conditions.
- Confirm that the AOM-based active stabilisation servo used in UWA's SKA Frequency Transfer System remains robust while deployed on overhead fibre links, over the longest planned SKA fibre links, and during extreme environmental conditions.
- Confirm that UWA's SKA Frequency Transfer System can utilise optical amplification to extend fibre-based frequency transfer beyond the longest planned SKA fibre links.

#### 4.1.1.8 Test(10) – SKA-LOW STFR.FRQ (UWA) v2 – Laboratory Demonstration

- SADT technical report: *Pre-CDR Laboratory Verification of UWA's SKA Synchronisation System* [RD13].

In this report, we present the experimental methods and frequency transfer performance results of the SKA-MID variant of UWA's *SKA Phase Synchronisation System* (SKA-LOW STFR.FRQ (UWA)). A subset of this information as well as the technical details of the frequency transfer technique are reported in the peer-reviewed publication [RD11].

The aims of the test were to describe the basic design and operation of the SKA-LOW STFR.FRQ (UWA) system and demonstrate its compliance with the applicable SKA engineering requirements. The key SKA-LOW STFR.FRQ (UWA) system characteristics specifically investigated in this report are:

- Frequency transfer stability — demonstrate the frequency transfer performance of the system and that it is compliant with the 1.9% coherence loss requirements at 1 s (REQ-3242) and 60 s (REQ-3243),
- Frequency transfer accuracy — demonstrate the frequency accuracy achieved with the system,
- Phase drift — demonstrate the phase drift of the transferred frequency and that it is compliant with REQ-3244 by being less than 1 rad over 600 s,
- Jitter — demonstrate that the jitter of the transferred signal is compliant with the requirement defined by EICD 100-0000000-026_03-SADTtoLFAA_ICD.

#### 4.1.1.9 Test(12) – SKA-LOW STFR.FRQ (UWA) v2 – Temperature and Humidity

- SADT technical report: *Pre-CDR Laboratory Verification of UWA's SKA Synchronisation System* [RD13].

SKA subsystems are expected to operate at their specified performance levels throughout a range of climatic conditions that they might reasonably be subjected to in their operational environments, including expected temperature and humidity ranges. The tests and results presented below demonstrate that the SKA-LOW STFR.FRQ (UWA) system meets the climatic requirements by maintaining the required coherence loss limits throughout the specified climate range.

The aims of these measurements were to demonstrate compliance of the SKA-LOW STFR.FRQ (UWA) system with the one-second (REQ-3242) and 60-second (REQ-3243) SKA coherence requirements under the following conditions:

- SAT.STFR.FRQ_REQ-105-075 — for all ambient temperatures of equipment located in the RPFs between 18°C and 26°C,
- SAT.STFR.FRQ_REQ-105-076 —for a range of non-condensing relative humidities in the RPFs between 40% and 60%,
- SAT.STFR.FRQ_REQ-105-079 — for all ambient temperatures of equipment located in the CPFs between 18°C and 26°C, and
- SAT.STFR.FRQ_REQ-105-080 — for a range of non-condensing relative humidities in the CPFs between 40% and 60%.

#### 4.1.1.10 Test(14) – SKA-LOW STFR.FRQ (UWA) v2 – Seismic Resilience

- SADT technical report: *Pre-CDR Laboratory Verification of UWA's SKA Synchronisation System* [RD13].

SKA subsystems are expected to continue to function at the required performance levels after a minor Earth tremor event. The tests and results presented below show that the SKA-LOW STFR.FRQ (UWA) system meets the requirements by demonstrating continued operation at the required coherence loss limits after experiencing ground accelerations up to 1 $m/s^2$.

The aims of these measurements were to demonstrate compliance of the SKA-LOW STFR.FRQ (UWA) system with the one-second (REQ-2268) and 60-second (REQ-2692) SKA coherence requirements after experiencing a seismic event resulting in a maximum peak ground acceleration of 1 $m/s^2$ (SADT.SAT.FRQ_REQ-2650-01-01).

#### 4.1.1.11 Test(16) – SKA-LOW STFR.FRQ (UWA) v2 – Electromagnetic Compatibility

- SADT technical report: *Pre-CDR Laboratory Verification of UWA's SKA Synchronisation System* [RD13].

The design considerations for the SKA aim to minimize the telescope's exposure and sensitivity to radio-frequency interference (RFI) that could impair the SKA's observational capabilities. This includes self-generated RFI, which is denoted as internal electro-magnetic interference (EMI). The tests and results presented below show that the EMI emitted by the SKA-LOW STFR.FRQ (UWA) system is within the designed limits and the system meet the SKA electromagnetic compatibility (EMC) requirements

The aims of this test were to demonstrate compliance of the SKA-LOW STFR.FRQ (UWA) system with the RFI/EMC requirements REQ-2462 and SADT.SAT.STFR.FRQ_REQ-2462-01-01 that SAT.STFR.FRQ components emitting electromagnetic radiation within frequency intervals for broad and narrow band cases shall be within the SKA RFI/EMI Threshold Levels as defined in SKA-TEL-SKO-0000202-AG-RFI-ST-01.

### 4.1.2 SKA-mid Design for Mass Manufacture Archetype

- SADT technical report: *Design-for-Manufacture of the SKA1-Mid Frequency Synchronisation System* [RD22].

The SKA1-mid frequency transfer system previously consisted of a field-deployable prototype (FDP). This prototype was used to test the performance of the system while maintaining the capacity for iteration through the use of discrete optical and electronic components. The performance of the system was astronomically verified through testing at the Australian Telescope Compact Array (ATCA) in 2017. As a result, the FDP can now progress to the DfM stage. The DfM process must work within the constraints placed on the SKA1-mid by its MW frequency reference to develop a miniaturised 'DfM archetype' of the SKA1-mid frequency transfer system. Throughout the system's development, the design will be continually tested to ensure the new system maintains the stability level established by the FDP.

This report outlines the development and testing of this DfM archetype. First it gives an overview of the relevant principles of frequency transfer, and describe the role of frequency transfer in the SKA. A summary of the previous work done at UWA on the SKA1-mid frequency transfer system will be given, including the successful astronomical verification of the FDP carried out earlier this year. The report then covers the DfM process. This design section will begin with the layout and redesign of the system on a schematic level, followed by the creation of several printed circuit boards (PCBs) to be incorporated into the encompassing DfM system for testing. The results of this testing is then be summarised, and the report concludes with an outline of the current state of the DfM archetype, as well as a discussion of further applications for the system.

## 4.2 Independent Assessment of Solution

The detailed design of UWA's *SKA phase synchronisation system* as described in a draft version of this report, as well in the nine journal papers and seven SADT reports provided in Section 2.6, was critically assesses by three independent domain expert in phase-synchronisation systems for radio telescopes arrays.

The primary purpose of the review was to conduct a critical assessment of the principles of UWA's *SKA phase synchronisation system* and comment on its suitability as a system for the SKA. The terms of reference were for the reviewers were to attempt to identify any weaknesses/risks/limitations with our system that we were either unaware of (in which case the detailed design may be able to make updated), or that only look like weaknesses/risks/limitations because we did not cover the material sufficiently well in the draft documentation (in which case we would simply look to improve the documentation).

The deadline for the reviews was Monday the 17th of July (two weeks prior to the submission of this document to the SADT office). No major issues were raised and no significant changes to our detailed design had to be implemented. However, this report was significantly updated in response to the many comments made by all three reviewers. In addition, some key suggestions were incorporated into recommendation and further work listed in Section 5.2. These three independent assessment helped us build confidence in our detailed design but also helped us improve our design so that we are ultimately able to deliver the best phase synchronisation system possible for the SKA telescope.

### 4.2.1 ASTRON Netherlands Institute for Radio Astronomy – Gijs Schoonderbeek

This review focused on two elements of UWA's *SKA phase synchronisation system*: The short term stability, dominated by the oven-controlled clean-up oscillator located within the Receiver Module at the antenna site; and the long term stability, which is driven by the quality of the control loop of the phase-synchronisation system. The report is attached as Appendix 7.10.1.

### 4.2.2 Jet Propulsion Laboratory – Larry D'Addario

The comments in this review concerned three main areas. Firstly, the review called for a budget of noise limits for the transfer performance, to help inform whether the performance could be improved, or to help identify any components that could be degraded without degrading the transfer performance (unfortunately, in the time available we were unable to conduct the detailed measurements required to support such a noise budget, but it is something we will certainly be doing outside the scope of the SKA in the future). The second request was to provide more detailed explanations of design choices relating to selection of specific signal frequencies and power levels of the various components within our system (notably the AOMs and DDS reference frequencies). And finally, the review called for a discussion on robustness of our system during distributions in normal operation (e.g. from power outages). The report is attached as Appendix 7.10.2.

### 4.2.3 Square Kilometre Array South Africa – Johan Burger

The comments from the final review can be classed into three main areas. It starts with comments relating to the minimising risk and complexity by locating complex equipment at the central site, and having only simple and robust equipment located at the antenna site. The next few comments relate the signal to noise budget of the entire system and how this impacts the dynamic range, with the view to increase robustness through an increase in operational dynamic range. Finally, the review called for more information relating to the key component of Microwave Shift, the dual-parallel Mach-Zehnder modulator. The report is attached as Appendix 7.10.3.

# 5 CONCLUSIONS/RECOMMENDATIONS

## 5.1 Conclusions

To enable the SKA to meet its goal of being a transformational telescope in the 50 MHz to 14 GHz frequency range, active stabilisation of the phase-coherent reference signals transmitted to the antenna sites is required to overcome the environmental disturbances acting to degrade the transmissions. In this report we have described the detailed design of UWA's *SKA phase synchronisation system* for the SKA1-low telescope developed by the University of Western Australia (UWA).

The solution is based on the successful stabilised frequency reference distribution system employed by the ALMA telescope, incorporating recent advances made by the international frequency metrology community, and innovations developed by researchers at UWA, optimised for the needs of the SKA1-low.

The system receives an electronic reference signal from SAT.CLOCKS at the SKA1-low Central Processing Facility (CPF), and transfers the full stability of the reference signal across the SAT network to each Remote Prcoseeing Facility (RPF). At the RPF, an electronic copy of the reference signal is provided to LFAA. The system is controlled and monitored using SAT.LMC with the required local infrastructure provided by SADT.LINFRA.

A complete solution description has been presented, detailing the design justification and all aspects of the detailed design from hardware selection to integration. These details have been used to develop a comprehensive cost model.

An evaluation of the design has been carried out, including an overview of the testing of the design and verification of its performance and compliance with the SKA design requirements. This evaluation is supported by independent assessments from domain experts at the ASTRON Netherlands Institute for Radio Astronomy, the Jet Propulsion Laboratory, and Square Kilometre Array South Africa. The design has been shown to be fully complaint with all SKA design requirements.

Remaining tests and work required to take this system forward into the SKA1 Construction Phase are detailed below in Section 5.2.

## 5.2 Recommendations and Further Work

### 5.2.1 AAVS1 Integration Field Trial

Previous astronomical verification work with ASKAP [RD19] and ATCA [RD18, 20] is described in Section 4.1.1. We have €14.4k of funding to continue SKA1-low integration tests with AAVS1. UWA and Curtin University have arranged joint MRO field trial. Our plan calls to integrate our system with the SKA1-low AAVS1.

### 5.2.2 Continue work on Design for Mass Manufacture

UWA researchers in partnership with MeerKAT and UoM electronic engineers, progressed the detailed designs described in this document, into a set of mass-manufacture prototypes, effectively getting a head-start at addressing manufacturing issues that may be encounter by contractors during SKA Construction. The first set of mass-manufacture prototypes s of the Transmitter Modules and Receiver Modules for SKA1-low were completed in Q2, 2016 [RD21]; and for SKA1-mid in Q1, 2017 [RD22]. However, these systems still comprise multiple PCBs that were developed independently to aid with rapid development and debugging. However, we now ready to move onto building the final archetypes for not just the Transmitter Modules and Receiver Modules, but the other bespoke elements of UWA's *SKA phase synchronisation system* (notably Microwave Shift). If the down-select is successful, we have secured €33.6k in funding (not including salaries and on-costs) that will become available for this task.

### 5.2.3 Employ a hybrid passive/active frequency transfer system

As UWA's *SKA phase synchronisation system* exceed the SKA1-mid Functional Performance Requirement by at least two orders of magnitude across most Normal Operating Conditions (see Section 3.3.2), one could consider a hybrid system that incorporates passive frequency transfer for the shorter links, and only uses our active stabilised system for the longest links. The obvious down-side is the additional complexity in procurement, operation and maintenance. UWA initially considered proposing such a system. However, our preliminary calculations showed that while there is some cost gains to be made in equipment capital expenditure, this is rapidly offset in labour capital expenditure, and then also across both equipment and labour operational expenditure.

### 5.2.4 Using SKA-mid system for both mid and low

As described in Section 3.2.2, we have designed a variant of UWA's *SKA phase synchronisation system* specifically optimised for SKA1-low, as well the variant for SKA1-mid described in this report. In Section 3.3.2, we show that the SKA1-mid variant exceed the SKA1-mid Functional Performance Requirement by at least two orders of magnitude across most Normal Operating Conditions. If the SKA1-mid variant were used as the phase synchronisation system for SKA1-low, it would exceed Functional Performance Requirement by another order of magnitude. This seems like extreme overkill; however, having one system for both telescopes would undoubtedly generate significant cost savings capital and operation expenditure (with the down-side of having to add the relatively complex Microwave Shift element to the SKA1-low system).

### 5.2.5 Reducing cost by ignoring the possibility of reflections

One of the key design parameters of UWA's *SKA phase synchronisation system* is designing for maximum robustness. To this end significant cost and effort has gone into ensuring the system will be immune from unwanted optical reflections that are inevitably present on real-world optical fibre links (see Section 3.3.5.1). Inexplicably, the need to be able to operate on anything less-than-ideal links was not formulated as a requirement, but achieving this capability requires the addition of two (of the total) thee AOMs per fibre link. Modifying our design to remove the two anti-reflections AOMs would cut 25% from the capital expenditure cost of the system.

## 5.3 Acknowledgements

A large amount of work has gone into the development of the detailed design of UWA's *SKA phase synchronisation system* presented in this report and the authors are indebted to the many people who have supported this work.

We would like to thank our University of Manchester/SADT consortium colleagues; Keith Grainge, Althea Wilkinson, Bassem Alachkar, Rob Gabrielczyk, Mike Pearson, Jill Hammond, Michelle Hussey, Samantha Lloyd, Richard Oberland, Paul Boven, and Richard Whitaker for their efforts and support. We would especially like to acknowledge the aid of Richard Whitaker in designing the PCB electronics, and Paul Boven for helping us to understand many of the SAT systems, concepts, and requirements.

We have received a great deal of support from the International Centre for Radio Astronomy Research and in particular would like to thank Peter Quinn, Lister Staveley-Smith, Andreas Wicenec, and Tom Booler. We are very grateful to Franz Schlagenhaufer for conducting the EMC test and compiling the EMC reports. We are also very grateful for the input provided by Richard Dodson and Maria Rioja.

We have also received a great deal of support from Ian McArthur and Jay Jay Jegathesan at the UWA School of Physics and Astrophysics. We are indebted to Michael Tobar, Eugene Ivanov, and the other members of the Frequency and Quantum Metrology group for lending us equipment, allowing us to raid their supplies, and discussing important metrology concepts. We also wish to thank Gesine Grosche for use of her group's dual-drive Mach-Zehnder modulator.

Through ICRAR and the School of Physics and Astrophysics we have had the assistance of a number of undergraduate students who have made significant contributions to the project. A huge thank you to Julian Bocking, Gavin Siow, Benjamin Courtney-Barrer, Ben Stone, and Maddy Sheard for the efforts they put into this project. Another very big thank you goes to Simon Stobie who has advanced this project both as an undergraduate research intern, and later by working on the microwave electronics for his honours project. Simon's efforts and contribution are evidenced by the fact he was awarded first-class honours for the work he completed in this project.

The authors wish to thank *AARNet* for the provision of light-level access to their fibre network infrastructure. This infrastructure has proven invaluable to the development and verification of UWA's design.

For our astronomical field trials with ATCA, we are very grateful for the help we received from staff at the CSIRO Paul Wild Observatory and CSIRO Astronomy and Space Science, especially Tasso Tzioumis for helping to arrange the visit. We would like to thank Mike Hill, Brett Lennon, Jock McFee, Peter Mirtschin, and Jamie Stevens for their efforts, without which these field trials would not have been a success. At the other end of the continent, we would like to thank Aidan Hotan, Maxim Voronkov, Suzy Jackson, and Mark Leach for their help and support during the ASKAP field trials at the Murchison Radio Observatory.

For our South African SKA site overhead fibre field trials, we are very grateful for all the local help we received from Bruce Wallace, Jaco Müller, Romeo Gamatham, Tim Gibbon, Roufurd Julie, and Johan Burger. Without their contribution, these outcomes would not have been possible. Thanks also to Charles Copley for providing the C-BASS weather station data. Also with SKA South Africa, we would like to thank Sias Malan for his mass-manufacture design help.

We are very grateful to Larry D'Addario from JPL, Gijs Schooderbeek from ASTRON, and Johan Burger from SKA South Africa for taking the time to review this solution and an early draft of this report.

Apologies to anyone who should be on this list, but we have accidently neglected to mention.

This report describes work being carried out for the SKA Signal and Data Transport (SaDT) consortium as part of the Square Kilometre Array (SKA) project. The SKA project is an international effort to build the world's largest radio telescope, led by the SKA Organisation with the support of 10 member countries.

# 6 STATEMENT OF COMPLIANCE

We confirm that the UWA's SKA Phase Synchronisation System is compliant with all the relevant SKA Level 1 (rev 10) Requirements and associated ECPs. Detailed compliance is shown in the table below for all functional and performance requirements. Compliance to other requirements may be found in the SADT Compliance Matrix.

Furthermore, how the UWA solution complies with the requirements related to manufacturability and extensibility to SKA2 (SKA1-SYS_REQ-2433, SKA1-SYS_REQ-2462, SKA1-SYS_REQ-2559, SKA1-SYS_REQ-2562, SKA1-SYS_REQ-2594, SKA1-SYS_REQ 2599) are discussed in the *Design-for-Manufacture of the SKA1-Mid Frequency Synchronisation System* report [RD22] (Appendix 7.3.7).

## 6.1 Functional Performance Requirements

The SKA1-low functional performance requirements, as defined in [AD1], are listed in Table 15.

**Table 15 – SKA1-low Functional Performance Requirements**

| Requirement # | Requirement Description | Compliance | Reference |
|---|---|---|---|
| • Coherence requirements | | | |
| SADT.SAT.STFR.FRQ_REQ-3242 | SAT.STFR.FRQ shall distribute a frequency reference with no more than 1.9% maximum coherence loss, within a maximum integration period of 1 second, and up to an operating frequency of 350 MHz. | Compliant | SADT Rep. **380** (2017) §3.3.2.1<br>Submitted PTL (2017) [RD11]<br>SADT Rep. **620** (2017) §3.4.1<br>SADT Rep. **617** (2015) §5.4 |
| SADT.SAT.STFR.FRQ_REQ-3243 | SAT.STFR.FRQ shall distribute a frequency reference with no more than 1.9% maximum coherence loss, within a maximum integration period of 1 minute, and up to an operating frequency of 350 MHz. | Compliant | SADT Rep. **380** (2017) §3.3.2.1<br>Submitted PTL (2017) [RD11]<br>SADT Rep. **620** (2017) §3.4.1<br>SADT Rep. **617** (2015) §5.4 |
| • Phase drift requirement | | | |
| SADT.SAT.STFR.FRQ_REQ-3244 | SAT.STFR.FRQ shall distribute a reference frequency to a performance allowing a maximum of 1 radian phase drift for intervals up to 10 minutes, and up to an operating frequency of 350 MHz. | Compliant | SADT Rep. **380** (2017) §3.3.2.2<br>SADT Rep. **620** (2017) §3.4.2<br>SADT Rep. **617** (2015) §5.4 |
| • Jitter requirement | | | |
| EICD *SADTtoLFAA_ICD* | Jitter shall be equal to or less than 74 femtoseconds for SKA1-low as defined by EICD *100-0000000-026_03-SADTtoLFAA_ICD*. | Compliant | SADT Rep. **380** (2017) §3.3.2.3<br>SADT Rep. **620** (2017) §3.4.3<br>SADT Rep. **617** (2015) §5.5 |

## 6.2 Normal Operating Conditions

The SKA1-low normal operating conditions, as defined in [AD1], are divided into two classes; with the environmental conditions listed in Table 16 and design conditions in Table 17.

**Table 16 – SKA1-low Environmental Conditions**

| Requirement # | Requirement Description | Compliance | Reference |
|---|---|---|---|
| • Ambient Temperature and Humidity Requirements | | | |
| SADT.SAT.STFR.FRQ_REQ-105-075 | SAT.STFR.FRQ components sited within the Remote Processing Facility (RPF) shall withstand, and under normal operating conditions operate within specification, a fluctuating thermal environment between +18°C and +26°C. | Compliant | SADT Rep. **380** (2017) §3.3.3.1.1<br>SADT Rep. **620** (2017) §5.4.1 [RD13] |
| SADT.SAT.STFR.FRQ_REQ-105-076 | SAT.STFR.FRQ components sited within the Remote Processing Facility (RPF) shall withstand, and operate within, a fluctuating non-condensing, relative humidity environment between 40% and 60%. | Compliant | SADT Rep. **380** (2017) §3.3.3.1.1<br>SADT Rep. **620** (2017) §5.4.2 [RD13] |
| SADT.SAT.STFR.FRQ_REQ-105-079 | SAT.STFR.FRQ components sited within the Central Processing Facility (CPF) shall withstand, and under normal operating conditions operate within specification, a fluctuating thermal environment between +18°C and +26°C. | Compliant | SADT Rep. **380** (2017) §3.3.3.1.2<br>SADT Rep. **620** (2017) §5.4.1 [RD13] |
| SADT.SAT.STFR.FRQ_REQ-105-080 | SAT.STFR.FRQ components sited within the Central Processing Facility (CPF) shall withstand, and operate within, a fluctuating non-condensing, relative humidity environment between 40% and 60%. | Compliant | SADT Rep. **380** (2017) §3.3.3.1.2<br>SADT Rep. **620** (2017) §5.4.2 [RD13] |
| SADT.SAT.STFR.FRQ_REQ-3070 | SAT.STFR.FRQ equipment and fibre located in non-weather protected locations shall be sufficiently environmentally protected to survive, and perform to specification for all ambient temperatures of between -5°C and +50°C. | Compliant | SADT Rep. **380** (2017) §3.3.3.1.3<br>SADT Rep. **620** (2017) §5.4 [RD13] |
| SADT.SAT.STFR.FRQ_REQ-3070 | SAT.STFR.FRQ equipment and fibre located in non-weather protected locations shall be sufficiently environmentally protected to survive, and perform to specification for rates of change of ambient temperature of up to ±3°C every 10 minutes. | Compliant | SADT Rep. **380** (2017) §3.3.3.1.3<br>SADT Rep. **620** (2017) §5.4 [RD13] |
| • Wind Speed Requirement | | | |
| SADT.SAT.STFR.FRQ_REQ-3070 | SAT.STFR.FRQ equipment and fibre located in non-weather protected locations shall be sufficiently environmentally protected to survive and perform to specification under normal SKA telescope operating wind conditions up to wind speeds of 40km/hr. | Compliant | SADT Rep. **380** (2017) §3.3.3.2<br>Submitted TUFFC (2017) [RD16]<br>SADT Rep. **109** (2015) §2.5 [RD14] |
| • Seismic Resilience Requirement | | | |
| SADT.SAT.STFR.FRQ_REQ-2650 | SAT.STFR.FRQ components shall be fully operational subsequent to seismic events resulting in a maximum instantaneous peak ground acceleration of 1 $m/s^2$. Note: Seismic events include underground collapses in addition to earthquakes. | Compliant | SADT Rep. **380** (2017) §3.3.3.3<br>SADT Rep. **620** (2017) §7.4 [RD13] |

**Table 17 – SKA1-low Design Conditions**

| Requirement # | Requirement Description | Compliance | Reference |
|---|---|---|---|
| • Telescope Configuration Requirement | | | |
| SADT.SAT.STRF.FRQ_REQ-2142 | SAT.STFR.FRQ shall disseminate the LOW Reference Frequency (the "Disseminated Reference Frequency") to 36 Remote Processing Facilities (RPFs) located on the LOW Spiral Arms as defined by *SKA-TEL-SKO-0000422 SKA1_Low Configuration Coordinates*. | Compliant | SADT Rep. **380** (2017) §3.3.3.4<br>In Prep. AJ (2017) [RD5] |

## 6.3 Key Additional Requirements

The remaining key additional requirements for SKA1-low are given in Table 18.

**Table 18 – SKA1-low Key Additional Requirements**

| Requirement # | Requirement Description | | |
|---|---|---|---|
| • Monitoring Requirement | | | |
| SADT.SAT.STFR.FRQ_REQ-2280 | At least the following STFR component parameters shall be monitored: the Lock signal (indicating that the STFR system is functioning correctly); the Control Voltage (giving an indication of how much control is still available to keep the STFR locked); and the Phase measurement (showing the corrections which have been applied to the frequency to compensate for the effects of changes in the fibre connecting the transmit and receive units of the STFR.FRQ system). | Compliant | SADT Rep. **380** (2017) §3.3.4.1<br>SADT Rep. **380** (2017) §3.5.2 |
| • Radio Frequency Interference Requirements | | | |
| SADT.SAT.STFR.FRQ_REQ-2462 | SAT.STFR.FRQ components emitting electromagnetic radiation within frequency intervals for broad and narrow band cases shall be within the SKA RFI/EMI Threshold Levels as defined in *SKA-TEL-SKO-0000202-AG-RFI-ST-01*. | Compliant | SADT Rep. **380** (2017) §3.3.4.2<br>SADT Rep. **620** (2017) §9.4 [RD13] |
| • Space Requirements | | | |
| SADT NWA Model SKA-low Rev. 03 | The Candidate's solution shall meet the following maximum space requirements as defined by the space allocated to SAT.STFR.FRQ by the SADT NWA model *SKA-TEL-SADT-0000522-MOD_NWAModelLow Revision 3.0*. | Compliant | SADT Rep. **380** (2017) §3.3.4.3<br>SADT Rep. **380** (2017) §3.4.1 |
| • Availability Requirement | | | |
| SADT.SAT.STFR.FRQ_REQ-3245 | SAT.STFR.FRQ (end-to-end system excluding fibre) shall have 99.9% "Inherent Availability." | Compliant | SADT Rep. **380** (2017) §3.3.4.4 |
| • Power Requirements | | | |
| SADT.SAT.STFR.FRQ_REQ-105-148 | The total power consumption of combined SAT.STFR.FRQ components located in Central Processing Facility (CPF) (LOW) shall be no more than 1.2 kW. | Compliant | SADT Rep. **380** (2017) §3.3.4.5<br>Appendix 7.9.2<br>SADT Rep. **522** (2017) 'CPF' tab [RD26] |
| SADT.SAT.STFR.FRQ_REQ-105-149 | The total power consumption of combined SAT.STFR.FRQ components located in each Remote Processing Facility (RPF) (LOW) shall be no more than 70 Watts. | Compliant | SADT Rep. **380** (2017) §3.3.4.5<br>Appendix 7.9.2<br>SADT Rep. **522** (2017) 'RPF' tab[RD26] |

# 7 APPENDICES

The appendices include information relevant to UWA's *SKA phase synchronisation system* both SKA1-low and SKA1-mid, to maintain consistency between this document, and the equivalent detailed design report for SKA1-mid [RD30].

## 7.1 Concept Documents

### 7.1.1 Concept document – Time and Frequency Dissemination for the Square Kilometre Array

Appendices\Concept document - Time and Frequency Dissemination for the Square Kilometre Array.pdf

### 7.1.2 Concept document – Transfer of microwave-frequency reference signals over optical fibre links

Appendices\Concept document - Transfer of microwave-frequency reference signals over optical fibre links.pdf

## 7.2 Journal Papers

### 7.2.1 Journal paper – Square Kilometre Array: The Radio Telescope of the XXI Century

Appendices\Journal paper - Square Kilometre Array The Radio Telescope of the XXI Century.pdf [RD4].

### 7.2.2 Journal paper – A Clock for the Square Kilometre Array

Appendices\Journal paper - A Clock for the Square Kilometre Array.pdf [RD23].

### 7.2.3 Journal paper – A Phase Synchronization System for the Square Kilometre Array

Appendices\Journal paper - A Phase Synchronization System for the Square Kilometre Array.pdf [RD5].

### 7.2.4 Journal paper – Simultaneous transfer of stabilized optical and microwave frequencies over fiber

Appendices\Journal paper - Simultaneous transfer of stabilized optical and microwave frequencies over fiber.pdf [RD24].

### 7.2.5 Journal paper – Simple Stabilized Radio-Frequency Transfer with Optical Phase Actuation

Appendices\Journal paper - Simple Stabilized Radio-Frequency Transfer with Optical Phase Actuation.pdf [RD11].

### 7.2.6 Journal paper – Stabilized microwave-frequency transfer using optical phase sensing and actuation

Appendices\Journal paper - Stabilized microwave-frequency transfer using optical phase sensing and actuation.pdf [RD10].

### 7.2.7 Journal paper – Characterization of Optical Frequency Transfer over 154 km of Aerial Fiber

Appendices\Journal paper - Characterization of optical frequency transfer over 154 km of aerial fiber.pdf [RD15].

### 7.2.8 Journal paper – Stabilized Modulated Photonic Signal Transfer Over 186 km of Aerial Fiber

Appendices\Journal paper - Stabilized Modulated Photonic Signal Transfer Over 186 km of Aerial Fiber.pdf [RD16].

### 7.2.9 Journal paper – Astronomical verification of a stabilized frequency reference transfer system for the Square Kilometre Array

Appendices\Journal paper - Astronomical verification of a stabilized frequency reference transfer system for the SKA.pdf [RD17].

## 7.3 SADT Reports

### 7.3.1 SADT report – Pre-PDR Laboratory Verification of UWA's SKA Synchronisation System

Appendices\SKA-TEL-SADT-0000616 - Pre-PDR Laboratory Verification of UWAs SKA Synchronisation System.pdf [RD12].

### 7.3.2 SADT report – Pre-CDR Laboratory Verification of UWA's SKA Synchronisation System

Appendices\SKA-TEL-SADT-0000620 - Pre-CDR Laboratory Verification of UWAs SKA Synchronisation System.pdf [RD13].

### 7.3.3 SADT report – UWA South African SKA Site Long-Haul Overhead Fibre Field Trial Report

Appendices\SKA-TEL-SADT-0000109 - UWA South African SKA Site Long-Haul Overhead Fibre Field Trial Report.pdf [RD14].

### 7.3.4 SADT report – SKA-low Astronomical Verification

Appendices\SKA-TEL-SADT-0000617 - SKA-low Astronomical Verification.pdf [RD19].

### 7.3.5 SADT report – SKA-mid Astronomical Verification

Appendices\SKA-TEL-SADT-0000524 - SKA-mid Astronomical Verification.pdf [RD18].

### 7.3.6 SADT report – Notes on Calculating the Relationship between Coherence Loss and Allan Deviation

Appendices\SKA-TEL-SADT-0000619 - Notes on Calculating the Relationship between Coherence Loss and Allan Deviation.pdf [RD25].

### 7.3.7 SADT report – Design-for-Manufacture of the SKA1-Mid Frequency Synchronisation System

Appendices\SKA-TEL-SADT-0000618 - Design-for-Manufacture of the SKA1-Mid Frequency Synchronisation System.pdf [RD22].

## 7.4 Detailed Design Overview Files

### 7.4.1 SKA-mid Detailed Design – Overview

Appendices\SKA-mid Detailed Design - Overview.pptx

### 7.4.2 SKA-low Detailed Design – Overview

Appendices\SKA-low Detailed Design - Overview.pptx

## 7.5 Mechanical Detailed Design Files

### 7.5.1 SKA-mid Detailed Design – Solid Edge computer aided design files

Appendices\SKA-mid Detailed Design - Solid Edge computer aided design files.zip

### 7.5.2 SKA-low Detailed Design – Solid Edge computer aided design files

Appendices\SKA-low Detailed Design - Solid Edge computer aided design files.zip

## 7.6 Optical Detailed Design Files

### 7.6.1 SKA-mid Detailed Design – Optical Schematics

Appendices\SKA-mid Detailed Design - Optical Schematics.pptx

### 7.6.2 SKA-low Detailed Design – Optical Schematics

Appendices\SKA-low Detailed Design - Optical Schematics.pptx

## 7.7 Electronic Detailed Design Files

### 7.7.1 SKA-mid Detailed Design – Electronic Schematics

Appendices\SKA-mid Detailed Design - Electronic Schematics.pptx

### 7.7.2 SKA-low Detailed Design – Electronic Schematics

Appendices\SKA-low Detailed Design - Electronic Schematics.pptx

### 7.7.3 SKA-mid Detailed Design – Design Spark circuit schematic and PCB layout files

Appendices\SKA-mid Detailed Design - Design Spark circuit schematic and PCB layout files.zip

### 7.7.4 SKA-low Detailed Design – Design Spark circuit schematic and PCB layout files

Appendices\SKA-low Detailed Design - Design Spark circuit schematic and PCB layout files.zip

## 7.8 Modelling Software

### 7.8.1 Modelling software – Java interactive tool for modelling frequencies in optical fibre networks

Appendices\Modelling software - Java interactive tool for modelling frequencies in optical fibre networks.zip

### 7.8.2 Modelling software – SKA Excel Frequency Calculator

Appendices\Modelling software - SKA Excel Frequency Calculator.xlsx

## 7.9 Bill of Materials and Cost Model Database

### 7.9.1 SKA-mid Detailed Cost Model – Capex Equipment

Appendices\SKA-mid Detailed Cost Model - Capex Equipment.xlsx

### 7.9.2 SKA-low Detailed Cost Model – Capex Equipment

Appendices\SKA-low Detailed Cost Model - Capex Equipment.xlsx

### 7.9.3 SKA-mid Detailed Cost Model – Capex Labour

Appendices\SKA-mid Detailed Cost Model - Capex Labour.xlsx

### 7.9.4 SKA-low Detailed Cost Model – Capex Labour

Appendices\SKA-low Detailed Cost Model - Capex Labour.xlsx

### 7.9.5 Light Touch Solutions – Capex Labour Costing Analysis report

Appendices\SKA-mid and SKA-low Labour cost review.pdf

## 7.10 Independent Assessment of Solution

### 7.10.1 ASTRON Netherlands Institute for Radio Astronomy – Gijs Schoonderbeek

Appendices\Independent Assessment - ASTRON Netherlands Institute for Radio Astronomy.pdf

### 7.10.2 Jet Propulsion Laboratory – Larry D'Addario

Appendices\Independent Assessment - Jet Propulsion Laboratory.pdf

### 7.10.3 Square Kilometre Array South Africa – Johan Burger

Appendices\Independent Assessment - Square Kilometre Array South Africa.pdf